\documentclass[aps,prb,reprint,superscriptaddress,eqsecnum,citeautoscript]{revtex4-1}
\usepackage{amsmath,amssymb,graphicx,color}

\usepackage[colorlinks=true,linkcolor=black,citecolor=blue,urlcolor=blue]{hyperref}

\newcommand{\vc}[1]{\boldsymbol{#1}}

\begin{document}
\title{Spin-orbit-entangled electronic phases in 4$d$ and 5$d$ transition-metal compounds}
\author{Tomohiro Takayama}
\affiliation{Max Planck Institute for Solid State Research, Heisenbergstrasse
  1, 70569 Stuttgart, Germany}
\affiliation{Institute for Functional Matter and Quantum Technologies, University of Stuttgart, Pfaffenwaldring 57, 70550 Stuttgart, Germany}
\author{Ji\ifmmode \check{r}\else \v{r}\fi{}\'{\i} Chaloupka}
\affiliation{Department of Condensed Matter Physics, Masaryk University, Brno, Czech Republic}
\affiliation{Central European Institute of Technology, Masaryk University, Brno, Czech Republic}
\author{Andrew Smerald}
\affiliation{Max Planck Institute for Solid State Research, Heisenbergstrasse
  1, 70569 Stuttgart, Germany}
\author{Giniyat Khaliullin}
\affiliation{Max Planck Institute for Solid State Research, Heisenbergstrasse
  1, 70569 Stuttgart, Germany}
\author{Hidenori Takagi}
\affiliation{Max Planck Institute for Solid State Research, Heisenbergstrasse
  1, 70569 Stuttgart, Germany}
\affiliation{Institute for Functional Matter and Quantum Technologies, University of Stuttgart, Pfaffenwaldring 57, 70550 Stuttgart, Germany}
\affiliation{Department of Physics,
  University of Tokyo, 7-3-1 Hongo, Tokyo 113-0033, Japan}

\date{\today}
\begin{abstract}
Complex oxides with $4d$ and $5d$ transition-metal ions recently emerged as a new paradigm in correlated electron physics, due to the interplay between spin-orbit coupling and electron interactions. For $4d$ and $5d$ ions, the spin-orbit coupling, $\zeta$, can be as large as 0.2-0.4 eV, which is comparable with and often exceeds other relevant parameters such as Hund's coupling $J_{\rm H}$, noncubic crystal field splitting $\Delta$, and the electron hopping amplitude $t$. This gives rise to a variety of spin-orbit-entangled degrees of freedom and, crucially, non-trivial interactions between them that depend on the $d$-electron configuration, the chemical bonding, and the lattice geometry. Exotic electronic phases often emerge, including spin-orbit assisted Mott insulators, quantum spin liquids, excitonic magnetism, multipolar orderings and correlated topological semimetals. This paper provides a selective overview of some of the most interesting spin-orbit-entangled phases that arise in $4d$ and $5d$ transition-metal compounds.
\end{abstract}

\maketitle

\section{Introduction}

In the 1960's and 70's, correlated-electron physics in transition-metal oxides was already an active field of research, and a major topic in condensed matter science. The basic picture of spin and orbital ordering and the interplay between them was unveiled during this period, and collected into the Kanamori-Goodenough rules. However, the exploration of exotic electronic phases beyond conventional magnetic ordering was stymied by a lack of materials and theoretical tools.  

In 1986, high-$T_{\rm c}$ superconductivity was discovered in the layered 3$d$ Cu oxides, which accelerated both experimentally and theoretically the search for novel spin-charge-orbital coupled phenomena produced by electron correlations. The major arena of such exploration was complex oxides with 3$d$ transition metal ions from Ti to Cu,  which led to the discoveries of unconventional superconductivity, colossal magneto-resistance, multiferroics and exotic spin-charge-orbital orderings. 4$d$ and 5$d$ transition metal oxides were also studied, but not as extensively as 3$d$ transition metal oxides, except for 4$d$ Sr$_2$RuO$_4$, where possible $p$-wave superconductivity was discussed. This was at least partially due to the less prominent effect of correlations in 4$d$ and 5$d$, which arises from the wavefunctions being more spatially extended than in 3$d$.

In the late 2000's, the layered perovskite Sr$_2$IrO$_4$ was identified as a weak Mott insulator, and the crucial role of spin-orbit coupling (SOC) in stabilizing the Mott state awoke a growing interest in 5$d$ Ir oxides and other 4$d$ and 5$d$ compounds. 
For Ir$^{4+}$ ions, there are five 5$d$-electrons, which reside in the $t_{2g}$ manifold, and therefore have an effective orbital moment $L$ = 1 and form a spin-orbit-entangled $J$ = 1/2 pseudospin.
These $J$ = 1/2 pseudospins behave in some ways like $S$ = 1/2 spins, but have an internal spin-orbital texture where the up/down spin states reside on different orbitals.
The resulting spin-orbital entanglement gives rise to non-trivial interactions between the $J$ = 1/2 pseudospins, and this led to the proposal that the Kitaev model, with bond-dependent Ising interactions, can be realized on edge-shared honeycomb networks of $J$ = 1/2 pseudospins.
In consequence, $J$ = 1/2 honeycomb magnets made out of 5$d^5$ Ir$^{4+}$ and 4$d^5$ Ru$^{3+}$ ions have been extensively studied over the last ten years, in an effort to discover the expected quantum spin-liquid state and associated Majorana fermions. 

Despite the considerable excitement, the $d^5$ $J$ = 1/2 Mott state is not the only spin-orbit-entangled state of interest among the many 4$d$ and 5$d$ transition metal compounds. With different filling of $d$-orbitals and different local structures, a rich variety of spin-orbit-entangled states can be formed, which are characterized not only by dipolar moments but also by multipoles such as quadrupolar and octupolar moments.  Exotic states of such spin-orbit-entangled matter can be anticipated, reflecting the internal spin-orbital texture and the lattice symmetry, and including excitonic magnetism and multipolar liquids. 

4$d$ and 5$d$ transition metal compounds also form interesting itinerant states of matter, with one prominent example being the topological semimetal.
This arises from the interplay of lattice symmetry and SOC, and, in contrast to typical topological semimetals, 4$d$ and 5$d$ semimetals tend to also have strong electron correlations.
Thus they provide an arena in which to study the overlap between topological physics and strong correlation.

Spin-orbit-entangled phases are also formed in 4$f$ and 5$f$ electron systems, which have been extensively studied, and it is worth spelling out what makes 4$d$ and 5$d$ electron systems distinct. 
One obvious difference is that they interact through exchange processes with a much larger energy scale, making them more accessible to experiment, and thus increasing the variety of phenomena that can be effectively probed.
Also, the spin-orbit-entangled $J$ states in 4$d$ and 5$d$ systems are much less localized than those of 4$f$, and can often be itinerant, opening up, for example, the exploration of correlated topological semimetals, and SOC driven exotic states formed near metal-insulator transitions.
The $d$- and the $f$- electron systems thus clearly play a complementary role in the exploration of spin-orbit-entangled phases.

This review is intended to provide readers with a broad perspective on the emerging plethora of 4$d$ and 5$d$ transition metal oxides and related compounds. We would like to address two basic questions: 1) What kind of exotic phases of spin-orbit-entangled matter are expected in 4$d$ and 5$d$ compounds? 2) To what extent are the proposed concepts realized? We limit our discussion to the compounds with octahedrally coordinated 4$d$ and 5$d$ transition metal ions accommodating less than 6 electrons in their $t_{2g}$ orbitals (low-spin configuration), where the effect of the large SOC is prominent due to the smaller crystal field splitting of $t_{2g}$ orbitals as compared to $e_g$.  As there are many reviews of pseudospin-1/2 $d^5$ Mott insulators, here we will discuss spin-orbit-entangled states in 4$d$ and 5$d$ transition metal compounds from a broader perspective, covering, in addition to $d^5$ compounds, $d^1$, $d^2$ and $d^4$ insulators, as well as itinerant systems with strong SOC.

\section{Concept of spin-orbit-entangled states and materials overview}
\label{sec:concept}
%{{{1

In Mott insulators, charge fluctuations are frozen, and the low-energy physics is
driven by the spin and orbital degrees of freedom of the constituent ions. In
$3d$ compounds, orbital degeneracy and hence orbital magnetism is largely
quenched by noncubic crystal fields and the Jahn-Teller (JT) mechanism, and magnetic
moments are predominantly of spin origin (with some exceptions mentioned
below). The large SOC in $4d$ and $5d$ transition metal ions competes with and
may dominate over crystal field splitting and JT effects, and the revived orbital magnetism becomes a source
of unusual interactions and exotic phases. We first discuss the spin-orbital
structure of the low-energy states of transition metal ions in a most
common, cubic crystal field environment, and then proceed to their
interactions and collective behavior.

\subsection{Spin, orbital, and pseudospin moments in Mott insulators}
%{{{2

\begin{figure*}[tb]
\begin{center}
\includegraphics[scale=0.95]{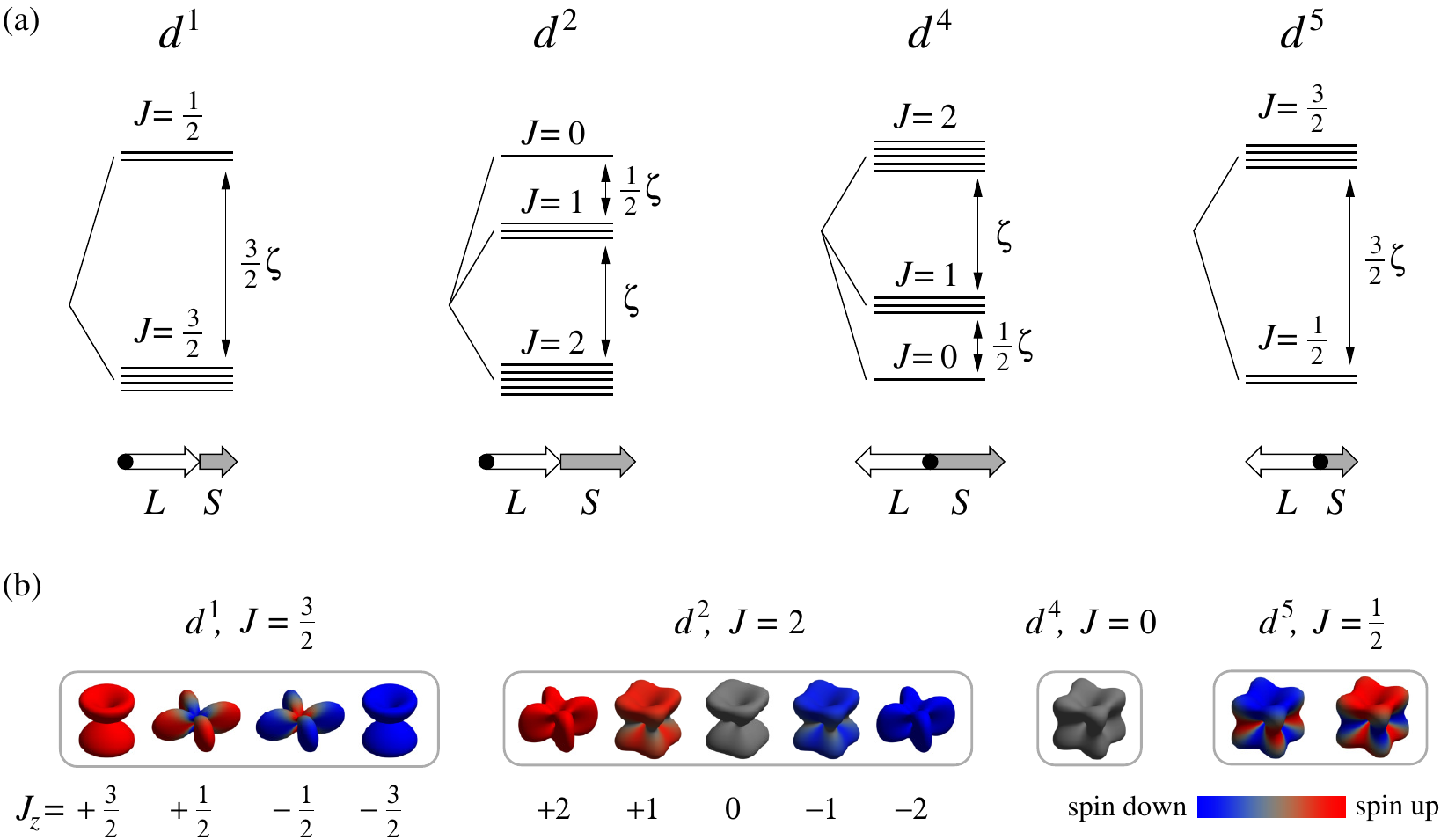}
\end{center}
\caption{
(a) Low-energy levels of $d^1$, $d^2$, $d^4$, and $d^5$ ions in cubic crystal
field. 
The degeneracy of the levels is shown by the number of close lines.
For less than half-filled $t_{2g}$ shell, the SOC aligns the effective orbital
angular momentum $L$ and spin $S$ to form larger total angular momentum:
$J=3/2$ quartet in $d^1$ case and $J=2$ quintuplet in $d^2$ case,
respectively. In the case of more than half-filled $t_{2g}$ shell, $L$ and
$S$ are antialigned, leading to $J=0$ singlet ground state for the $d^4$
configuration while the $d^5$ one hosts pseudospin $J=1/2$.
(b) Orbital shapes corresponding to the ground-state $J$-levels. Only the angular distribution of the electron density is considered. It is represented by
a surface plot where the distance to the origin is proportional to the integral
density in the corresponding direction.  The color of the surface indicates 
normalized spin polarization
$(\rho_\uparrow-\rho_\downarrow)/(\rho_\uparrow+\rho_\downarrow)$ 
taking values in the range $[-1,+1]$. It is shown for electrons in the case of
$d^1$ and $d^2$ states and for the holes in the $t_{2g}^6$ configuration in
the case of $d^{4}$ and $d^5$ states.
}
\label{fig:d1245}
\end{figure*}

The valence wavefunctions of $4d$ and $5d$ ions are spatially extended and
the Hund's coupling is smaller than the cubic crystal field splitting, $10Dq$, between
$t_{2g}$ and $e_g$ orbitals. Low-spin ground states
are therefore stabilized for $4d$ and $5d$ electron configurations more frequently
than in the $3d$ case, where the Hund's coupling typically overcomes the
$t_{2g}$-$e_g$ crystal field splitting. In the low-spin state, electrons occupy 
$t_{2g}$ orbitals forming total spin-$1/2$ ($d^1$, $d^5$), spin-$1$ ($d^2$, $d^4$), and spin-$3/2$ ($d^3$). In all of them (except $d^3$
orbital singlet not discussed here), the orbital sector is threefold degenerate,
formally isomorphic to the $p$-orbital degeneracy, and can thus be described in
terms of an effective orbital angular momentum $L=1$. 
(The effective orbital angular momentum is often distinguished by a special
mark, see, e.g., the ``ficticious angular momentum''
$\tilde{l}$ in the textbook by Abragam and Bleaney \cite{Abr70}, but here we
conveniently choose a simpler notation $L$.)
By a direct calculation of matrix elements of the physical orbital momentum $L_d=2$ of $d$-electrons
within the $t_{2g}$ manifold, one finds a relation $\vc L_d=-\vc L$ between
the angular momentum operators.
By employing the effective $\vc L$ operator, the SOC reads as 
$H=\mp\lambda \vc S\vc L$, 
where the negative (positive) sign refers to a less (more) than
half-filled $t_{2g}$ shell of $d^1$, $d^2$ ($d^4$, $d^5$) configurations, and
$\lambda$ is related to the single-electron SOC strength, $\zeta$, via
$\lambda=\zeta/2S$. The resulting spin-orbital levels are shown in
Fig.~\ref{fig:d1245}(a). Apparently, SOC breaks particle-hole symmetry:
due to the above sign change, the levels are mutually inverted within the
pairs of complementary electron/hole configurations such as $d^1$/$d^5$ and
$d^2$/$d^4$. The ground states thus have completely different total angular
momentum $\vc J=\vc S+\vc L$. Its constituents $\vc S$ and $\vc L$ align in
parallel (antiparallel) fashion for the less (more) than half-filled case.  The
corresponding ``shapes'' of the ground-state electron densities are depicted
in Fig.~\ref{fig:d1245}(b). Their nonuniform spin polarization with a coherent mixture of spin-up and down densities clearly
shows the coupling between the spin and the orbital motion of electrons.

%- FIGURE -----------------------------------------------------------------
%--------------------------------------------------------------------------

The observed variety of ionic ground states among $d^n$ configurations brings
about a distinct physics for each of the $d^n$ ions. $d^1$ and $d^2$ ions with $J > 1/2$
host higher order magnetic multipoles, and may lead to unconventional
high-rank order parameters ``hidden'' magnetically. Nominally nonmagnetic $J = 0$ 
$d^4$ ions may develop unusual magnetism due to condensation of
the excited $J=1$ level. $d^5$ ions with Kramers doublet $J = 1/2$ are formally akin to
spin one-half quantum magnets which are of special interest in the context of
various quantum ground states.

When evaluating the magnetic properties, the orbital component of the magnetic
moment needs to be incorporated. In terms of the effective $\vc L$, the magnetic
moment operator reads as $\vc M=2\vc S-\vc L$. In principle, $\vc L$ comes
with the so-called covalency factor $\kappa$ but we omit it for simplicity. For
$d^1$ with parallel $L=1$ and $S=1/2$, one has $L=2S$ and thus $M=0$, i.e.
the $J=3/2$ quartet has zero $g$-factor and is thus nonmagnetic. In reality,
$\kappa<1$ makes the compensation only partial, resulting in a small magnetic
moment. For $d^2$ with $S=L=J/2$, one finds $\vc M=(1/2)\vc J$, i.e. $g=1/2$.
$d^4$ with $J=0$ is a nonmagnetic singlet. $d^5$ with antiparallel $S=1/2$ and
$L=1$ has $g$-factor $g=-2$, i.e.  it is of opposite sign relative to the electron
$g$-factor. The above $g$-factors strongly deviating from pure spin $g=2$ are
the fingerprints of large orbital contribution to magnetism. In the $d^5$ case, e.g., one finds
that $\vc S=(-1/3)\vc J$ only, while $\vc L=(4/3)\vc J$; that
is, magnetism of $d^5$ compounds is predominantly of orbital origin.

%- FIGURE -----------------------------------------------------------------
\begin{figure}[tb]
\begin{center}
\includegraphics[scale=0.95]{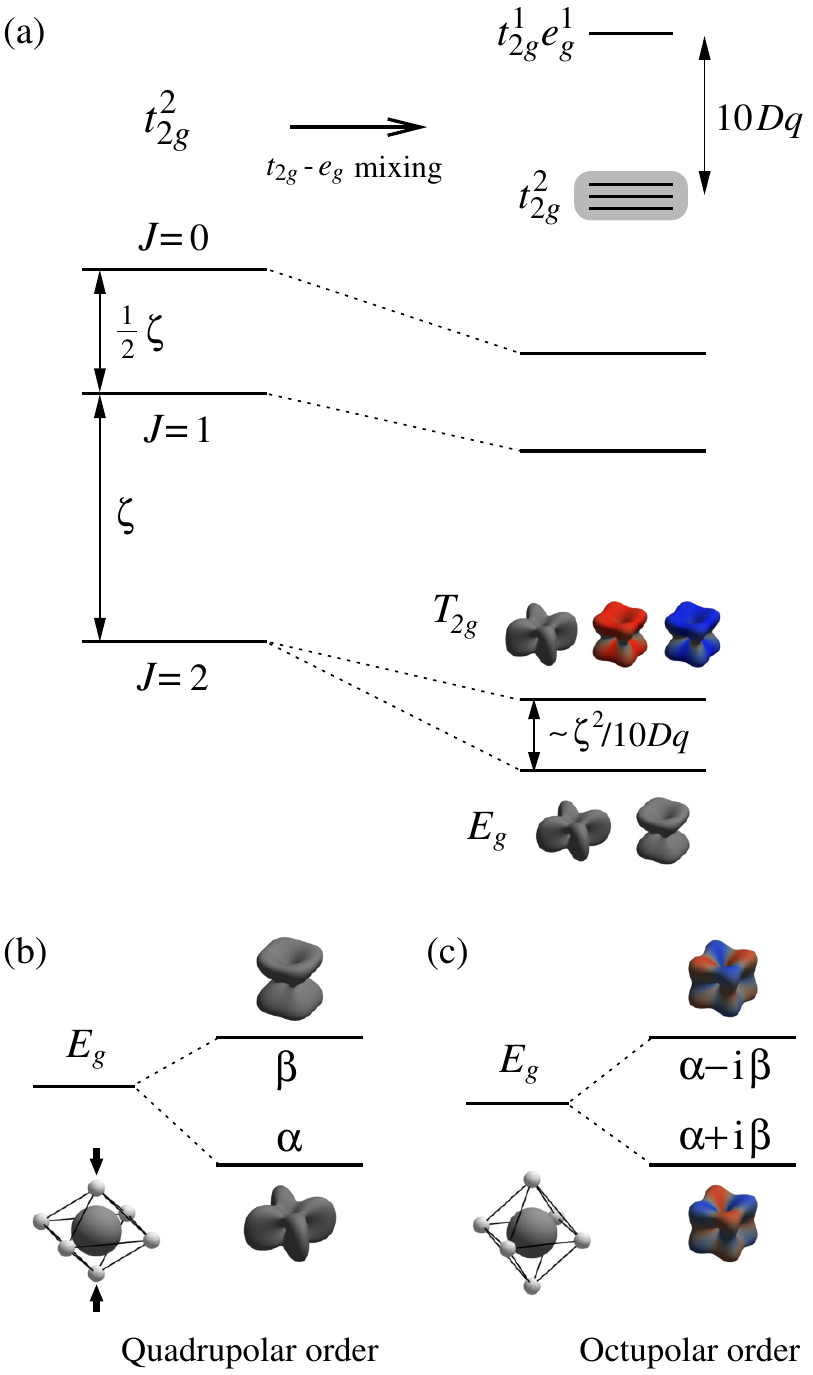}
\caption{
(a)~Shifts and splitting of the $J=0,1,2$ levels of $d^2$ ion when considering
mixing of the ground state $t_{2g}^2$ configuration with $t_{2g}^1e_g^1$ states by virtue of SOC.
Focusing on the lowest levels, we that find the originally five-fold degenerate
$J=2$ states split into an $E_g$ doublet and $T_{2g}$ triplet. Evaluated perturbatively for
$\zeta\ll 10Dq$, the splitting comes out proportional to $\zeta^2/10Dq$.
(b)~Tetragonal compression leads to an increased repulsion of $d$-electrons from
apical oxygen ions and further singles out ``planar'' states from $E_g$ and
$T_{2g}$ sets. The quadrupolar moment hosted by the $E_g$ doublet gets pinned
this way.
(c)~Complex combinations of the $E_g$ doublet states that expose the octupolar
moment of ``cubic'' shape.
}
\label{fig:d2split}
\end{center}
\end{figure}
%--------------------------------------------------------------------------

Figure~\ref{fig:d2split} focuses in detail on the particular case of the $d^2$
configuration. The $J=2$ ground state is special because it is isomorphic to a $d$-electron with orbital moment $L_d=2$, and thus
it has to split under cubic crystal field into a triplet of $T_{2g}$ and a doublet
of $E_g$ symmetries (often denoted as $\Gamma_5$ and $\Gamma_3$ states, respectively). The corresponding wavefunctions in the basis of $J_z$
eigenstates $|J_z\rangle$ are \cite{Abr70}: $|\pm 1\rangle$, and
$(|+2\rangle-|-2\rangle)/\sqrt{2}$ for $T_{2g}$, and $|0\rangle$ and
$(|+2\rangle +|-2\rangle)/\sqrt{2}$ for $E_g$ states, just like for single
electron $d$-orbitals of $t_{2g}$ and $e_g$ symmetries, as required by the above isomorphism. Physically, this
splitting arises, e.g., due to the admixture of the $t_{2g}^1e_g^1$ configuration into
$t_{2g}^2$ by SOC. More specifically, second-order energy corrections to $T_{2g}$ and $E_g$
levels are different, which gives a splitting of the $J=2$ level by
$\sim\zeta^2/10Dq$. With SOC parameter $\zeta=0.2\:\mathrm{eV}$ and
$10Dq=3\:\mathrm{eV}$, typical for $4d$ and $5d$ ions, one obtains a sizable
splitting of $20\:\mathrm{meV}$, with the $E_g$ doublet being the lower one. It is
evident from the above wavefunctions that the $E_g$ state has no dipolar moment and is therefore
magnetically silent. Instead, the $E_g$
doublet hosts quadrupolar and octupolar moments. This is again analogous to
$e_g$ electrons, which are quadrupole active, and it has been discussed in
the context of manganites that they may host an octupolar moment as 
well \cite{Tak00,Maezono2000, Bri01}. While this effect was not observed in real $e_g$ 
orbital systems, the spin-orbit-entangled $E_g$ doublet may show octupolar order driven by intersite exchange interactions, unless 
JT coupling to lattice stabilizes quadrupolar order instead.

Concerning the JT activity of the ionic ground states, the $d^4$ singlet and
$d^5$ Kramers doublet possess no orbital degeneracy and are thus ``JT-silent''
in the first approximation. However, both $d^1$ and $d^2$ are JT active ions, and
 structural phase transitions as in usual $3d$ systems can be expected. As
shown in Fig.~\ref{fig:d1245}(b), the shapes of the $\pm 3/2$ and $\pm 1/2$ states
of the $d^1$ ion are different; therefore, they will split under tetragonal
lattice distortions. In fact, these two Kramers doublets can be regarded as an
effective $e_g$ orbital, so JT coupling would read exactly as for the $e_g$
orbital, albeit with an effective JT coupling constant reduced by $1/\sqrt3$, as
a result of SOC unification of the Hilbert space. Similarly, the $E_g$ doublet
of the $J=2$ manifold in the case of a $d^2$ ion should experience JT coupling.  Overall, the JT effect
(and related structural transition) is still operative, but it affects both
spin and orbital degrees of freedom simultaneously as a result of the spin-orbit
transformation of the wavefunctions, and conventional JT orbital ordering
is converted into a magnetic quadrupolar ordering of $J=3/2$
or $J=2$ moments. 
An important consequence of spin-orbital entanglement is that, distinct from
usual orbital order in 3$d$ systems, magnetic quadrupolar order breaks not
only the point-group symmetry of a crystal but also the rotational symmetry in
magnetic space, resulting in anisotropic, non-Heisenberg-type magnetic
interactions, such as XY or Ising models. In other words, JT coupling in spin-orbit-entangled
systems has a direct and more profound influence on magnetism.

The above discussion is based on the $LS$-coupling scheme, which is adequate for obtaining the
ground state quantum numbers. However, the corresponding wavefunctions, and hence effective $g$-factors, as
well as excited-state energy levels, may get some corrections to $LS$-coupling
results. This is important for the interpretation of the experimental data.
Similarly, the admixture of the $t_{2g}^{n-1}e_g^1$ configuration into the ground state
$t_{2g}^n$ wavefunctions by SOC and multielectron Coulomb interactions is
present for all $d^n$, and may renormalize the $g$-factors and wavefunctions \cite{Thornley1968, Stamokostas2018}.
However, these effects cannot split the ground state $J$-levels, except the
$J=2$ level of the $d^2$ configuration, as discussed above.

In general, the ground state manifold of transition metal ions in Mott insulators is
conveniently described in terms of effective spin (``pseudospin'') $\tilde{S}$, where $2\tilde{S} + 1$ is the degeneracy of this manifold. For low-spin $d^n$ ions in a cubic symmetry,
pseudospin $\tilde{S}$ formally corresponds to effective total angular momentum $J$ (often called $J_{\rm eff}$), with the
exception of the $d^2$ case with an $E_g$ doublet hosting a pseudospin
$\tilde{S}=1/2$. One has to keep in mind however, that even in the case of cubic symmetry,
the pseudospin wavefunctions are different from pure $J$ states because of
various corrections (deviations from $LS$ scheme, admixture of $e_g$ states,
etc.) discussed above. This is even more so when noncubic crystal fields are
present and become comparable to SOC. We will occasionally use both
$\tilde{S}$ and $J$ pseudospin notations, depending on convenience (e.g.,
reserving $J$ for the Heisenberg exchange constant in some cases). 

In strong spin-orbit-entangled systems, the notion of pseudospins remains useful even
in doped systems, at least at low doping where Mott correlations, and hence the
ionic spin-orbit multiplets, are still intact locally. In highly doped systems, a
weakly-correlated regime, a conventional band picture emerges, where SOC operates on a
single-electron level.

%}}}

\subsection{Pseudospin interactions in Mott insulators}
%{{{2

%- FIGURE -----------------------------------------------------------------
\begin{figure}[tb]
\begin{center}
\includegraphics[scale=0.95]{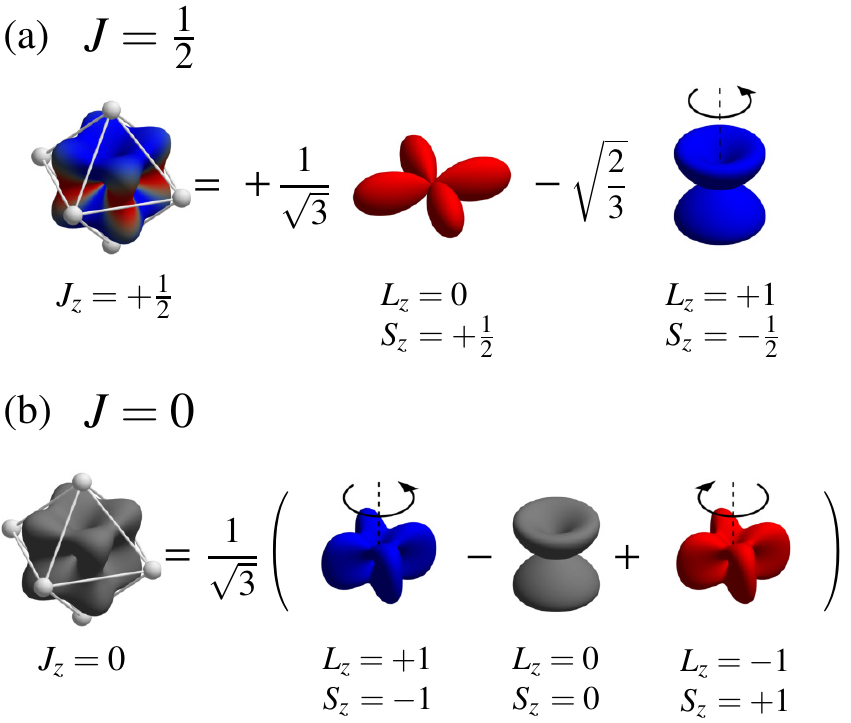}
\caption{
(a)~Decomposition of the $J=1/2$ Kramers doublet state of $d^5$ into
$|L_z,S_z\rangle$ components of the single hole in $t_{2g}^6$ configuration.
The effective angular momentum is indicated by the rotating arrow, spin by the color
following the convention of Fig.~\ref{fig:d1245}. For both contributions,
$L_z$ and $S_z$ sum up to $J_z=+1/2$.
(b)~Similar decomposition of the $J=0$ two-hole ground state of $d^4$. The
total orbital angular momentum $L_z$ and spin $S_z$ may be combined in three
ways here. The internal compensation in $L$ and $S$ creates a cubic-shaped
object showing no spin polarization.
}
\label{fig:d12wf}
\end{center}
\end{figure}
%--------------------------------------------------------------------------

The key element when considering the interactions among pseudospins is the
entanglement of spin and orbital degrees of freedom. In the pseudospin state,
various $|L_z,S_z\rangle$ combinations are superposed, forming a composite
object. Figure~\ref{fig:d12wf} shows two important examples for the $d^5$ and
$d^4$ cases that will be extensively discussed later.
Mixing the spins and orbitals in a coherent way, pseudospins do experience all
the interactions that operate both in the spin and orbital sectors, which have
very different symmetry properties. Electron exchange processes conserve total
spin, and hence the spin interactions are of isotropic Heisenberg ($\vc S_i\vc S_j$) 
form. The orbital interactions are however far more complex --
they are anisotropic both in real and magnetic spaces \cite{Kug82,Kha01,Kha03}.
In high-symmetry crystals, orbitals are strongly frustrated, because they 
are spatially anisotropic and hence cannot simultaneously satisfy all the
interacting bond directions. Via the spin-orbital entanglement, the bond-directional and frustrating nature of the orbital interactions are transferred to the pseudospin interactions \cite{Kha05}.

Moreover, apart from the exchange interactions driven by virtual electron
hoppings, there are other contributions to the orbital interactions, especially in the $d^1$ and $d^2$ cases. These are mediated by the orbital-lattice JT coupling to the virtual
JT-phonons, and by electrostatic multipolar interactions between $d$-orbitals
on different sites \cite{Kanamori60,Che10}. In low-energy effective Hamiltonians, these
interactions transform into pseudospin multipolar couplings, driving both
structural and ``spin-nematic'' transitions breaking cubic symmetry in real
and pseudospin spaces. The JT-driven interactions are also important in $d^4$
systems, as they split the excited $J=1$ levels, and hence promote magnetic
condensation \cite{Liu19}. In the case of $d^5$ systems with
Kramers-degenerate $J=1/2$ pseudospins, the effective Hamiltonians are
predominantly of exchange origin, as in usual spin-$1/2$ systems, although JT
orbital-lattice coupling still shows up in the fine details of pseudospin
dynamics, in the form of pseudospin-lattice coupling \cite{Liu19}.

In general, the low-energy pseudospin Hamiltonians may take various forms
depending on the electron configuration $d^n$ and symmetry of the crystal
structure. Sensitivity of orbital interactions to bonding geometry is a
decisive factor shaping the form of the pseudospin Hamiltonians. We illustrate
this by considering spin-orbital exchange processes in two different cases -- when
metal(M) -oxygen(O) octahedra MO$_6$ share the corners, and when they share the
edges. These two cases are common in transition-metal compounds and are referred to as
$180^\circ$ and $90^\circ$ bonding geometry, reflecting the approximate angle
of the M-O-M bonds.  For simplicity, we limit ourselves to the case of $d^5$
ions with wavefunctions [c.f. Fig.~\ref{fig:d12wf}(a)]:
\begin{align}
|f_{\tilde{\uparrow}}\rangle &= 
 +\sin\vartheta\, |0,\uparrow \rangle
 -\cos\vartheta\, |+1,\downarrow \rangle , \\
|f_{\tilde{\downarrow}}\rangle &= 
 -\sin\vartheta\, |0,\downarrow \rangle 
 +\cos\vartheta\, |-1,\uparrow \rangle ,
\end{align}
where we represent the pseudospin-$1/2$ Kramers doublet by $f$-fermion that
is associated with a hole in the full $t_{2g}^6$ configuration. The
spin-orbit mixing angle is determined by
$\tan2\vartheta = 2\sqrt{2}/(1+2\Delta/\lambda)$, where $\Delta$ is tetragonal splitting of the $t_{2g}$ orbital level. 
 In the cubic limit of $\Delta = 0$ shown in Fig.~\ref{fig:d12wf}(a), one has $\sin\vartheta=1/\sqrt{3}$, $\cos\vartheta=\sqrt{2/3}$.

%- FIGURE -----------------------------------------------------------------
\begin{figure}[tb]
\begin{center}
\includegraphics[scale=0.95]{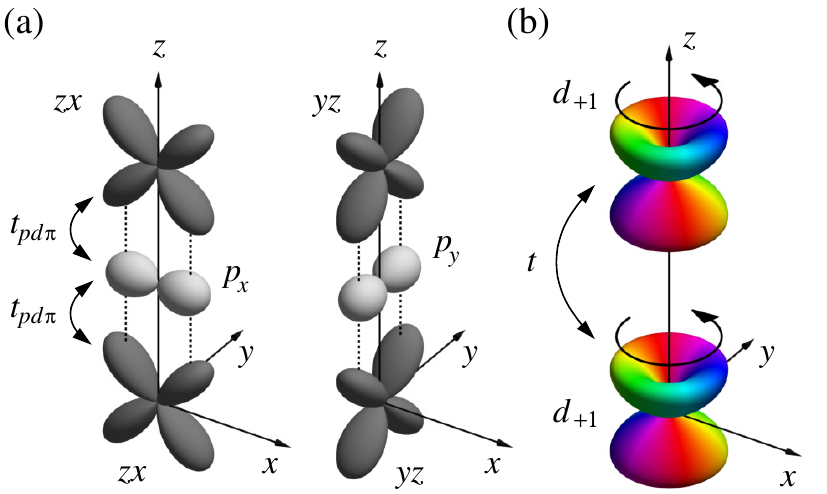}
\caption{
(a)~Hopping via oxygen in the case of $180^\circ$ \mbox{M-O-M} bonds. The
orbital label of the two active $t_{2g}$ orbitals (here $zx$ and $yz$) is
conserved. The third orbital ($xy$, not shown) cannot connect to the oxygen $p$ states.
(b)~Combining orbitals into $L_\alpha$ eigenstates with $\alpha$ determined by
the bond direction, the above rules lead to a conservation of $L_\alpha=\pm1$
while the $L_\alpha=0$ $xy$-orbital is inactive.
}
\label{fig:hop180}
\end{center}
\end{figure}
%--------------------------------------------------------------------------

%- FIGURE -----------------------------------------------------------------
\begin{figure}[tb]
\begin{center}
\includegraphics[scale=0.95]{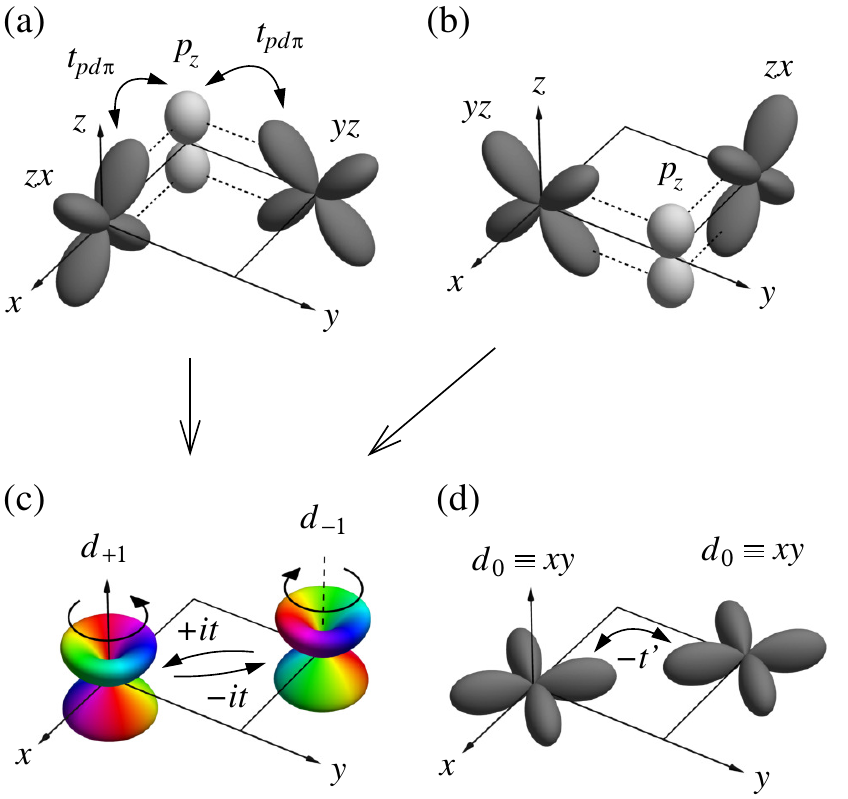}
\caption{
Hopping in the case of $90^\circ$ M-O-M bonding geometry. 
(a),~(b)~Two $t_{2g}$ orbitals $zx$ and $yz$ are interconnected by the hopping via oxygen 
$p$ orbitals. They are selected by the orientation of the M$_2$O$_2$ plaquette. 
(c)~The complementarity of the orbital labels connected in the M-O-M bridge
results in orbital moment non-conserving hopping when considering $L_\alpha$
eigenstates. Here $\alpha=z$ so that $L_z=+1$ is flipped to $L_z=-1$ and vice
versa. The corresponding hopping amplitudes are imaginary: $\pm it$.
(d)~Direct overlap of $d$ orbitals opens an additional hopping channel where
the remaining $L_z=0$ $xy$-orbital is active.
}
\label{fig:hop90}
\end{center}
\end{figure}
%--------------------------------------------------------------------------

The individual orbital components of the pseudospins are subject to distinct
hopping processes, as dictated by the orbital symmetry combined with the 
particular bonding geometry (see Figs.~\ref{fig:hop180} and \ref{fig:hop90}).
For the corner-sharing $180^\circ$ case presented in Fig.~\ref{fig:hop180}, 
the nearest-neighbor (NN) hopping Hamiltonian takes the form
\begin{equation}\label{eq:Hhop180}
\begin{split}
H(180^\circ)  &= -t 
( a^\dagger_{i\sigma} a^{\phantom{\dagger}}_{j\sigma}
+ b^\dagger_{i\sigma} b^{\phantom{\dagger}}_{j\sigma} + \mathrm{H.c.} ) \\
              &= -t 
( d^\dagger_{+1i\sigma} d^{\phantom{\dagger}}_{+1j\sigma}
+ d^\dagger_{-1i\sigma} d^{\phantom{\dagger}}_{-1j\sigma} + \mathrm{H.c.} ) .
\end{split}
\end{equation}
A summation over the spin index $\sigma=\:\uparrow,\downarrow$ is assumed.
Two of the three $t_{2g}$ orbitals participate in oxygen-mediated hopping 
with the amplitude $t=t_{pd\pi}^2/\Delta_{pd}$; the active pair \{$a$, $b$\}
 is selected by the bond direction $\alpha$ and may be combined 
into effective orbital moment $L_\alpha=\pm 1$ eigenstates $|d_{\pm 1}\rangle$.
For example, the $z$ bond considered in Fig.~\ref{fig:hop180} picks up 
$|a\rangle\equiv|yz\rangle$ and $|b\rangle\equiv|zx\rangle$ that form 
$L_z=\pm 1$ eigenstates 
$|d_{\pm 1}\rangle =\mp(|yz\rangle\pm i|zx\rangle)/\sqrt2$. 
The third orbital
$|c\rangle\equiv|xy\rangle\equiv|d_0\rangle$ cannot couple to the mediating 
$p$-orbitals of oxygen for symmetry reasons.
Since the NN hopping preserves both spin and orbital in this case, pseudospin
is also a conserved quantity. Projecting $H(180^\circ)$ onto
the \mbox{$f$-doublet} subspace defined above, one indeed observes pseudospin-conserving hopping 
$H=-t_f (f^\dagger_{i\uparrow} f^{\phantom{\dagger}}_{j\uparrow} 
+ f^\dagger_{i\downarrow} f^{\phantom{\dagger}}_{j\downarrow} + \mathrm{H.c.})$, 
which should therefore lead to isotropic Heisenberg exchange
$H=J (\vc{\tilde S}_i \vc{\tilde S}_j)$ with $J=4 t_f^2/U$.

%- FIGURE -----------------------------------------------------------------
\begin{figure*}[tb]
\begin{center}
\includegraphics[scale=1.05]{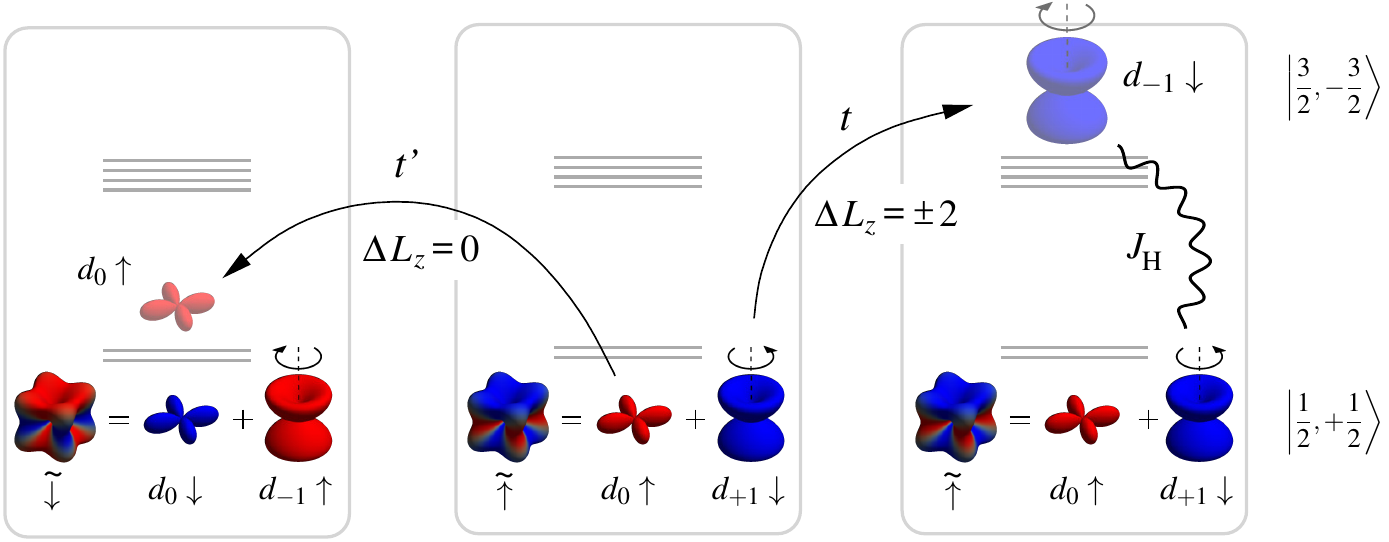}
\caption{
Virtual processes generating the effective interactions among pseudospins
$J=1/2$ of $d^5$ configurations in $90^\circ$ bonding geometry. An M$_2$O$_2$
plaquette perpendicular to the $z$ axis as in Fig.~\ref{fig:hop90} is assumed.
Compared to Fig.~\ref{fig:d12wf}(a), here we scale the $d_0$, $d_{\pm 1}$
orbitals to visually hint on their relative contributions to the pseudospin
wavefunctions.
(left part) $L_z$-conserving direct hopping $t'$ uses the $d_0$ part of the
hole wavefunction and leads to a conventional Heisenberg exchange $\tilde{\vc S}_{i}\tilde{\vc S}_{j}$ following the
Pauli exclusion principle for the $d_0$ orbital. 
(right part) Hopping via oxygen $t$ takes the $d_{+1}$ part of the hole
wavefunction and by the $L_z$ flip creates a virtual $d^4$ configuration
combining the original $d^5$ hole and $J_z=-3/2$ quartet hole. The only option
to reach a final state with two $J=1/2$ pseudospins by the second $t$ hopping
is to remove this $J_z=-3/2$ quartet state again, leaving the initial
pseudospin directions unchanged during the exchange process. Therefore, the
effective interaction is of Ising $\tilde{S}^z_i \tilde{S}^z_j$ type. Hund's
exchange $J_H$ between the major $d_{\pm 1}$ parts of the two holes in the
virtual $d^4$ configuration prefers aligned pseudospins on the two sites which
results in {\em ferromagnetic} Kitaev interaction $K \tilde{S}^z_i
\tilde{S}^z_j$ with $K<0$.
}
\label{fig:Kitaev}
\end{center}
\end{figure*}
%--------------------------------------------------------------------------

The situation in the case of $90^\circ$ bonding geometry is completely different.  As shown in
Fig.~\ref{fig:hop90}, two bond-selected $t_{2g}$ orbitals \{$a$, $b$\} spanning
the $|d_{\pm 1}\rangle$ subspace are again active in the oxygen-mediated hopping
$t=t_{pd\pi}^2/\Delta_{pd}$, but their labels get interchanged during the hopping: $a \leftrightarrow b$, i.e. $yz \leftrightarrow zx$ for $z$ bond.
The third orbital $c$ corresponding to $|d_0\rangle$ participates in direct
hopping $t'$. The two hopping channels are captured by the NN Hamiltonian
\cite{Kha05}:
\begin{multline}
H(90^\circ) =
t\,( a^\dagger_{i\sigma} b^{\phantom{\dagger}}_{j\sigma}
    +b^\dagger_{i\sigma} a^{\phantom{\dagger}}_{j\sigma})
 -t' c^\dagger_{i\sigma} c_{j\sigma} + \mathrm{H.c.} \\
=it\,( d^\dagger_{+1i\sigma} d^{\phantom{\dagger}}_{-1j\sigma} 
      -d^\dagger_{-1i\sigma} d^{\phantom{\dagger}}_{+1j\sigma} ) 
  -t'  d^\dagger_{0i\sigma}  d^{\phantom{\dagger}}_{0j\sigma} + \mathrm{H.c.}
\end{multline}
In contrast to the $180^\circ$ case discussed above, the $t$-hopping term does
not conserve $L_z$, but changes it by $\Delta L_z=\pm 2$; $|d_{+1}\rangle \leftrightarrow |d_{-1}\rangle$. Due to spin
conservation, the total angular momentum projection has to change by the same
amount, i.e. $\Delta J_z=\pm 2$. However, such hopping cannot connect
pseudospin-$1/2$ states (maximal $\Delta J_z=\pm 1$ can be reached by hopping $f^\dagger_{i\uparrow} f^{\phantom{\dagger}}_{j\downarrow} + \mathrm{H.c.}$); indeed, projection of the $t$-term in
$H(90^\circ)$ onto pseudospin $f$-space gives simply zero. This implies that
the pseudospin wavefunctions cannot form $d$-$p$-$d$ bonding states, and thus
the conventional pseudospin exchange term $4t^2/U$ due to hopping $t$ via
oxygen ions is completely suppressed \cite{Kha05,Jac09}. The
situation is similar to the $e_g$ orbital exchange in the $90^\circ$ bonding
geometry, where $e_g$ orbitals cannot form $d$-$p$-$d$ bonding states and thus
no spin-exchange process is possible. As in the $e_g$ case, the pseudospin
interactions in the edge-shared geometry are generated by various corrections
to the above picture (a direct overlap of pseudospins due to the $t'$-term,
electron hopping to higher spin-orbital levels, corrections to pseudospin
wavefunctions due to non-cubic crystal fields, etc.).  
The resulting pseudospin Hamiltonians are typically strongly anisotropic, and
the most important and actually leading term in real compounds is
bond-dependent Ising coupling. Figure~\ref{fig:Kitaev} illustrates how such
an interaction emerges due to the $t$-hopping from ground state $J=1/2$ to higher
spin-orbit $J=3/2$ levels and subsequent Hund's coupling of the excited
electrons in the virtual state. The resulting exchange interaction reads as
$H=K \tilde{S}^z_i \tilde{S}^z_j$, and the corresponding coupling constant
$K\propto -(J_H/U)\, 4t^2/U$ is of ferromagnetic sign \cite{Jac09}. Considered
on honeycomb lattices, this interaction generates the famous Kitaev model where the Ising axis is not global but bond dependent, taking the mutually orthogonal directions $x$, $y$, and $z$ on three different NN bonds\cite{Kit06}. This results in strong frustration and a spin-liquid ground state. On the other hand, a direct hopping $t'$, which conserves both
orbital and spin angular momentum, leads to conventional AF Heisenberg coupling
$\propto t'^2/U$.       

We will later discuss the pseudospin interactions in more detail in the
context of some representative compounds, after a brief materials overview. 

%}}}

\subsection{Materials overview}

As discussed above, the interactions between spin-orbit-entangled pseudospins critically depend on the bonding geometry, and the ground states are determined by the network of each bonding unit, namely crystal structures. Before discussing the properties of representative materials, it would be instructive to overview the crystal structures that are frequently seen in the 4$d$ and 5$d$ transition-metal compounds. We will introduce crystal structures comprising the corner-sharing or edge-sharing network of MO$_6$ octahedra.

\subsubsection{Corner-sharing network of MO$_6$ octahedra}

\begin{figure}[t]
\begin{center}
\includegraphics[scale=0.43]{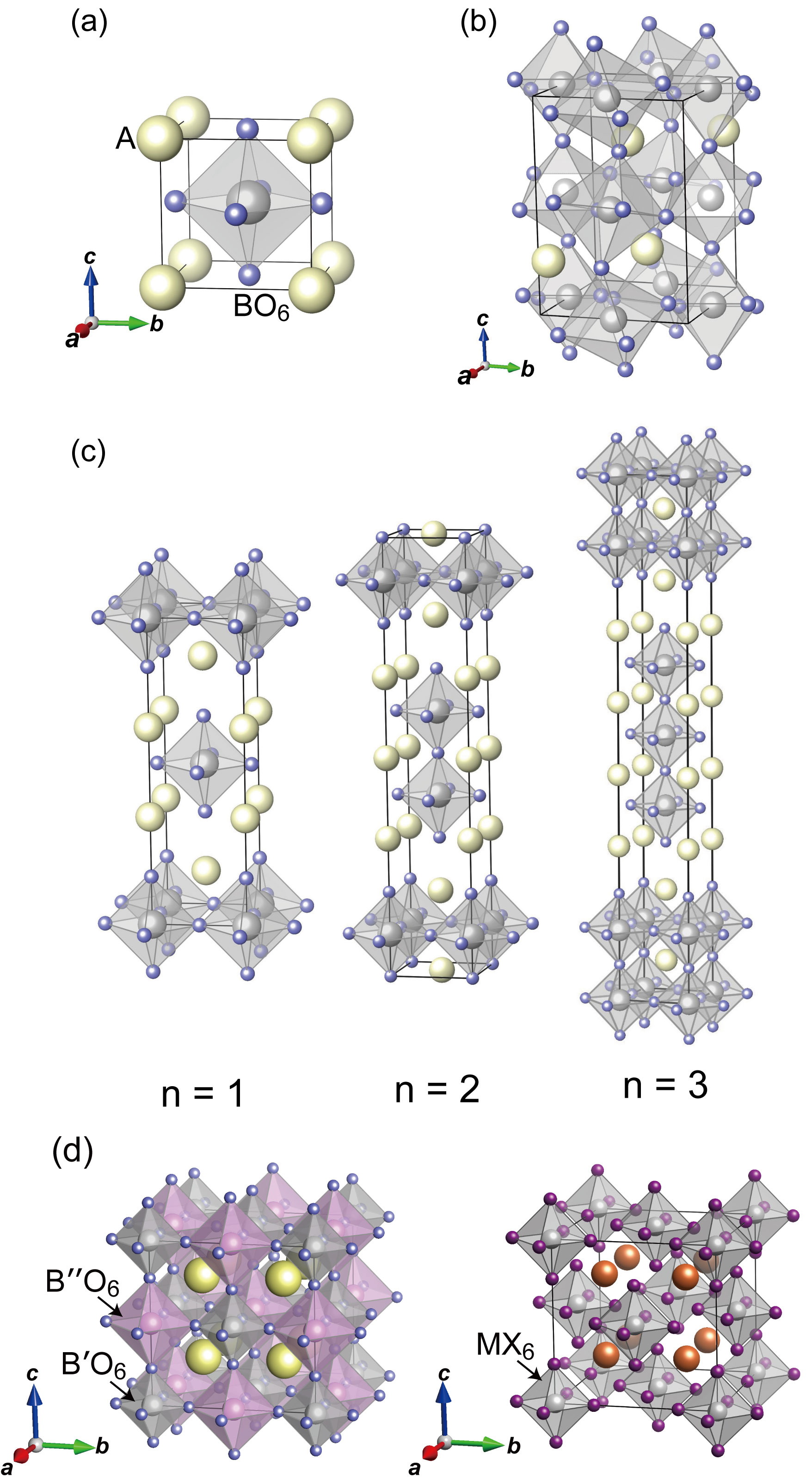}
\caption{Crystal structures of perovskite and its derivatives. (a) Cubic perovskite ABO$_3$. (b) Orthorhombic perovskite with the GdFeO$_3$-type distortion. (c) Ruddlesden-Popper series layered perovskite A$_{\rm n+1}$B$_{\rm n}$O$_{\rm 3n+1}$. (d) Double-perovskite A$_2$B$^{\prime}$B$^{\prime \prime}$O$_6$ (left) and A$_2$MX$_6$-type halides (right). The crystal structures are visualized by using VESTA software \cite{Momma11}.
}
\label{fig:perovskite}
\end{center}
\end{figure}

The most representative structure with corner-sharing MO$_6$ octahedra is the perovskite structure with a chemical formula of ABO$_3$ (A and B are cations). Small transition-metal ions are generally accommodated into the B-site, and the BO$_6$ octahedra form a three-dimensional corner-sharing network. The stability of the perovskite structure is empirically evaluated by the Goldschmidt tolerance factor $t$ = $(r_{\rm A} + r_{\rm O})$/$\sqrt{2}(r_{\rm B} + r_{\rm O})$ where $r_{\rm A}$, $r_{\rm B}$ and $r_{\rm O}$ are the ionic radius of A, B and oxygen ions, respectively. Note that the perovskite structure consists of alternate stacking of AO layer and BO$_2$ layer. $t$ = 1 means that the ionic radii are ideal to form a cubic perovskite structure [Fig.~\ref{fig:perovskite}(a)] with the perfect matching of the spacings of constituent ions for AO and BO$_2$ layers. As the ionic radius $r_{\rm B}$ for 4$d$ and 5$d$ transition-metal ions is relatively large, the tolerance factor $t$ of 4$d$ and 5$d$ perovskites are normally less than 1, giving rise to lattice distortions to compensate the mismatch between AO and BO$_2$ layers. A distorted perovskite structure frequently found in 4$d$ and 5$d$ transition-metal oxides is the orthorhombic GdFeO$_3$-type (Space group $Pbnm$) [Fig.~\ref{fig:perovskite}(b)]. The BO$_6$ octahedra rotate about the $c$-axis and tilt around the [110] direction (Glazer notation: $a^{-}a^{-}c^{+}$ \cite{Glazer1972}). Because of this distortion, the B-O-B angle is smaller than 180$^{\circ}$. Many 4$d$ and 5$d$ transition-metal perovskites such as CaRuO$_3$, NaOsO$_3$ and SrIrO$_3$ crystallize in the GdFeO$_3$-type structure \cite{YGShi2009,Bensch1990, Longo1971}.

In addition to the three-dimensional network, the quasi-two-dimensional analogue with 180$^{\circ}$ bonding geometry is realized in the layered derivatives of perovskite structure. The square lattice of octahedrally-coordinated transition-metal ions is seen in the K$_2$NiF$_4$-type (A$_2$BO$_4$) layered structure, where the alternate stacking of (AO)$_2$-BO$_2$ layers along the $c$-axis is formed. Generally, the BO$_6$ octahedra are tetragonally distorted in the layered perovskites such as Sr$_2$VO$_4$. As in the ABO$_3$-type perovskite, the mismatch of ionic radii of A and B cations results in the rotation and the tilting distortion of BO$_6$ octahedra, making the B-O-B angle less than 180$^{\circ}$. For example, the layered iridate Sr$_2$IrO$_4$ possesses the rotation of IrO$_6$ octahedra about the $c$-axis \cite{Crawford1994}, whereas Ca$_2$RuO$_4$ hosts both a rotation and tilting distortion of RuO$_6$ octahedra \cite{Friedt2001}. 

The K$_2$NiF$_4$(A$_2$BO$_4$)-type perovskite is an end member of a Ruddlesden-Popper series with a general chemical formula of A$_{\rm n+1}$B$_{\rm n}$O$_{\rm 3n+1}$. This formula can be rewritten as AO(ABO$_3$)$_{\rm n}$ [= AO(AO-BO$_2$)$_{\rm n}$], which makes it easier to view the crystal structures; there are n-layers of BO$_6$ octahedra, sandwiched by the double rock-salt-type AO layers as illustrated in Fig.~\ref{fig:perovskite}(c). Generally, with increasing number of layers, $n$, the electronic structure becomes more three-dimensional and hence the bandwidth increases, which may induce a metal-insulator transition as a function of $n$. 

The double-perovskite structure is a derivative of perovskite, and also called rock-salt-ordered perovskite. In the double-perovskites, two different cations, B$^{\prime}$ and B$^{\prime \prime}$, occupy the octahedral site alternately and form a rock-salt sublattice. The ordered arrangement of two different B cations is usually seen when the difference of valence state of the two cations is more than 2. Both B$^{\prime}$ and B$^{\prime \prime}$ ions comprise a face-center-cubic (FCC) sublattice. Note that there is neither a direct B$^{\prime}$-O-B$^{\prime}$ bond nor a B$^{\prime \prime}$-O-B$^{\prime \prime}$ bond. The FCC lattice with $d^1$ or $d^2$ ions has been proposed to be a possible realization of multipolar ordering of $d$ electrons, and the double-perovskite oxides with magnetic B$^{\prime}$ and nonmagnetic B$^{\prime \prime}$ ions have been studied intensively as will be discussed in Section V. The FCC lattice of $d^1$ or $d^2$ ions is also realized in the series of transition-metal halides with a chemical formula A$_2$MX$_6$ (A$^{+}$: alkali ion, M$^{4+}$: transition-metal ion, and X$^-$: halogen ion) \cite{Armstrong1980}, where MX$_{6}^{2-}$ octahedra and A$^{+}$ ions form anti-fluorite-like arrangement. This structure can be viewed as a B$^{\prime \prime}$-site deficient double-perovskite A$_2$M$\Box$X$_6$, where $\Box$ denotes a vacancy. A$_2$MX$_6$ crystallizes in a cubic structure with a large A ion such as Cs$^+$. M$^{4+}$ ions with an ideal cubic environment form a FCC sublattice, but also form a distorted structure when the size of A ion is small. A wide variety of 4$d$ and 5$d$ transition-metal elements can be accommodated into this structure.

\begin{figure}[t]
\begin{center}
\includegraphics[scale=0.4]{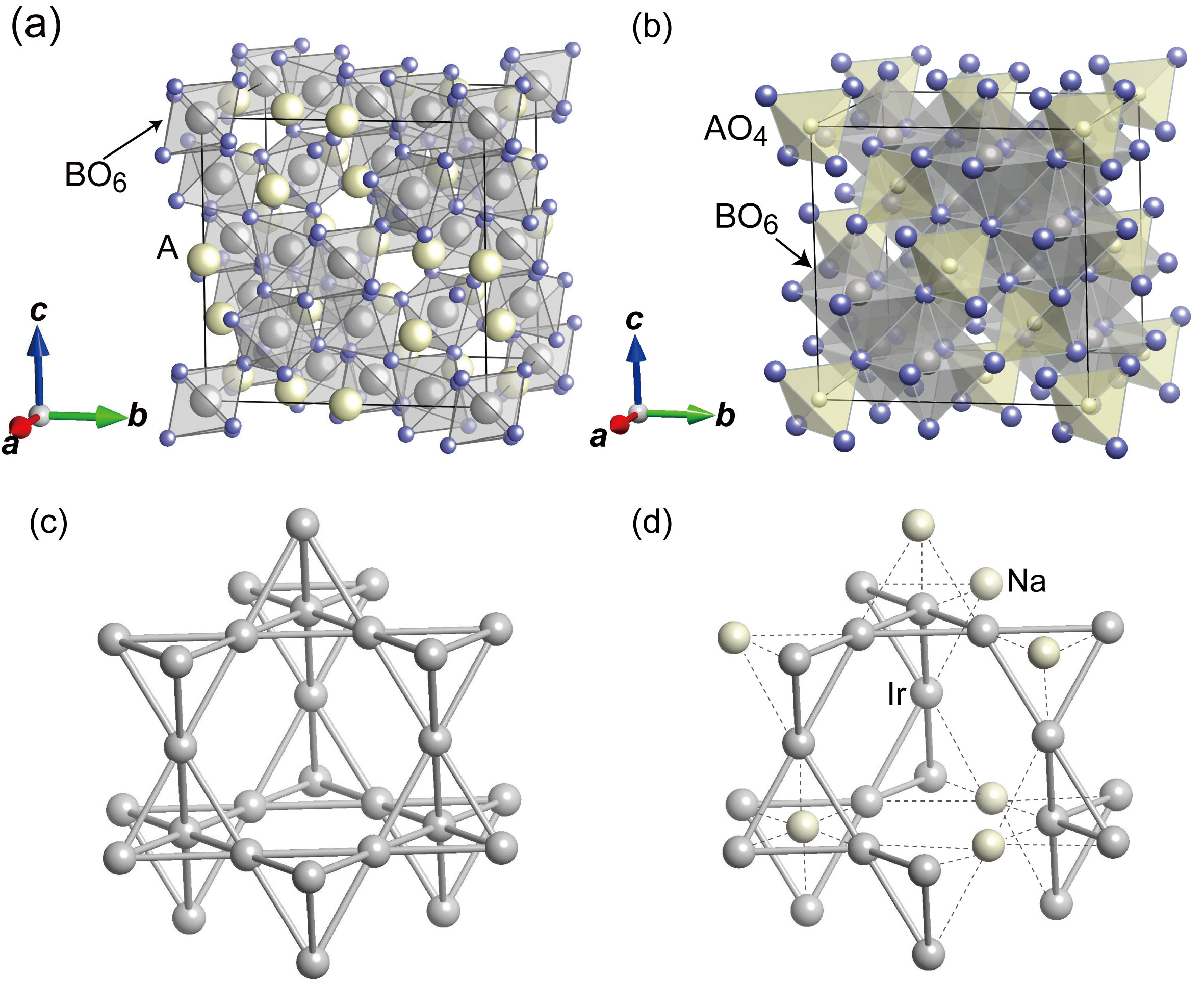}
\caption{Crystal structures of (a) pyrochlore oxide A$_2$B$_2$O$_7$ and (b) spinel oxide AB$_2$O$_4$. (c) Pyrochlore sublattice comprised by B (or A) atoms of pyrochlore oxide or by B atoms of spinel oxide. (d) Hyperkagome sublattice of Ir atoms found in Na$_4$Ir$_3$O$_8$. The pyrochlore sublattice is shared by 3:1 ratio of Ir and Na atoms in an ordered manner.
}
\label{fig:pyrochlore}
\end{center}
\end{figure}

Another important class of materials with corner-sharing MO$_6$ octahedra is the pyrochlore oxide with a general formula A$_2$B$_2$O$_7$ (more specifically A$_2$B$_2$O$_6$O$^{\prime}$ where O and O$^{\prime}$ represent two different oxygen sites) [Fig.~\ref{fig:pyrochlore}(a)]. Transition-metal ions are accommodated into the B-cation site, which forms a BO$_6$ octahedron, whereas the A-cation is surrounded by six O and two O$^{\prime}$ atoms in a distorted cubic-like environment. The sublattice of the B-cations, as well as that of A-cations, is a network of corner-shared tetrahedra called the pyrochlore lattice [Fig.~\ref{fig:pyrochlore}(c)]. The pyrochlore lattice is known to provide geometrical frustration when magnetic moments of constituent ions interact antiferromagnetically. There are many ways to view the pyrochlore structure as described in Ref.~[\onlinecite{Subramanian1983}]. Most conventionally, the B atom is located at the origin of unit cell (Wyckoff position 16$c$) for the space group $Fd\overline{3}m$ (No. 227, origin choice 2). In this setting, the only tuneable parameters are the lattice constant and the $x$ coordinate of the O site \cite{Gardner2010}. With $x$ = 0.3125, the BO$_6$ forms an ideal octahedron and the B-O-B angle is $\sim$141$^{\circ}$. In 4$d$ and 5$d$ transition-metal pyrochlore oxides, $x$ is usually larger than 0.3125, and the BO$_6$ octahedra are compressed along the [111] direction pointing to the center of B-tetrahedra. The compressive distortion gives rise to a trigonal crystal field on B ions and decreases the B-O-B angle between the neighboring octahedra from 141$^{\circ}$, which reduces the hopping amplitude and thus bandwidth. When the ionic radius of the A atom becomes smaller, the trigonal distortion is enhanced. A metal-insulator transition is seen as a function of the size of A ions in pyrochlore oxides such as molybdates A$_2$Mo$_2$O$_7$ and iridates A$_2$Ir$_2$O$_7$ (A: trivalent ions such as rare-earth or Y$^{3+}$) \cite{Moritomo2001, Matsuhira2011}. The trigonal crystal field which splits the $t_{2g}$ manifold potentially competes with SOC.

\subsubsection{Edge-sharing network of MO$_6$}

As discussed above, the edge-sharing, namely 90$^{\circ}$ M-O-M, bonding geometry provides magnetic interactions distinct from those in 180$^{\circ}$ bonds. With the edge-sharing network of MO$_6$ octahedra, one can realize a variety of lattice structures of interest, such as the triangular lattice in ABO$_2$, the honeycomb lattice in A$_2$BO$_3$ and the pyrochlore lattice in AB$_2$O$_4$ spinels. They can be constructed from the rock-salt structure.

\begin{figure}[t]
\begin{center}
\includegraphics[scale=0.45]{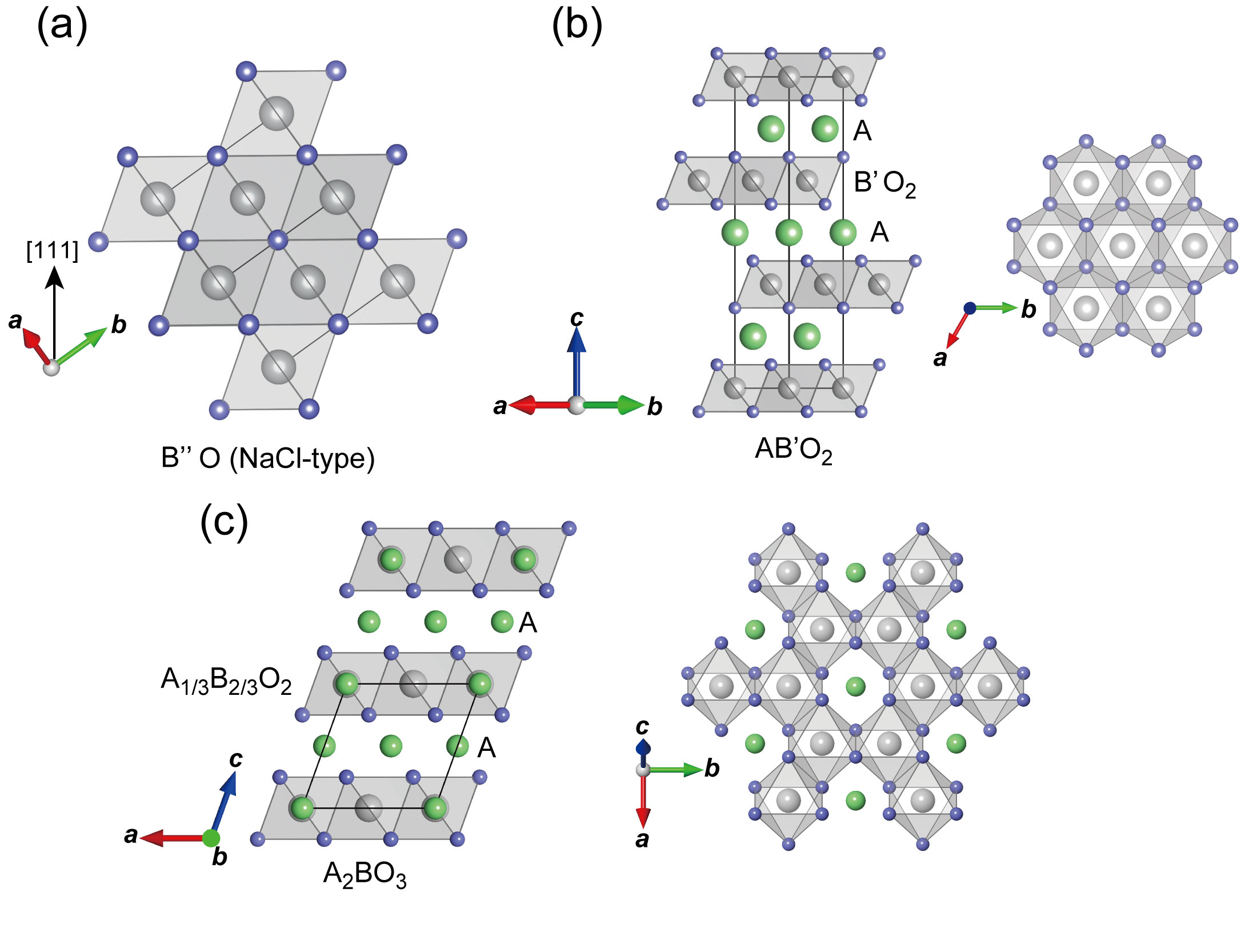}
\caption{(a) Rock-salt structure of B$^{\prime \prime}$O. (b) AB$^{\prime}$O$_2$-type structure formed by replacing B$^{\prime \prime}$ with A$^+$ and B$^{\prime 3+}$ ions which stack alternately along the $c$-axis. (c) Layered honeycomb structure of A$_2$BO$_3$. The triangular layer of B$^{\prime 3+}$ ions in (b) is substituted by the 2:1 ratio of B$^{4+}$ and A$^+$ ions forming a honeycomb lattice.
}
\label{fig:NaCl}
\end{center}
\end{figure}

To derive the layered triangular and honeycomb structures from the rock-salt-type B$^{\prime \prime 2+}$O$^{2-}$ (B$^{\prime \prime}$: transition-metal atom), first consider the rock-salt structure viewed along the cubic [111] direction [Fig.~\ref{fig:NaCl}(a)]. It consists of an alternating stack of the triangular B$^{\prime \prime 2+}$ planes and the triangular O$^{2-}$ planes. By replacing every pair of adjacent B$^{\prime \prime 2+}$ planes with an A$^+$ plane and B$^{\prime 3+}$ plane, we have the layered AB$^{\prime}$O$_2$-type structure with triangular layers of A$^+$ and B$^{\prime 3+}$ [Fig.~\ref{fig:NaCl}(b)]. The B$^{\prime}$O$_6$ octahedra form the edge-shared triangular lattice. The trivalent B$^{\prime 3+}$ can be replaced by a 2:1 ratio of B$^{4+}$ and A$^+$ ions. The large difference of valence states between A$^+$ and B$^{4+}$ cations facilitates the ordered arrangement of two cations in the triangular plane. As a result, the A$_{1/3}$B$_{2/3}$ layers contain a honeycomb network of BO$_6$ octahedra connected by three of their six edges [Fig.~\ref{fig:NaCl}(c)]. The alternate stacking of an A$^+$-cation layer and an A$^{+}_{1/3}$B$^{4+}_{2/3}$ layer corresponds to the chemical formula A$_2$BO$_3$ [= A(A$_{1/3}$B$_{2/3}$)O$_2$] as found in Na$_2$IrO$_3$ and Li$_2$RuO$_3$ \cite{Mather2000}. The three-dimensional honeycomb structure of $\beta$-Li$_2$IrO$_3$ and $\gamma$-Li$_2$IrO$_3$ can be derived from the rock-salt structure as well, but the ordering pattern of Li$^+$ and Ir$^{4+}$ ions are different from the [111] ordering described above \cite{Takayama2015, Modic2014}. Those 4$d$ and 5$d$ transition-metal oxides with a honeycomb network are attracting attention as a realization of exotic quantum magnetism.

In the ordered rock-salt structures described above, all octahedral voids created by oxygen atoms are filled by cations. The rock-salt structure also has tetrahedral voids that can be occupied by cations. By partially filling the tetrahedral and octahedral voids, a spinel structure AB$_2$O$_4$ can be constructed [Fig.~\ref{fig:pyrochlore}(b)]. In the spinel structure, B cations form a network of corner-shared tetrahedra as in the pyrochlore oxides. The crucial difference from A$_2$B$_2$O$_7$ pyrochlore oxides is that the BO$_6$ octahedra in the spinel structure are connected by edge-sharing bonds. The number of spinel oxides containing 4$d$ or 5$d$ transition-metal atoms is rather limited. When multiple cations occupy the pyrochlore B-sublattice, they may form an ordered arrangement. The prominent example is hyperkagome iridate Na$_4$Ir$_3$O$_8$ \cite{Okamoto2007}. In Na$_4$Ir$_3$O$_8$, the B-site pyrochlore lattice of the spinel is shared in a 3:1 ratio of Ir and Na atoms. The Ir sublattice is viewed as corner-shared triangles in three dimensions, which has been dubbed the hyperkagome lattice [Fig.~\ref{fig:pyrochlore}(d)]. The properties of hyperkagome iridate will be discussed in Sections III.C and VI.C.

%}}}

\section{Pseudospin-1/2 magnetism in $\vc d^\mathbf{5}$ compounds}
%{{{1

The collective behavior of $d^5$ ions with Kramers doublet ground states can 
be described in terms of a pseudospin-$1/2$
Hamiltonian. Thanks to the spin one-half algebra, pairwise interactions
between pseudospins are reduced to various bi-linear terms. While the forms
 of these terms are dictated by lattice symmetry, the corresponding
coupling constants may vary broadly, depending on the details of the local
chemistry of a given material \cite{Goo63}. In Sec.~\ref{sec:concept}, we
emphasized the difference between the $180^\circ$ and $90^\circ$ bonding
geometry that lead to either conventional Heisenberg interaction or strongly
frustrated bond-selective interactions of the Kitaev type. Focusing on these two
cases, we now consider a few representative examples of $d^5$ compounds realizing pseudospin-$1/2$ physics.

\subsection{180$^\circ$ M-O-M bonding, perovskites}
%{{{2

The perovskite iridate \mbox{Sr$_2$IrO$_4$} has emerged as a model system for
understanding the spin-orbit-entangled magnetism of $5d$ electrons.
The first experimental evidence for the $J=1/2$ state was provided by a combination of
angle-resolved photoemission spectroscopy (ARPES), optical spectroscopy, and x-ray absorption measurement \cite{Kim08} and
later by resonant elastic x-ray scattering \cite{Kim09}, which confirmed the
complex structure of the $d^5$ hole.

The relevant pseudospin-$1/2$ model may be anticipated based on the
corner-shared IrO$_6$ octahedra in the perovskite structure with approximately
$180^\circ$ Ir-O-Ir bonds (the bonds are not completely straight due to
$11^\circ$ in-plane octahedra rotations \cite{Hua94}). Since the pseudospin
wavefunctions overlap well in the $d$-$p$-$d$ hopping channel, and hopping is
pseudospin conserving as discussed in Sec.II.B, the dominant interaction is represented by
Heisenberg coupling. However, there are additional terms which arise due to
hoppings to higher level orbitals, tetragonal distortions and octahedral rotations, and these lead to the following NN-interaction Hamiltonian: 
\begin{equation}
H=J(\vc{\tilde{S}}_i\vc{\tilde{S}}_j) + D(\vc{\tilde{S}}_i\times\vc{\tilde{S}}_j) 
+ A (\vc{\tilde{S}}_i\vc r_{ij})(\vc{\tilde{S}}_j\vc r_{ij})+ J_z \tilde{S}_i^z \tilde{S}_j^z.
\end{equation}
Here, the $D$ term is an antisymmetric Dzyaloshinskii-Moriya (DM) interaction caused by octahedral rotations and $A$ and $J_z$ represent symmetric anisotropy terms. The $J_z$ term is derived from the combined effect of the octahedral rotations and tetragonal fields, while the A term with dipole-dipole-coupling type bond-directional structure is symmetry allowed even in an ideal cubic structure \cite{VanVleck1937}.
The coupling constants have been calculated in Ref.~[\onlinecite{Jac09}], and
vary as a function of tetragonal crystal field, rotation angle etc. This Hamiltonian
nicely accounts for a number of properties of \mbox{Sr$_2$IrO$_4$}, including
nearly Heisenberg spin dynamics akin to the cuprates \cite{Fuj12,Kim12}.

A closer look in magnon data \cite{Kim12} reveals that the model above has to
be extended, including longer-range interactions $J_2$ and $J_3$, which turn
out to be much larger than in cuprates. This might be related to the fact that
$5d$ electrons are more extended spatially, and to the relatively small Mott gap \cite{Kim08}. More recently, it has been
shown that the pseudospins in iridates couple to lattice degrees of freedom
via a dynamical admixture of higher-lying spin-orbital levels to the ground
state wavefunctions \cite{Liu19}, which explains the observed in-plane magnon
gaps \cite{Por19}, and predicts sizable magnetostriction effects breaking
tetragonal symmetry below $T_N$. 

In addition to elastic x-ray scattering \cite{Kim09}, the spin-orbit-entangled
nature of the $d^5$ ions in \mbox{Sr$_2$IrO$_4$} was detected by resonant inelastic x-ray scattering (RIXS) experiments that observed \cite{Kim12} transitions from $J=1/2$ to $J=3/2$
levels, directly confirming the level structure shown in Fig.~\ref{fig:d1245}. These excitations, dubbed ``spin-orbit exciton'', formally behave as a
doped hole in a quantum antiferromagnet moving in a crystal by emitting and
absorbing magnons. The resulting exciton-magnon continuum, as well as the
expected quasiparticle peak below it, have been indeed observed \cite{Kim14}, similar
to doped holes in antiferromagnetic cuprates. Also like in the hole-doped cuprates, 
spin-excitation spectra obtained by RIXS on La-doped \mbox{Sr$_2$IrO$_4$} \cite{Gre16,Liu16} 
revealed paramagnons persistent well into the metallic phase. 

Encouraged by the above analogies with cuprates, doped \mbox{Sr$_2$IrO$_4$} samples have
been studied in the search for unconventional superconductivity. However, experiments on doped \mbox{Sr$_2$IrO$_4$} are severely impeded by difficulties in obtaining clean samples. Techniques beyond
conventional chemical doping such as La-substitution producing electron doped
Sr$_{2-x}$La$_x$IrO$_4$ have to be employed. For example, surface electron
doping achieved by potassium deposition on the surface of parent
\mbox{Sr$_2$IrO$_4$} enabled ARPES and scanning tunneling microscopy (STM) studies
\cite{Kim14doped,Kim16,Yan15}; another promising route is ionic liquid gating
\cite{Lu15,Rav16}. Although no clear evidence for superconductivity was so far
detected, Fermi surface and pseudogap phenomena as in cuprates have been
observed in ARPES \cite{Kim14doped,Kim16} and STM experiments \cite{Yan15}. For more a detailed account on doped \mbox{Sr$_2$IrO$_4$}, we recommend the recent review \cite{Ber19}.

Next, we briefly discuss the bilayer iridate \mbox{Sr$_3$Ir$_2$O$_7$}. Being
``in-between'' quasi-two-dimensional insulator \mbox{Sr$_2$IrO$_4$} and three-dimensional metal \mbox{SrIrO$_3$}, this compound
is close to the Mott transition, with a small insulating gap \cite{Moo08}.
Nonetheless, pseudospin-$1/2$ magnons, as well as $J=3/2$ spin-orbit excitons,
have been observed in RIXS experiments \cite{Kim12bilRXRD,Kim12bilRIXS}
showing that the spin-orbit-entangled nature of low-energy states remains
largely intact. A remarkable observation is that the magnetic moment direction
and magnon spectra in this compound are radically different from those in the sister compound
\mbox{Sr$_2$IrO$_4$}. Possible explanations for this have been
offered -- based on enhanced anisotropic pseudospin couplings
\cite{Kim12bilRIXS} and on dimer formation on the links connecting the two
layers \cite{Mor15}.

%}}}

\subsection{90$^\circ$ M-O-M bonding, honeycomb lattice}
%{{{2

The case of $J=1/2$ pseudospins on a honeycomb lattice has sparked a broad
interest after the proposal of Ref.~[\onlinecite{Jac09}] that the corresponding
materials, such as Na$_2$IrO$_3$ (see Fig.~\ref{fig:NaCl}), may realize a Kitaev honeycomb model \cite{Kit06}. Since there is
already a vast literature on this topic, including several review articles
\cite{Rau16,Win17,Tre17,Her18,Tak19,Mot20}, we will make just a few remarks
concerning the ``unwanted'' (i.e. non-Kitaev) exchange terms that are present
in the Kitaev-model candidate materials studied so far.

As explained in Sec.II.B above, the pseudospin-$1/2$ wavefunctions cannot communicate with
each other via the oxygen ions -- the corresponding hopping integral is zero in the
cubic limit. Finite interactions (albeit not as strong as in $180^\circ$ case)
do originate from higher order processes involving spin-orbit $J$ = 3/2 virtual states, or from communication of the pseudospins via a direct
$t'$ hopping, as illustrated in Fig.~\ref{fig:Kitaev}.  Hoppings to higher lying $e_g$ states with subsequent Hund's coupling, as well as $pd$ charge-transfer excitations do also contribute. Phenomenologically, symmetry considerations dictate the
following general form of NN interactions \cite{Kat14,Rau14a,Rau14b}
\begin{multline}
H=K \tilde{S}_i^z \tilde{S}_j^z + J(\vc{\tilde{S}}_i \vc{\tilde{S}}_j) +
\Gamma (\tilde{S}_i^x \tilde{S}_j^y+\tilde{S}_i^y \tilde{S}_j^x) \\
+ \Gamma'(\tilde{S}_i^x \tilde{S}_j^z + \tilde{S}_i^z \tilde{S}_j^x 
+ \tilde{S}_i^y \tilde{S}_j^z + \tilde{S}_i^z \tilde{S}_j^y) ,
\end{multline}
which is expressed here for an M$_2$O$_2$ plaquette perpendicular to the cubic
$z$ axis, as shown in Fig.~\ref{fig:hop90}. The first term represents the Kitaev interaction. According to perturbative calculations \cite{Rau14a,Rau14b,Win16}, the
off-diagonal exchange $\Gamma$ arises from combined $t$ and $t'$ hoppings,
while the $\Gamma'$ term is generated by trigonal crystal fields that modify
the pseudospin wavefunctions. Trigonal field also suppresses the
bond-dependent nature of the pseudospin interactions \cite{Kha05,Cha15}, so
the cubic limit is desired to support the Kitaev coupling $K$ and to suppress
the $\Gamma'$ term. However, the $J$ and $\Gamma$ terms, generated by a direct
hopping $t'$ and other possible hopping channels, remain finite even in the cubic limit. The crucial parameter here is the
M-M distance that controls the magnitude of the direct overlap of the $d$-wavefunctions. 

Apart from NN $J$, $\Gamma$, and $\Gamma'$ terms, the longer-range (second/third NN $J_{2/3}$)
interactions are likely present in many Kitaev materials. These
couplings, which are detrimental to the Kitaev spin liquid, are also related to the spatial extension of the $4d$ and $5d$ orbitals, and to the lattice structure ``details'' such as the
presence/absence of cations (e.g. Li or Na) within or near the honeycomb
planes [see Fig.~\ref{fig:NaCl}(c)], opening additional hopping channels. 

Nevertheless, the Kitaev-type couplings appear to be dominant in $5d$-iridates
and also $4d$-ruthenium chloride, as evidenced by a number of experiments (see, e.g.,
the review [\onlinecite{Tak19}]), although they are not yet strong enough to
overcome the destructive effects of non-Kitaev terms discussed above. So the efforts 
to design materials with suppressed ``unwanted'' interactions have to be
continued. An interesting perpective in this context may be the employment of
$J=1/2$ Co$^{2+}$ ions, as has been proposed recently
\cite{Liu18,San18,Liu20}. Although the energy scales for the pseudospin
interactions are smaller in this case, the less-extended nature of $3d$ wavefunctions may reduce the longer-range couplings, improving thus the conditions
for the realization of the Kitaev model.

The non-Heisenberg, bond-dependent nature of the pseudospin interactions is
expected to survive in weakly doped compounds, and affect their metallic
properties. In particular, it has been suggested that they should lead to an
unconventional $p$-wave pairing \cite{Kha04,Kha05,Hya12,You12,Oka13}.
However, experimental data on doped $d^5$ compounds with $90^\circ$ bonding
geometry is scarce, because a ``clean'' doping of Mott insulators is a
challenge in general. 

%}}}

\subsection{Pseudospins-1/2 on frustrated lattices}

A combination of geometrical frustration with spin-orbital frustration may
open an interesting pathway to exotic magnetism. In fact, the bond-dependent
pseudospin interactions have been first discussed in the context of
geometrically frustrated triangular\cite{Kha05} %(GK 2005)
and hyperkagome\cite{Che08} % (Chen 2008)
lattices. Here we discuss some spin-orbit-entangled pseudospin-1/2 systems with geometrically
frustrated lattices.       

\subsubsection{Hyperkagome iridate Na$_4$Ir$_3$O$_8$}

Among complex iridium oxides, Na$_4$Ir$_3$O$_8$ appears to be the first example of an exotic quantum magnet. In Na$_4$Ir$_3$O$_8$, Ir atoms share the B-site pyrochlore lattice of the spinel with Na atoms, and form a network of corner-shared triangles dubbed the hyperkagome lattice, as described in Sec.II.C.2. The hyperkagome lattice is geometrically frustrated if the Ir moments couple antiferromagnetically.

\begin{figure}[t]
\begin{center}
\includegraphics[scale=0.7]{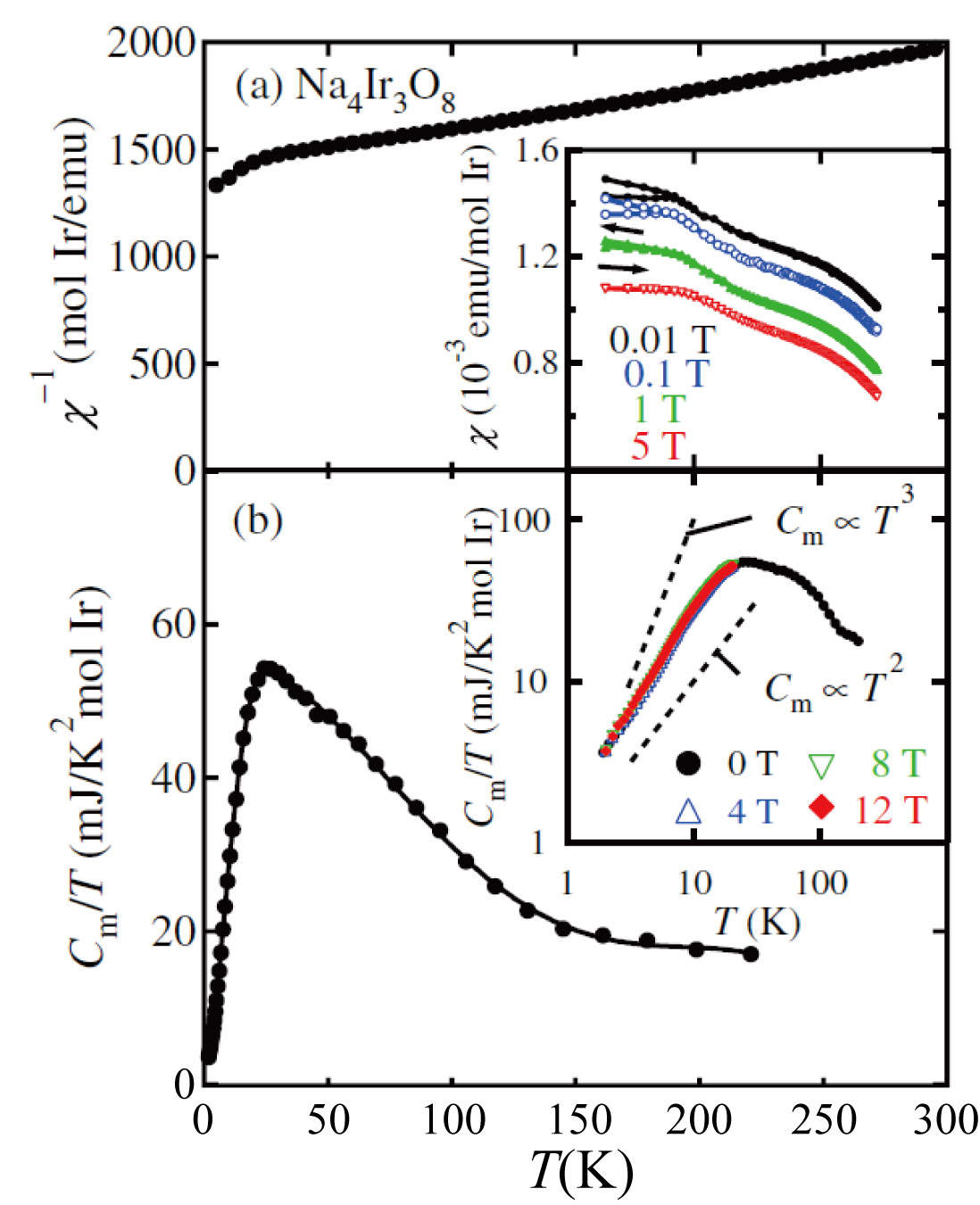}
\caption{Temperature dependence of (a) inverse magnetic susceptibility $\chi^{-1}(T)$, and (b) magnetic specific heat $C_m$ divided by temperature. Insets: (a) Temperature dependence of $\chi(T)$ at various magnetic fields. (b) $C_m/T$ at various magnetic fields, showing a power law behavior $C_m(T)\;\sim\;T^n$ (2 $< n <$ 3) at low temperatures. The figure is reproduced with permission from Ref.~[\onlinecite{Okamoto2007}] ($\copyright$2007 the American Physical Society).}
\label{fig:Na4Ir3O8}
\end{center}
\end{figure}

Indeed, Na$_4$Ir$_3$O$_8$ displays a strong antiferromagnetic interaction inferred from the large negative Weiss temperature ($|\theta_{\rm CW}|$ $\sim$ 650 K) in the magnetic susceptibility $\chi(T)$ \cite{Okamoto2007}. Nevertheless, no sign of magnetic ordering has been seen down to 2 K both in $\chi(T)$ and the specific heat $C(T)$ as shown in Fig.~\ref{fig:Na4Ir3O8}. Na$_4$Ir$_3$O$_8$ thus appeared as the first candidate for a three-dimensional quantum spin liquid.

In the $\chi(T)$ of Na$_4$Ir$_3$O$_8$, a small bifurcation was seen at around 6 K [the inset to Fig.~\ref{fig:Na4Ir3O8}(a)]. It was originally interpreted as a glassy behavior due to a small amount of impurity/defects \cite{Okamoto2007}. However, it has been pointed out from the $^{23}$Na-NMR and $\mu$SR measurements that Na$_4$Ir$_3$O$_8$ exhibits a spin-glassy frozen state or quasistatic spin correlation \cite{Shockley2015,Dally2014}. We note that the presence of such a glassy state may be associated with the disorder of Na ions in the octahedral A-site. Since the successful growth of Na$_4$Ir$_3$O$_8$ single crystals was reported recently \cite{HZheng2018}, the understanding of its magnetic ground state is advancing.

After the discovery of the spin-liquid behavior in Na$_4$Ir$_3$O$_8$, a plethora of theoretical studies have been put forward. In the early days, most of the models have treated the Ir moments as $S$ = 1/2 and considered the antiferromagnetic Heisenberg model on the frustrated hyperkagome lattice. The classical model predicted the presence of nematic order \cite{Hopkinson2007}. For the quantum limit, the ground state has been discussed to be either a spin-liquid with spinon Fermi surface or a topological $Z_2$ spin liquid \cite{YZhou2008,Lawler2008Jun,Lawler2008Nov} .

Since the IrO$_6$ octahedra form an edge-sharing network, as in honeycomb-based iridates, the presence of anisotropic magnetic exchanges such as Kitaev-type coupling is anticipated. The electronic structure calculation showed that SOC of Ir gives rise to a split of the $t_{2g}$ orbitals into spin-orbit-entangled states with $J$ = 1/2 and 3/2 characters \cite{Norman2010}. In the pure Kitaev limit on a hyperkagome lattice, a nonmagnetic ground state, possibly spin-liquid or valence-bond-solid, has been postulated \cite{Kimchi2014}. In reality, as in honeycomb iridates, other magnetic exchanges are present, and the antiferromagnetic Heisenberg coupling seems to be the leading term \cite{Micklitz2010}. A microscopic model that takes into account both the Heisenberg term and  anisotropic exchanges, such as the DM interaction, predicts the emergence of $q$ = 0 noncoplanar order or incommensurate order depending on the magnitude of the anisotropic terms \cite{Che08,Shindou2016,Mizoguchi2016}. The spin-glass state of Na$_4$Ir$_3$O$_8$ may be associated with the presence of such competing magnetic phases.

\subsubsection{Pseudospin-1/2 on pyrochlore lattice}

In addition to the hyperkagome lattice, the pyrochlore lattice of pseudospin-1/2 states with an edge-shared bonding geometry is found in an A-site deficient spinel Ir$_2$O$_4$ \cite{Kuriyama2010}. Theoretically, Ir$_2$O$_4$ is discussed to host spin-ice-type ``2-in-2-out'' magnetic correlation and  a U(1) quantum spin-liquid state is predicted under tetragonal strain \cite{Onoda2019}. Ir$_2$O$_4$ has been obtained only in a thin-film form, and its magnetic properties remain yet to be investigated.

The frustrated magnetism of Ir$^{4+}$ moments is also realized in pyrochlore oxides A$_2$Ir$_2$O$_7$ (A: trivalent cation). The electronic ground state of A$_2$Ir$_2$O$_7$ depends on the size of the A-ion (ionic radius $r_{\rm A}$), which likely controls the bandwidth of Ir 5$d$ electrons \cite{Yanagishima2001}. With the largest $r_{\rm A}$ in the family of A$_2$Ir$_2$O$_7$, Pr$_2$Ir$_2$O$_7$ exhibits a metallic behavior down to the lowest temperature measured. With a slightly smaller $r_{\rm A}$ such as Nd$^{3+}$, Sm$^{3+}$ or Eu$^{3+}$, A$_2$Ir$_2$O$_7$ shows a metal-to-insulator transition as a function of temperature \cite{Matsuhira2007}, accompanied by a magnetic order. For an even smaller $r_{\rm A}$ than that of Eu$^{3+}$, A$_2$Ir$_2$O$_7$ remains insulating up to well above room temperature while magnetic ordering takes place only at a low temperature, pointing to the Mott insulating state \cite{Matsuhira2011}. We focus here on the magnetism of Ir pseudospin-1/2 moments in the Mott insulating ground state. The metallic states of the pyrochlore iridates will be discussed in Sec.VI.B.

For the pyrochlore iridates in the insulating limit, the local electronic state of Ir 5$d$ electrons is primarily of $J$ = 1/2 character, but a sizable mixing of the $J$ = 3/2 components is present \cite{Shinaoka2015}. The mixing was attributed to the presence of a trigonal crystal field, which is generated not only by the oxygen cage but also by the surrounding cations \cite{Hozoi2014}. Recently, it turned out that the inter-site hopping plays a dominant role in the $J$ = 3/2 mixing. In fact, by suppressing the hopping, a nearly pure $J$ = 1/2 state can be realized \cite{Krajewska2020}.  

\begin{figure}[t]
\begin{center}
\includegraphics[scale=0.58]{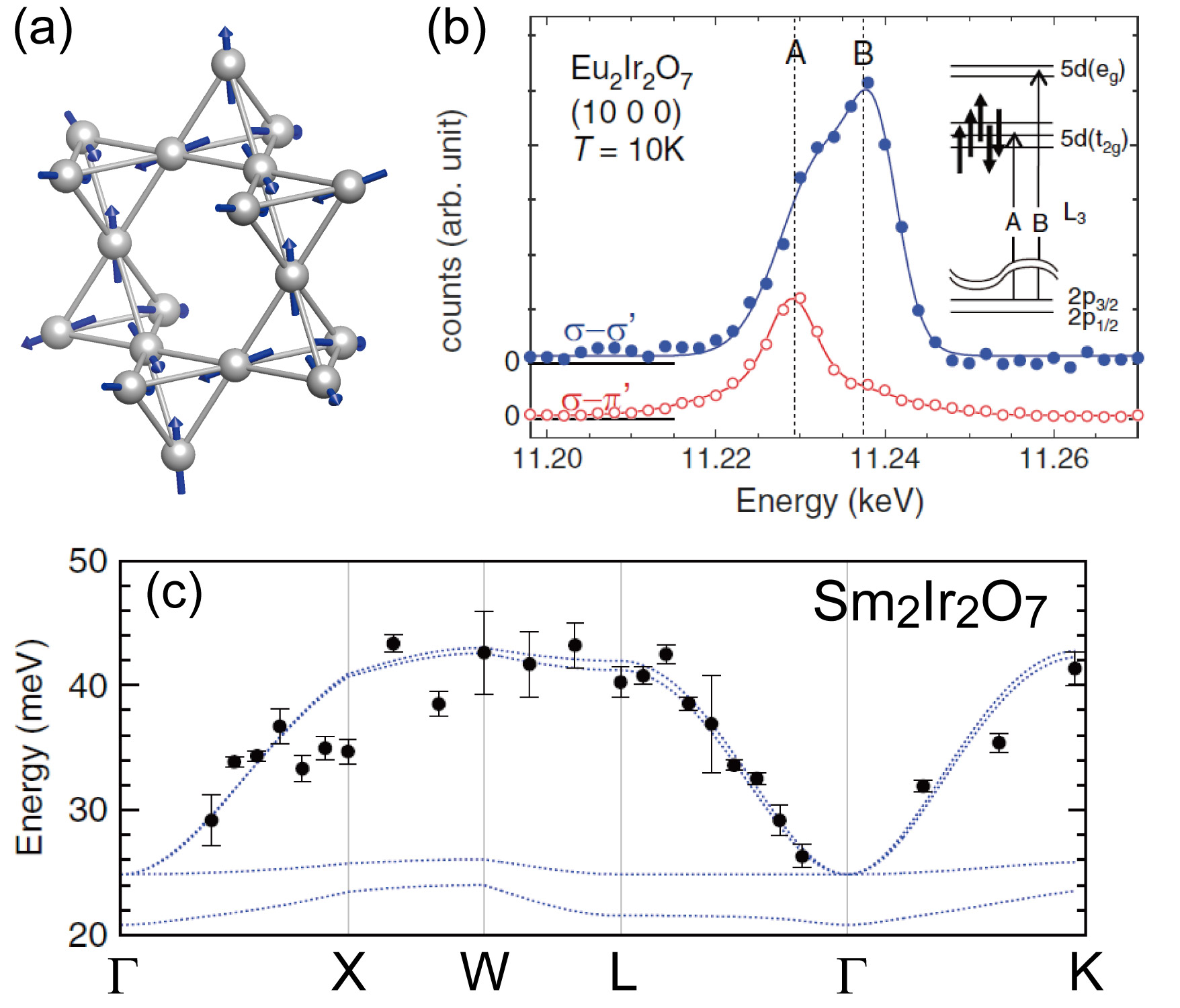}
\caption{(a) All-in-all-out (AIAO) magnetic ordering on a pyrochlore lattice. (b) The $q$ = 0 magnetic order revealed by resonant x-ray scattering in Eu$_2$Ir$_2$O$_7$ at the Ir $L_3$ absorption edge. The peaks at A(B) correspond to Ir 2$p_{3/2}$ $\rightarrow$ $t_{2g}$ (2$p_{3/2}$ $\rightarrow$ $e_g$) excitation, respectively. The strong resonant enhancement at A in the $\sigma$-$\pi^{\prime}$ channel points to a magnetic scattering. The resonant enhancements in the $\sigma$-$\sigma^{\prime}$ channel at both A and B originate from anisotropic tensor susceptibility (ATS) scattering. The figure is reproduced with permission from Ref.~[\onlinecite{Sagayama2013}] ($\copyright$2013 the American Physical Society). (c) The energy dispersion of magnetic excitation of Sm$_2$Ir$_2$O$_7$ obtained from RIXS. The black dots are the experimental data points and the blue dotted lines show the calculated magnon dispersion assuming AIAO magnetic order. The figure is reproduced with permission from Ref.~[\onlinecite{Donner2016}] ($\copyright$2016 the American Physical Society).
}
\label{fig:AIAO}
\end{center}
\end{figure}

The magnetic interaction between the $J$ = 1/2 pseudospins is predominantly attributed to the antiferromagnetic superexchange interaction via oxygen ions, where the Ir-O-Ir angle is approximately 130$^{\circ}$. Although the antiferromagnetic Heisenberg model on the pyrochlore lattice is predicted to show no magnetic ordering down to 0 K \cite{Moessner1998}, a DM interaction is present in the pyrochlore oxides as there is no inversion symmetry between the NN Ir atoms. It was shown in the spin Hamiltonian including Heisenberg and DM interaction on a pyrochlore lattice that the positive DM term gives rise to the all-in-all-out (AIAO) magnetic order, where all the magnetic moments on a tetrahedron of pyrochlore lattice are pointing inward or outward along the local [111] direction [Fig.~\ref{fig:AIAO}(a)] \cite{Elhajal2005}. On the other hand, when the DM term is negative, noncoplanar XY-type magnetic order appears where the moments lie in the plane perpendicular to the local [111] direction.

The magnetic structure of pyrochlore iridates has been studied by resonant x-ray scattering, and the $q$ = 0 magnetic order of Ir moments was revealed in Eu$_2$Ir$_2$O$_7$ [Fig.~\ref{fig:AIAO}(b)] \cite{Sagayama2013} . The $q$ = 0 propagation vector suggests the formation of the AIAO magnetic ordering or the non-coplanar XY antiferromagnetic order, as expected for the presence of DM interactions. In Sm$_2$Ir$_2$O$_7$ and Eu$_2$Ir$_2$O$_7$, the gapped magnon dispersion revealed by RIXS [Fig.~\ref{fig:AIAO}(c)] supports the AIAO order of Ir moments\cite{Donner2016,Hwan2018,EKHLee2013}. The AIAO magnetic order is discussed to give rise to a Weyl semimetallic state in the vicinity of the metal-insulator transition\cite{XWan2011}.

%}}}

\section{$\vc J$ = 0 systems: excitonic magnetism}
%{{{1

Perhaps the most radical impact of SOC on magnetism is
realized in compounds of $4d$ and $5d$ ions with $d^4$ configuration, such as
Re$^{3+}$, Ru$^{4+}$, Os$^{4+}$, and Ir$^{5+}$. For these ions with 
spatially extended $d$-orbitals, Hund's coupling is smaller than the
octahedral crystal field splitting 10$Dq$, so all four electrons occupy
$t_{2g}$ levels. The resulting $t_{2g}^4$ configuration has total spin
$S=1$ and threefold orbital degeneracy described by an effective orbital
moment $L=1$. Despite having well defined spin and orbital moments on every
lattice site, some $d^4$ compounds lack any magnetic order. This is because
SOC $\lambda \vc S\vc L$  with $\lambda>0$ binds $\vc S$ and
$\vc L$ moments into a local singlet state with zero total angular momentum
$J=0$, as shown in Fig.~1. 

Nonetheless, these nominally ``nonmagnetic'' ions may develop a collective
magnetism due to interaction effects \cite{Kha13, Meetei2015}. Although there are no preexisting local moments in the ionic ground state, the $J=1$ excitations become dispersive modes in a crystal,
and these mobile spin-orbit excitons may condense into a magnetically ordered
state. For this to happen, the exchange interactions should exceed a critical
value sufficient to overcome the energy gap $\lambda$ between $J=0$ and $J=1$
ionic states. The condensate wavefunction comprises a coherent superposition
of singlet and triplet states and carries a magnetic moment, whose length is
determined by the degree of admixture of triplets in the wavefunction.  Near the quantum critical point (QCP), the ordered
moment can be very small, and the magnetic condensate strongly fluctuates both in
phase and amplitude (i.e. rotation of moments and their length oscillations).
Formally, this is analogous to magnon condensation phenomenon in quantum dimer
systems~\cite{Gia08}, but the underlying physics and energy scales involved
here are different. While the spin gap in dimer models originates from antiferromagnetic exchange of two spins forming a dimer, the magnetic gap in $d^4$ systems is of intraionic nature and given by SOC. 

Spin-orbit-entangled $J=0$ compounds are interesting for possible novel phases near the
magnetic QCP, which can be driven by doping, pressure, and lattice control.
Here, the new element is that $J=1$ excitons are spin-orbit-entangled objects,
and, as we will see shortly, their interactions can be anisotropic and highly
frustrating. Thus, $J=0$ systems with ``soft'' moments can
naturally realize interplay between the two phenomena -- frustration and
quantum criticality -- a topic of current interest~\cite{Voj18}. As a general
property of orbitally degenerate systems, the symmetry and low-energy behavior of
spin-orbit excitonic models is dictated by chemical bonding geometry, and we
discuss two representative cases below.

%- FIGURE -----------------------------------------------------------------
\begin{figure}[tb]
\begin{center}
\includegraphics[scale=0.92]{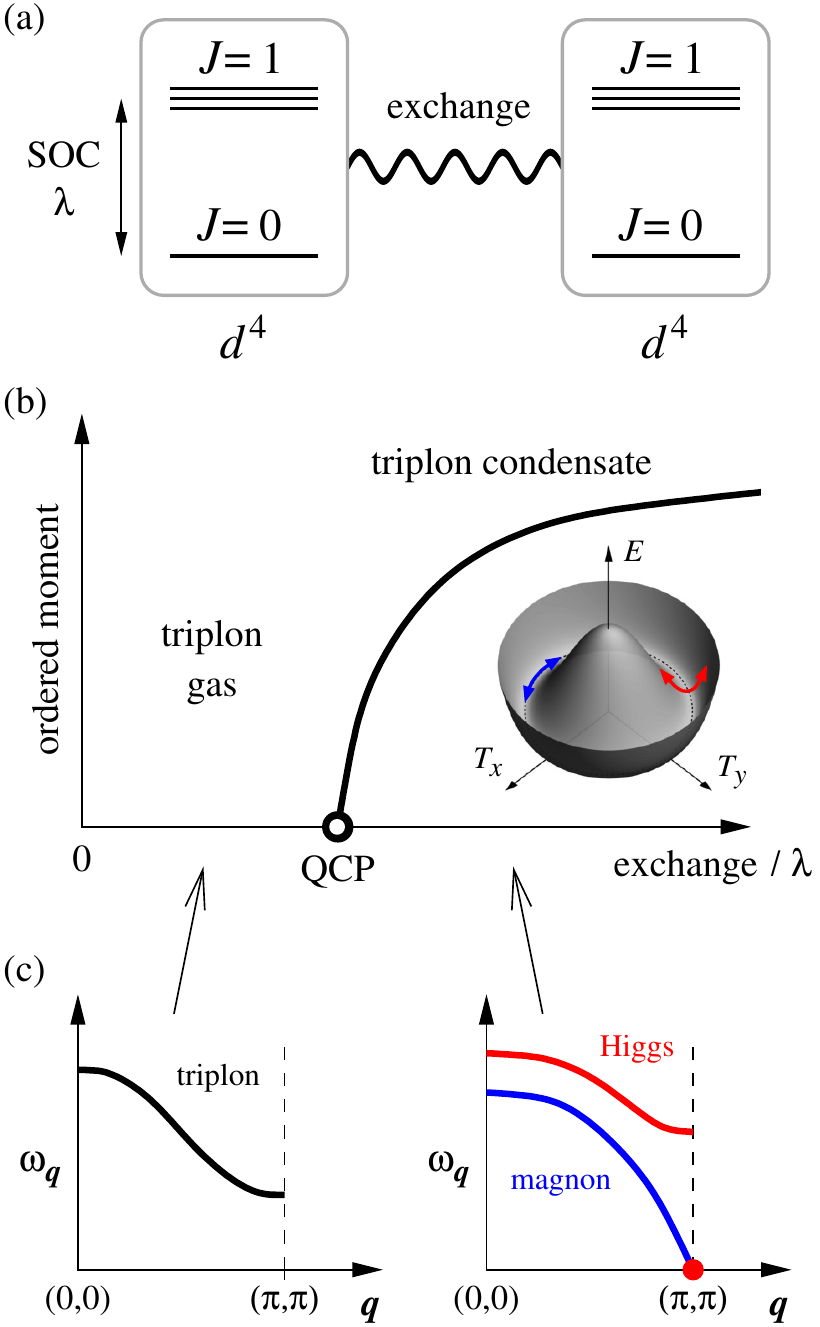}
\caption{
(a)~In the singlet-triplet model for $d^4$ systems with large SOC, each site
is supposed to host nonmagnetic $J=0$ ground state and low-lying $J=1$
triplet excitations at the energy $\lambda$. Competing with the intrasite SOC
gap $\lambda$ are various intersite exchange processes such as a transfer of a
$J=1$ excitation to a neighboring site or their pairwise creation and
annihilation.
(b)~Ordered moment in an excitonic magnet depending on the ratio of exchange
strength versus the local SOC gap $\lambda$. The quantum critical point (QCP)
separates the large-$\lambda$ phase where ``costly'' triplet excitations move
in an incoherent way and the phase where the condensate of triplets is
established.
(c)~Schematic dispersions of the elementary excitations. Before the
condensation, the elementary excitations are carried by triplons whose
dispersion softens near the AF momentum as the QCP is approached. Once the
triplon condensate is formed, the oscillations of its amplitude and the moment
direction become the new fundamental modes -- ``Higgs'' mode and magnons,
respectively. The red spot at ($\pi$, $\pi$) represents a magnetic Bragg point.
}
\label{fig:d4QCP}
\end{center}
\end{figure}
%--------------------------------------------------------------------------

\subsection{180$^\circ$ M-O-M bonding, perovskites}
%{{{2

Similarly to the pseudospin-$1/2$ $d^5$ case, the straight $180^\circ$ bond
geometry leads to a nearly isotropic model; in the following we thus first
focus on the isotropic model of $O(3)$ symmetry.  When approaching the excitonic
magnet formally, we need to properly reflect the ionic level structure with
nonmagnetic $J=0$ ground state and low-lying $J=1$ excitations. The most
natural way is to introduce hardcore bosons associated with the local
excitation $J=0\rightarrow 1$. These bosons, called here triplons $T$, come
with the energy cost $\lambda$ reflected by a local term $\lambda n_T = \lambda T^{\dagger}T$ and
experience various processes corresponding to the exchange interactions  between different sites.
In the second order in $T$ operators, they include triplon hopping and
creation or annihilation of triplon pairs in the symmetry-allowed combination
$\propto  ( T_{+1}T_{-1}-T_0T_0+T_{-1}T_{+1} )_{ij}$, where the indices ( $\pm 1, 0$ ) give the $J_z$ of
the triplons. Instead of $J_z$ eigenstates, it is convenient to use the
basis consisting of three triplon operators $T_\alpha$ of Cartesian flavors
(``colors'') $\alpha=x,y,z$ defined as
\begin{equation}\label{eq:Txyz}
T_x = \tfrac1{i\sqrt2}( T_1-T_{-1} ) , \;
T_y = \tfrac1{\sqrt2} ( T_1+T_{-1} ) , \;
T_z = i T_0, 
\end{equation}
which form a vector $\vc T$. The main contribution to the exchange $J$-Hamiltonian derived
in Ref.~[\onlinecite{Kha13}] then takes a manifestly $O(3)$ symmetric form:
\begin{equation}\label{eq:HTO3}
H=\lambda \sum_i \vc T^\dagger_i \vc T^{\phantom{\dagger}}_i
+\sum_{\langle ij\rangle} J_{ij} \left(
\vc T^\dagger_i 
\vc T^{\phantom{\dagger}}_j - 
\vc T^\dagger_i 
\vc T^\dagger_j + \mathrm{H.c.}\right) .
\end{equation}
The phase diagram of this model is determined by the competition of the
triplon cost $\lambda$ and superexchange coupling $J$ as schematically shown in
Fig.~\ref{fig:d4QCP}(b). At sufficient strength of the superexchange, triplons
undergo Bose-Einstein condensation like in spin-dimer systems \cite{Gia08}.
However, the physical meaning of triplons is very different here.
Since the magnetic moment of a $d^4$ ion resides primarily on the transition
between the $J=0$ and $J=1$ states, as described by $T$, the presence of a triplon
condensate with $\langle \vc T \rangle\propto i \mathrm{e}^{i\vc Q\vc R}$,
where $\vc Q$ is the ordering vector, directly translates to long-range
magnetic order. The resulting magnetic order is
characterized also by an unusual excitation spectra, see
Fig.~\ref{fig:d4QCP}(c). The condensation is preceded by softening of the
three-fold degenerate triplon modes near $\vc Q$. After condensation, the
modes split, giving rise to a two-fold degenerate magnon branch with $XY$-type
dispersion (i.e. with maximum at $q=0$) and the amplitude (Higgs) mode of the
condensate \cite{Kha13}. These two hallmarks of soft-spin magnetism can be
probed experimentally, as was done in the $J=0$ model system
\mbox{Ca$_2$RuO$_4$} with $d^4$ Ru$^{4+}$ ions \cite{Jai17,Sou17}.

%- FIGURE -----------------------------------------------------------------
\begin{figure}[tb]
\begin{center}
\includegraphics[scale=0.93]{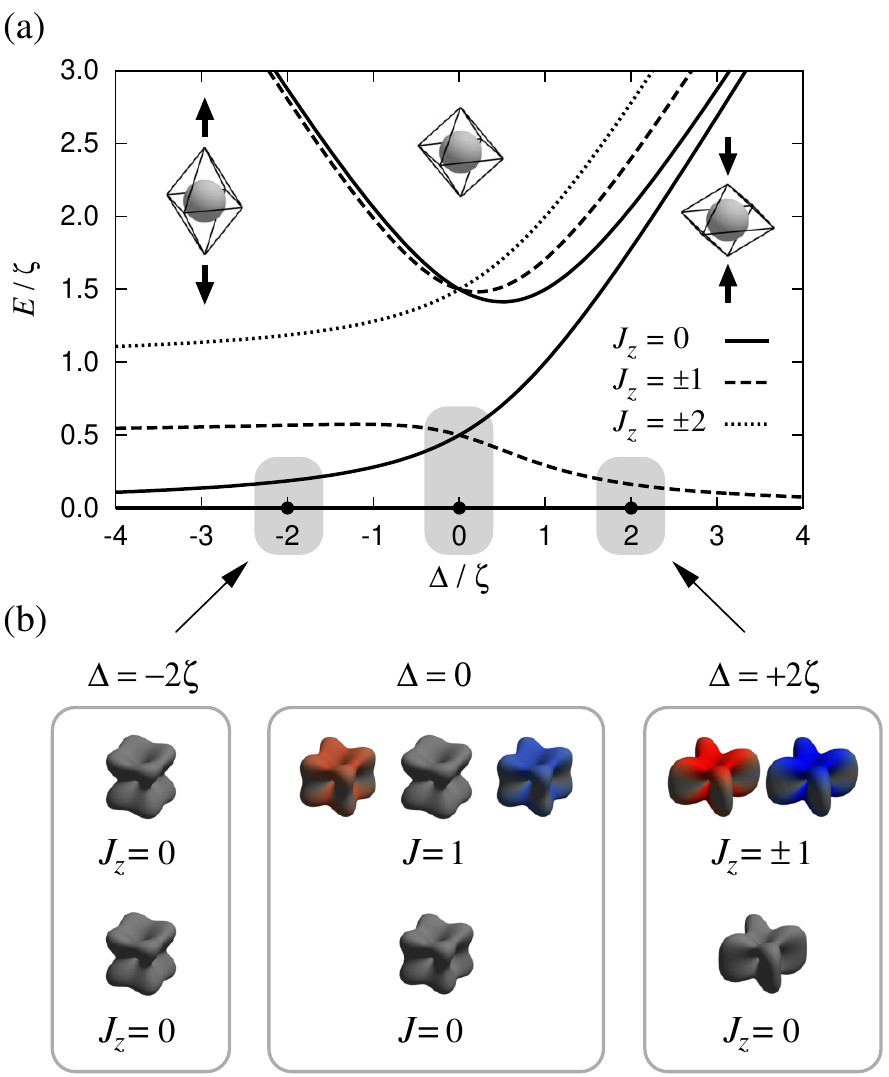}
\caption{
(a)~Splitting of $d^4$ levels in tetragonal crystal field (measured in units
of $\zeta=2\lambda$). Out of the three triplet excitations, MeO$_6$ octahedra
elongation/compression selects single $J_z=0$ state or the pair of $J_z=\pm 1$
states. Together with the $J_z=0$ ionic ground state (evolving from cubic
$J=0$ state), they form a local basis for an effective spin-1/2 (at
$\Delta<0$) or spin-1 (at $\Delta>0$) low-energy models.
(b)~Shapes of the relevant low-energy states represented in the same way as in
Fig.~\ref{fig:d1245} (electron density is used, not the hole one).
}
\label{fig:d4split}
\end{center}
\end{figure}
%--------------------------------------------------------------------------

The perovskite ruthenate Ca$_2$RuO$_4$ was identified as a Mott
insulator\cite{Nak97,Nak00} showing a metal-insulator transition around
$360\:\mathrm{K}$ \cite{Ale99} and antiferromagnetic order below $T_N\approx
110\:\mathrm{K}$ \cite{Nak97,Cao97,Bra98}. Early experiments \cite{Miz01}
revealed that SOC induces a substantial orbital angular momentum in Ru $4d$
levels, supporting the above $J=0$ picture.
In Ru$^{4+}$ ions, the SOC strength is roughly $\zeta\approx
150\:\mathrm{meV}$, giving the magnetic gap between $J=0$ and $1$ states of
the order of $\lambda=\zeta/2 \approx 75\:\mathrm{meV}$. Compared to
$\lambda$, exchange interactions are somewhat smaller in ruthenates and would
not be able to overcome such a gap. However, as shown in
Fig.~\ref{fig:d4split}(a), tetragonal distortion and the associated crystal
field $\Delta$ splits the $J=1$ excitation and may reduce the gap
significantly. The lower doublet $T_{x/y}$ (for the $\Delta>0$ case relevant for
Ca$_2$RuO$_4$) can then condense, giving rise to magnetic order, with the
ordered moment in the RuO$_2$ plane. This scenario points to an interesting
possibility of a lattice-controlled QCP (e.g. by strain) instead of by magnetic
field or high-pressure as in the dimer system TlCuCl$_3$ \cite{Rue03,Rue08}.
This also suggests the importance of the Jahn-Teller effect even in $J=0$ systems, which acts through the splitting of triplon levels and renormalization of the ground state wavefunction \cite{Liu19}.

%- FIGURE -----------------------------------------------------------------
\begin{figure}[tb]
\begin{center}
\includegraphics[scale=0.95]{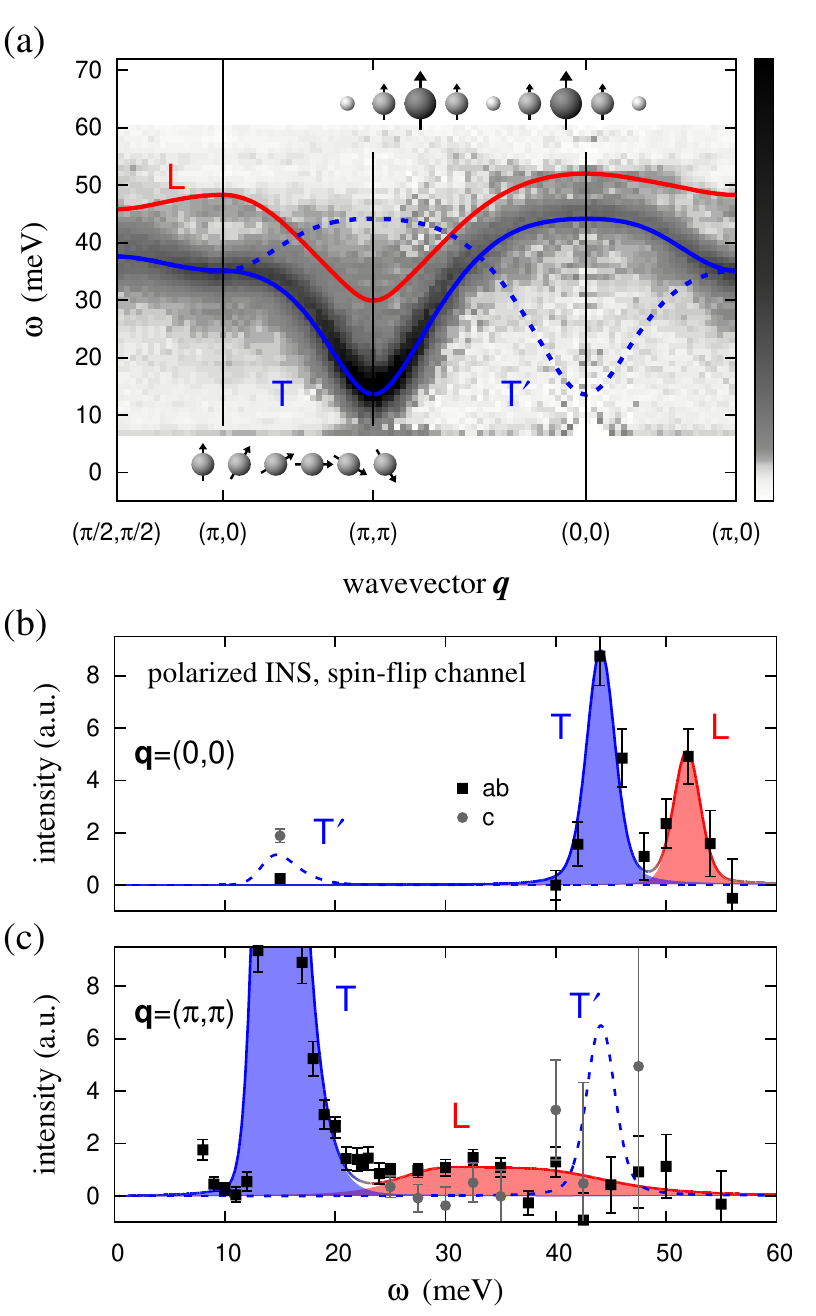}
\caption{
(a)~Magnetic excitations of \mbox{Ca$_2$RuO$_4$} mapped by inelastic neutron
scattering. The lines show model dispersions obtained within the model
of Ref.~[\onlinecite{Jai17}] (its simpler version is described in the text). 
The red line indicates longitudinal mode L corresponding to the Higgs mode, and the
blue lines represent the in-plane magnon T (solid line) and out-of-plane
magnon T' (dashed line).
Upper and lower insets are pictorial representations of the Higgs mode
(condensate amplitude oscillations) and magnons (rotations of
magnetic moments), respectively.
(b)~Magnetic response at zero wavevector $\vc q=(0,0)$ obtained by polarized INS.
In-plane polarization ($ab$, squares) and out-of-plane polarization ($c$,
circles) were resolved in the experiment. The experimental data are
overplotted on top of the theoretical magnetic response that is decomposed
according to the polarization of the modes.
(c) The same as in (b) for the ordering wavevector $\vc q=(\pi,\pi)$.
All the data are taken from Ref.~[\onlinecite{Jai17}] ($\copyright$2017 The Authors).
}
\label{fig:Ca2RuO4}
\end{center}
\end{figure}
%--------------------------------------------------------------------------

In Ca$_2$RuO$_4$ with $\Delta>0$, one of the $J=1$ states is lifted up by the
crystal field, and we are left with three low-energy states: ground state
singlet and the excited doublet, evolving from cubic $J=0$ and $J_z=\pm 1$ (or
$T_{x/y}$) states, respectively. These three states, whose wavefunctions are
shown in the right hand side of Fig.~\ref{fig:d4split}(b), can be used as a
local basis for an effective spin-1. The resulting $\tilde{S}=1$ Hamiltonian
obtained by mapping Eq.~\eqref{eq:HTO3} onto this basis has the form of the $XY$
model with a large single-ion anisotropy:
\begin{equation}
H=E \sum_i (\tilde{S}^z_i)^2
+J \sum_{\langle ij \rangle} \left( \tilde{S}_i^x \tilde{S}_j^x + \tilde{S}_i^y \tilde{S}_j^y \right) ,
\end{equation}
where $E$ denotes a singlet-doublet excitation gap that is smaller than
the singlet-triplet splitting $\lambda$ in Eq.~\eqref{eq:HTO3} due to a crystal
field effect. As a result, the exchange interaction $J$ may overcome the reduced
spin gap $E$ and induce magnetic order. 

The expected $XY$-type of magnon dispersion was indeed observed by inelastic
neutron scattering on \mbox{Ca$_2$RuO$_4$} \cite{Jai17}.  The experimental
dispersion presented in Fig.~\ref{fig:Ca2RuO4}(a) additionally features a
large magnon gap due to orthorhombicity, which is not included in the above
simplified model. The observation of the amplitude Higgs mode is to some
extent hindered by its strong decay into a 2D two-magnon continuum (as predicted
theoretically \cite{Pod11,Ros15}), which makes it a very broad feature in the
INS spectra near the AF wavevector $\vc Q=(\pi,\pi)$, see
Fig.~\ref{fig:Ca2RuO4}(c). On the other hand, the mode is relatively well
defined away from $\vc Q$ as visible in Figs.~\ref{fig:Ca2RuO4}(a),(b).
A more direct probe of the Higgs mode that enters INS spectra at momentum 
$\vc Q$ is the Raman scattering in the usually magnetically silent $A_g$ channel.
In Ref.~[\onlinecite{Sou17}], a Higgs mode in the scalar channel, ``unspoiled'' by the two-magnon
continuum, was identified in Raman spectra of \mbox{Ca$_2$RuO$_4$}, and found to
couple to phonons, giving them pronounced Fano-like lineshapes. Such
an interaction with lattice modes is natural for triplons, since they have a
``shape'' inherited from orbitals, and hence couple to lattice vibrations via
the Jahn-Teller mechanism as mentioned above.

Spin-orbit exciton condensation and related magnetic QCP are more likely
realized in $4d^4$ compounds such as ruthenates, where SOC  and exchange
interactions are of comparable scale and their competition can be tuned
experimentally. On the other hand, the $5d^4$ ion (Ir$^{5+}$ or Os$^{4+}$)
compounds are typically nonmagnetic, since spin-orbit $J=1$ excitations are
too high in energy, as evidenced by RIXS experiments in 5$d$-electron double
perovskites \cite{Dey16,Dav19,Yua17}. Weak magnetism detected in the $5d^4$
iridate has been explained as originating from the Ir$^{4+}$ and Ir$^{6+}$
magnetic defects, while the regular Ir$^{5+}$ sites remain indeed nonmagnetic
\cite{Fuc18}. 

%}}}

\subsection{90$^\circ$ M-O-M bonding, honeycomb lattice}
%{{{2

\begin{figure}[tb]
\begin{center}
\includegraphics[scale=0.95]{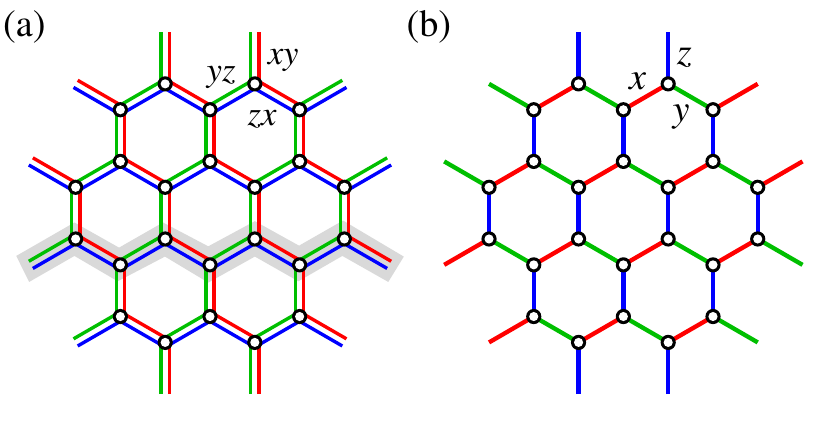}
\caption{
Patterns indicating active flavors (colors) for bond-selective triplon
interactions on honeycomb lattice: 
(a)~Oxygen mediated $t$-hopping case -- $XY$-type interactions -- two flavors
are active for a given bond directions (e.g., $T_x$ and $T_y$ on $z$-type
vertical bonds). Each color forms a system of separate zigzag chains (one of
the zigzag chains for $T_z$ boson is marked by shading). The symmetry
resembles famous compass models where each spin component interacts within its
own 1D chain. 
(b)~Direct $t'$-hopping case -- bond dependent Ising-type of interactions.
There is one-to-one correspondence between the active triplon color and the
bond direction ($T_z$ on $z$ bond, etc), establishing a bosonic analog of the
Kitaev model.
}
\label{fig:d4honey}
\end{center}
\end{figure}

When contrasting the $180^\circ$ and $90^\circ$ bonding geometries, we
encounter a situation analogous to the $J=1/2$ pseudospin case. While the
$180^\circ$ bonding geometry generates (in leading order) the isotropic $O(3)$
model of Eq.~\eqref{eq:HTO3}, discussed above, the bonds with $90^\circ$ oxygen
bridges are highly selective in terms of the active flavors for the triplon
interactions. Roughly speaking, when the oxygen-mediated hopping $t$
dominates, each bond allows exchange processes of the type contained in
Eq.~\eqref{eq:HTO3} for two triplon flavors only, depending on the bond
direction \cite{Kha13}. For instance, the $T_x$ and $T_y$ bosons are equally
active in $z$-type bonds, while the $T_z$ boson cannot move in that direction. For
the honeycomb lattice, the resulting pattern of active triplon pairs is
presented in Fig.~\ref{fig:d4honey}(a). On the other hand, the dominant direct
hopping $t'$ leads to the complementary Kitaev-like pattern of
Fig.~\ref{fig:d4honey}(b), with a direct correspondence between the bond
direction $\alpha$ and triplon flavor $T_\alpha$ active on that bond
\cite{Ani19,Cha19}. Each of these two cases is strongly frustrated; however,
the nature of the corresponding ground states is very different.

%- FIGURE -----------------------------------------------------------------
%--------------------------------------------------------------------------

In the interaction pattern of Fig.~\ref{fig:d4honey}(a), each bond shows an
$O(2)$ symmetry of the triplon exchange Hamiltonian, that is, 
\begin{equation}\label{eq:HTxy}
H^{(z)}_{ij} = J \sum_{\alpha=x,y} ( T^\dagger_{\alpha i} T^{\phantom{\dagger}}_{\alpha j} 
-T^\dagger_{\alpha i} T^\dagger_{\alpha j} + \mathrm{H.c.} ) 
\end{equation}
for a $z$-bond $\langle ij\rangle$. However, the global symmetry of the model
is only the discrete $C_3$ one. Namely, there are three zigzag chains (colored
differently), along which the individual triplon flavors can move.  This
arrangement does not support 2D long-range order but instead leads to
effective dimensionality reduction like in compass models \cite{Nus15}: $C_3$
symmetry is broken by selecting one particular triplon component, with
antiferromagnetic correlations along the corresponding 1D zigzag. Zigzag
chains interact via the hard-core constraint only (triplon density channel), so
there are no phase relations and  magnetic order between different chains. The
resulting magnetic correlations are highly anisotropic, both in real and spin
spaces. This is a combination of an orbital ordered and spin-nematic state, made
possible due to spin-orbital entanglement.  

The other limit illustrated by Fig.~\ref{fig:d4honey}(b) may be called a
bosonic Kitaev model, following the formal similarity of the triplon exchange
Hamiltonian, i.e.
$H^{(\alpha)}_{ij}=J ( T^\dagger_{\alpha i} T^{\phantom{\dagger}}_{\alpha j} 
-T^\dagger_{\alpha i} T^\dagger_{\alpha j} + \mathrm{H.c.} )$
for $\alpha$-type bonds, to the Kitaev interaction $K S^\alpha_i S^\alpha_j$.
As found in Ref.~[\onlinecite{Cha19}], the strong frustration of Kitaev-type
prevents a magnetic condensation at any strength of the exchange coupling $J$
relative to spin gap $\lambda$. Interestingly, the model shares a number of
other features with the spin-$1/2$ Kitaev model -- there is an extensive number
of $Z_2$ conserved quantities, magnetic correlations are strictly short-ranged
and confined to nearest-neighbor sites, and the excitation spectrum has a spin
gap. However, the strongly correlated triplon ``liquid'' ground state found in the
large exchange limit $J\gg\lambda$ is smoothly connected to dilute triplon gas
\cite{Cha19} and hence misses the defining characteristics (long-range
entanglement and emergent nonlocal excitations) of genuine spin liquids.
Consequently, no quasiparticle modes (like Majorana bands in spin-1/2 Kitaev
model) are present within the spin gap. Nonetheless, this strongly correlated
paramagnet is far from being trivial -- in contrast to what is conventional in pure
spin systems, magnetic correlations are highly anisotropic and strictly
short-ranged even in the limit where the spin gap is very small and the QCP is close
by. Magnetic order can be induced by subdominant (non-Kitaev type) triplon
interactions, as well as by doping, which suppresses the spin gap. Also, it has been found that triplon excitations acquire nontrivial band topology and protected edge states in a magnetic field \cite{Ani19}.

By mixing the above two complementary anisotropic limits with the
corresponding couplings $J\propto t^2/U$ and $J\propto t'^2/U$ in one-to-one
ratio, we recover an isotropic triplon model of Eq.~\eqref{eq:HTO3}. Since the
honeycomb lattice is not geometrically frustrated, the model shows the same
quantum critical behavior as in the square-lattice case, i.e. dispersing
triplons condense at a QCP and give rise to long-range antiferromagnetic order.
In this context, the ratio of the oxygen-mediated and direct hopping
amplitudes $t/t'$ turns out to be an important ``handle'' determining the
degree of frustration (as well as its type) of a singlet-triplet system with
$90^\circ$ bonding geometry.

On the materials side, the Ru-based honeycomb lattice compounds are potential
candidates to realize frustrated spin-orbit exciton models. In particular,
\mbox{Ag$_3$LiRu$_2$O$_6$} \cite{Kim10,Kum19,TakXX}, which is derived from
\mbox{Li$_2$RuO$_3$} by substituting Ag ions for Li ions between the honeycomb
planes, is of interest. While hexagonal symmetry is heavily broken by the
structural and spin-orbital dimerization in \mbox{Li$_2$RuO$_3$}
\cite{Miu07,Jac08,Kim14honeyRu}, \mbox{Ag$_3$LiRu$_2$O$_6$} avoids this
transition and thus may serve as a model system to study $J=0$ physics in a
nearly ideal honeycomb lattice. This compound shows no magnetic order
\cite{Kim10,Kum19,TakXX}, which implies that the triplon interactions are
either too weak to overcome the spin-orbit gap, or they are dominated by Kitaev-type couplings and thus highly frustrated. 

%}}}

To summarize this section, we note that physics of spin-orbit-entangled $J=0$ compounds
is still in its infancy, and indicate below a few directions for future
studies. 

(i) Frustrated spin-orbit exciton models, possible exotic phases and magnetic
QCP in these models; topological properties of spin-orbit excitations.
Experiments in various edge-shared structures and geometrically frustrated
lattices, pressure and strain control of magnetic and structural transitions. 

(ii) The nature of metallic states induced by electron doping, which injects
$J=1/2$ fermions moving in the background of $J=0$ states. Fermion hopping is
accompanied by creation and annihilation of spin-orbit excitons, which should
give rise to a strongly correlated metal. In the case of perovskite lattices with
$180^\circ$ bonding geometry, Ref.~[\onlinecite{Cha16}] suggested that electron
doping induces ferromagnetic correlations, and possible triplet pairing
mediated by spin-orbit excitations. On the experimental side, several studies
\cite{Cao00,Cao01,Nak13,Bos19} found doping driven ferromagnetic state in
\mbox{Ca$_2$RuO$_4$}; interestingly, the recent work has reported also on
signatures of superconductivity \cite{Nob20}. In compounds with $90^\circ$
bonding geometry, the hopping rules are different and interactions are
frustrated; studies of doping effects in such systems may bring some
surprises.  

%}}}

\section{Multipolar physics in $d$-electron systems with strong spin-orbit coupling}

Multipolar ordering in Mott insulators covers a whole host of phenomena, ranging from the relatively standard quadrupole ordering of $e_g$ electrons due to a cooperative Jahn-Teller effect \cite{Kug82,Kha05} to the formation of bond multipoles in highly quantum-entangled frustrated magnets \cite{andreev84,shannon06}.

In the absence of significant SOC, the orbital and spin degrees of freedom typically order at different temperatures.
At high temperature the cooperative Jahn-Teller effect drives both a structural distortion of the lattice and an associated orbital quadrupole order, while at lower temperature the exchange interaction causes the spins to order.  
%
%The ordering of the orbitals typically breaks the original cubic symmetry in a quadrupolar fashion, and so the intermediate phase has orbital quadrupole order.
%
Strong SOC ties together the spin and orbital degrees of freedom, negating the simple picture of separate transitions. 
As a consequence, the intermediate phase picks up a spin contribution to the quadrupole order, and at the same time the two transitions tend to get pushed closer together in temperature.

The introduction of strong SOC also opens up the possibility of unusual types of interactions, in particular higher-order biquadratic and triatic terms in the Hamiltonian.
These interactions can drive more unusual types of multipolar order, such as an octupolar ground state similar to those found in $f$-electron systems \cite{kuramoto09,santini09}, or, when combined with frustration, cause the multipolar order to melt away, leaving behind a multipolar spin-liquid.

%%%%%%%%%%%%%%%%%%%%%%%%%%%%%%%%%%%%%%%%%%%%%%%%%%%%%%%%%%

\subsection{Quadrupole ordering}

%%%%%%%%%%%%%%%%%%%%%%%%%%%%%%%%%%%%%%%%%%%%%%%%%%%%%%%%%%

The idea of tensor order parameters, familiar from the theory of classical multipole ordering, can be readily generalised to the quantum case.
Just as dipolar order is associated with a non-zero expectation value of the vector $\langle {\bf J} \rangle$, quadrupole order is associated with a finite expectation value of the rank-2 tensor,
 \begin{align}
 Q_i^{\mu\nu} = \frac{1}{2}\langle J_i^\mu  J_j^\nu +J_i^\nu  J_j^\mu \rangle - \frac{\langle {\bf J}_i \cdot {\bf J}_j \rangle}{3} \delta^{\mu\nu},
 \quad \mu,\nu \in \{ x,y,z\}
\end{align}
where the site indices $i,j$ can refer to the same or different sites.
%

%%%%%%%%%%%%%%%%
\subsubsection{Quadrupoles in $d^1$ systems}
%%%%%%%%%%%%%%%%

Quadrupole ordering is very common in strongly spin-orbit-entangled $d^1$ systems, since $d^1$ ions with a $J$ = 3/2 ground state are Jahn-Teller active, as discussed in Section II.A. 
The onset of quadrupole order occurs when the degeneracy of the $J=3/2$ quadruplet is split, selecting a low energy $J_z=\pm 1/2$ or $J_z=\pm 3/2$ doublet.
These doublets have a quadrupolar charge distribution (see Fig.~\ref{fig:d1245}), as opposed to the cubic charge distribution of the $J = 3/2$ quadruplet, and thus quadrupolar ordering occurs simultaneously with a structural transition in which the local symmetry of the $d^1$ ions is reduced.

\begin{figure}[h]
\centering
\includegraphics[width=0.4\textwidth]{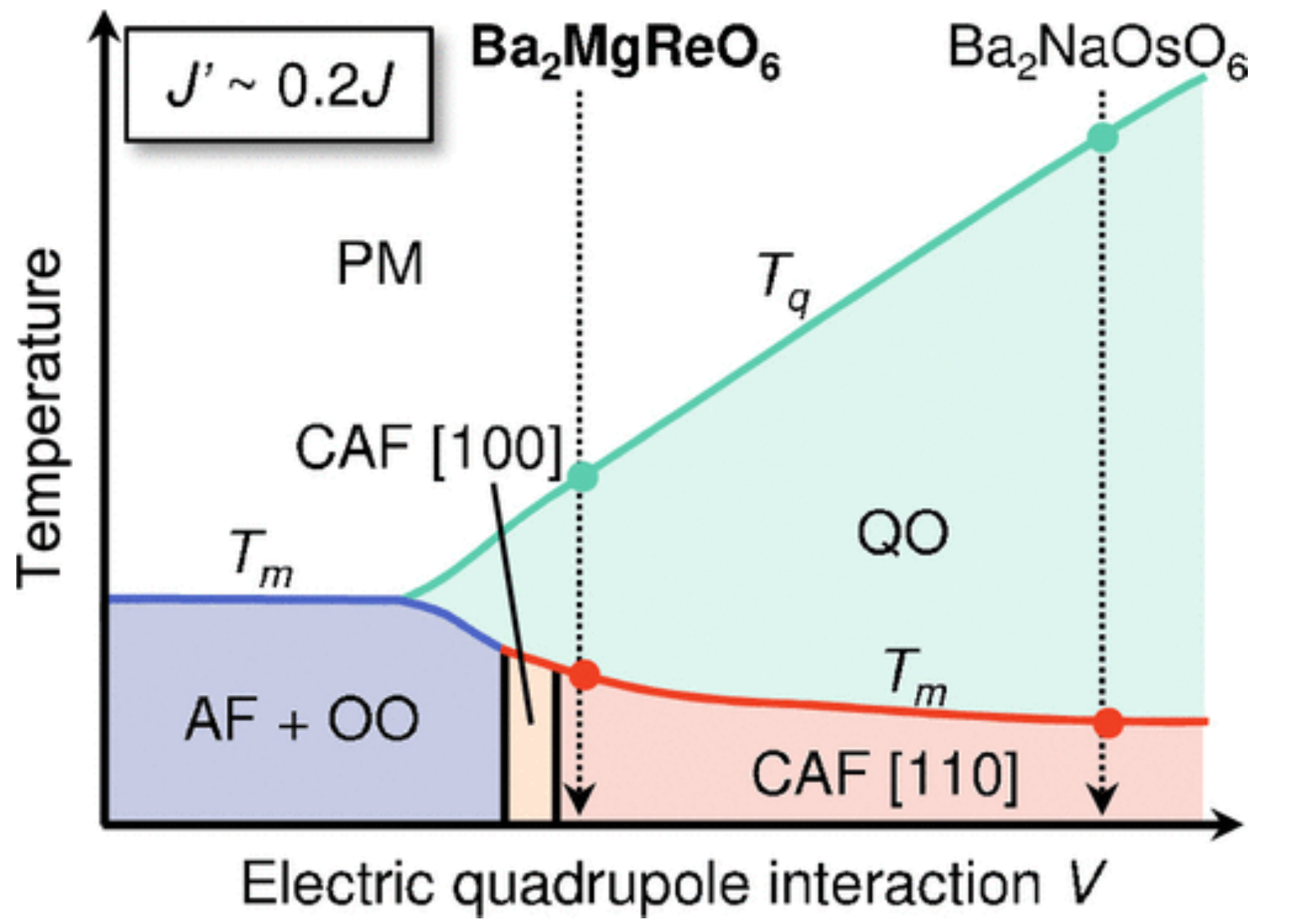}
\caption{
Schematic phase diagram proposed for strongly spin-orbit coupled $d^1$ ions.
For a large enough interaction $V$ there are two phase transitions as a function of temperature, with a high-temperature transition into a quadrupole ordered (QO) phase followed by a low-temperature transition into one of various dipolar phases, including antiferromagnetic (AF) and canted antiferrogmanetic (CAF[100], CAF[110]) orders.
The figure is reproduced with permission from Ref.~[\onlinecite{hirai19}] ($\copyright$2019 The Physical Society of Japan).
}
\label{fig:d1phasediag}
\end{figure}

The driving force for the quadrupole-ordering transition comes predominantly from electrostatic and Jahn-Teller interactions, with a helping hand from the exchange interaction.
A good way to see this theoretically is to project the well-known interactions for the 6-fold degenerate $t_{2g}$ manifold of electron configurations into the $J=3/2$ quadruplet \cite{chen10}. 
In addition to the usual bilinear interactions, the resulting effective Hamiltonian also contains large biquadratic interactions, such as $(J_i^z)^2(J_j^z)^2$, between neighbouring sites.
It has been known for a long time that these can be rewritten as quadrupole-quadrupole interactions \cite{blume69,chen71,papanicolaou88}, and so it is not surprising that they favour quadrupole ordering.

\begin{figure}[h]
\centering
\includegraphics[width=0.4\textwidth]{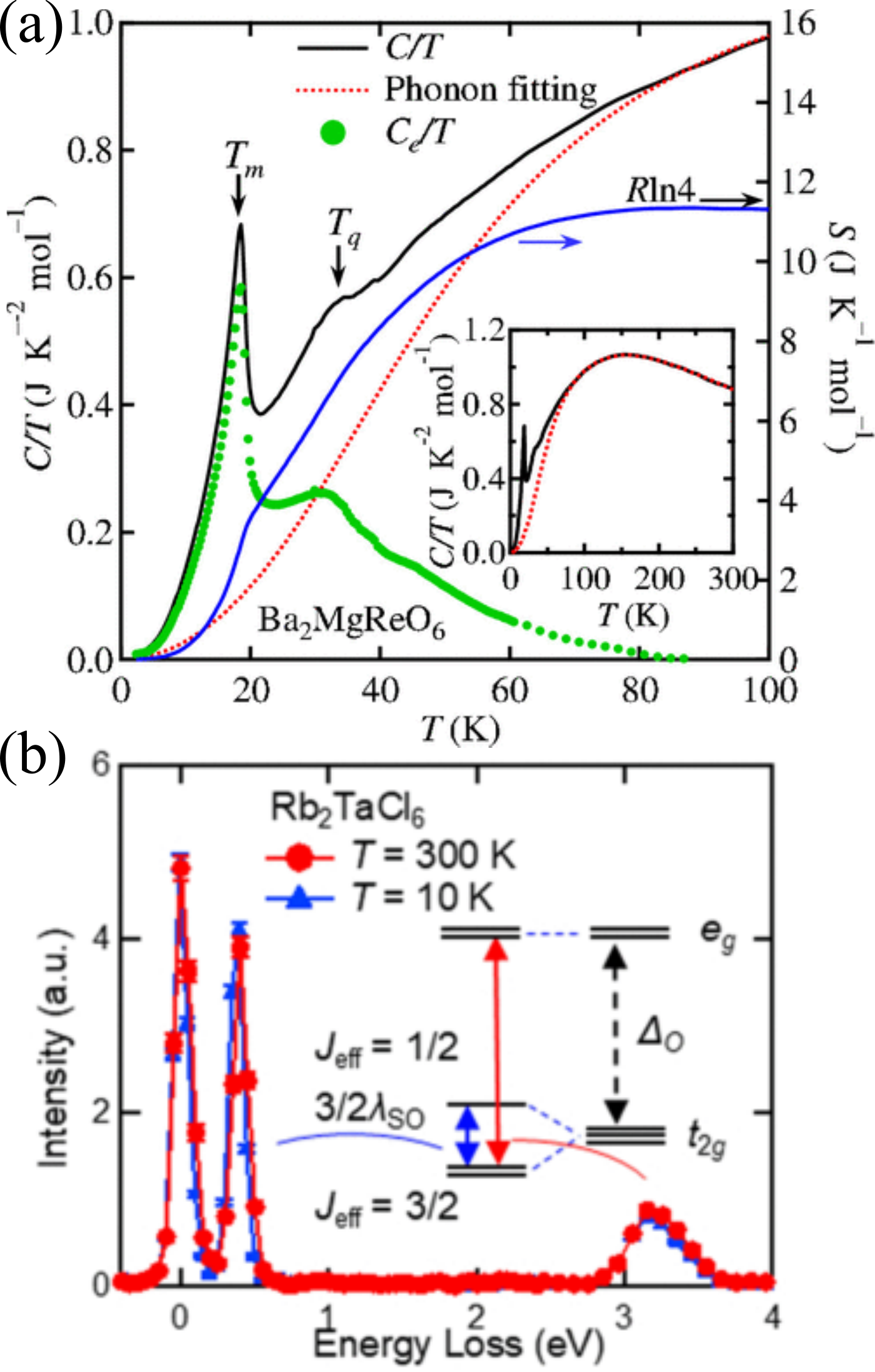}
\caption{
Evidence for the formation of a strongly spin-orbit-entangled $J=3/2$ state.
(a) Recovery of $R\ln$4 entropy in Ba$_2$MgReO$_6$ at high temperatures, taken with permission from Ref.~[\onlinecite{hirai19}] ($\copyright$2019 The Physical Society of Japan).
(b) Ta $L_3$-edge RIXS spectrum showing the splitting of the $t_{2g}$ level by SOC. Reproduced with permission from Ref.~[\onlinecite{ishikawa19}] ($\copyright$2019 the American Physical Society).
}
\label{fig:d1}
\end{figure}

Double-perovskite oxides [Fig.~\ref{fig:perovskite}(d)] provide some of the cleanest material realisations of spin-orbit-entangled $d^1$ Mott insulators.
The wide spacing of the magnetic ions makes them good Mott insulators with small intersite interactions, and allows a cubic ionic environment to be retained to low temperature. Figure \ref{fig:d1phasediag} illustrates a generic phase diagram proposed for $d^1$ systems with double-perovskite structure \cite{hirai19}.

While none of the known double-perovskite materials have a completely vanishing dipolar magnetic moment, as would be expected for isolated $J=3/2$ ions (see Sec.II.A), the magnetic moments are small, indicating only weak hybridisation with the surrounding oxygen ions. 
For example Ba$_2$NaOsO$_6$ has an effective moment of approximately $0.6\;\mu_{\sf B}$ \cite{erickson07}.
It also shows two transitions, with a higher temperature structural transition at $T_{\sf q}=9.5$ K suggestive of the onset of quadrupolar order, and a lower temperature transition at $T_{\sf m} \approx 7.5$ K into a magnetically ordered phase \cite{erickson07,lu17,willa19}.
However, since the symmetry above $T_{\sf q}$ is likely tetragonal rather than cubic, as suggested by the approximately $R$ln2 entropy recovery above $T_{\sf q}$ \cite{erickson07}, it is not clear how effectively the system explores the full $J=3/2$ manifold at higher temperatures.

A similar story is found in Ba$_2$MgReO$_6$, where there is an effective moment of about $0.7\;\mu_{\sf B}$, a high-temperature transition at $T_{\sf q} \approx 33$ K and a low-temperature transition at $T_{\sf m} \approx 18$ K to a similar magnetically ordered state to Ba$_2$NaOsO$_6$ \cite{marjerrison16,hirai19}.
However, unlike Ba$_2$NaOsO$_6$ the high temperature structure is cubic, and heat capacity measurements reveal that the full $R \ln 4$ entropy of the $J=3/2$ multiplet is obtained above about 80 K, as shown in Fig.~\ref{fig:d1}(a) \cite{hirai19}.
A small distortion of ReO$_6$ octahedra was observed below $T_{\sf q}$, which is consistent with quadrupole ordering \cite{hirai20}. 
%Questions remain about the presence or absence of the expected structural transition at  $T_{\sf q}$, with one suggestion being that antiferro-type ordering of quadrupoles may result in a small distortion that is difficult to detect \cite{hirai19}.

The closely related A$_2$TaCl$_6$ (A = Cs, Rb) family appears to provide a particularly good realisation of the $J=3/2$ state, as can be seen from the small  effective magnetic moment of $0.2-0.3\;\mu_{\sf B}$ \cite{ishikawa19}.
The suitability of the $J=3/2$ description has been confirmed by RIXS experiments [Fig.~\ref{fig:d1}(b)] and the recovery of the expected   $R \ln 4$ entropy at high temperature.
As with the double perovskite oxides, two transitions are observed, with the upper transition at $T_{\sf q}\approx 30$ K for Cs and $T_{\sf q}\approx 45$ K for Rb and the lower transition at  $T_{\sf m}\approx 5$ K for Cs and $T_{\sf m}\approx 10$ K for Rb.
The upper transition is associated with a structural transition from cubic to compressed tetragonal, and is suggestive of a ferro-quadrupolar phase forming via selection of the $J_z=\pm1/2$ doublet.

%%%%%%%%%%%%%%%%
\subsubsection{Quadrupoles in $d^2$ systems}
%%%%%%%%%%%%%%%%

Quadrupolar order for $d^2$ ions can be expected either from ordering of the low-lying nonmagnetic $E_g$ doublet (see Fig.~\ref{fig:d2split}), or driven by a combination of electrostatic, Jahn-Teller and exchange interactions acting within the full $J=2$ quintuplet \cite{chen11}.
However, there is currently a lack of materials showing the type of double quadrupolar and magnetic transitions observed in many $d^1$ compounds.

%%%%%%%%%%%%%%%%%%%%%%%%%%%%%%%%%%%%%%%%%%%%%%%%%%%%%%%%%%

\subsection{Octupole ordering}

%%%%%%%%%%%%%%%%%%%%%%%%%%%%%%%%%%%%%%%%%%%%%%%%%%%%%%%%%%

Octupole phases involve the ordering of the rank-3 tensor,
\begin{align}
O_i^{\mu\nu\xi} = \langle \overline{ J^\mu_i  J^\nu_j J^\xi_k } \rangle,  \quad \mu,\nu, \xi \in \{ x,y,z\},
\end{align}
in the absence of dipolar or quadrupolar order, where the bar indicates symmetrisation over the superscripts.

%%%%%%%%%%%%%%%%
\subsubsection{Octupoles in $d^1$ systems}
%%%%%%%%%%%%%%%%

\begin{figure}[h]
\centering
\includegraphics[width=0.48\textwidth]{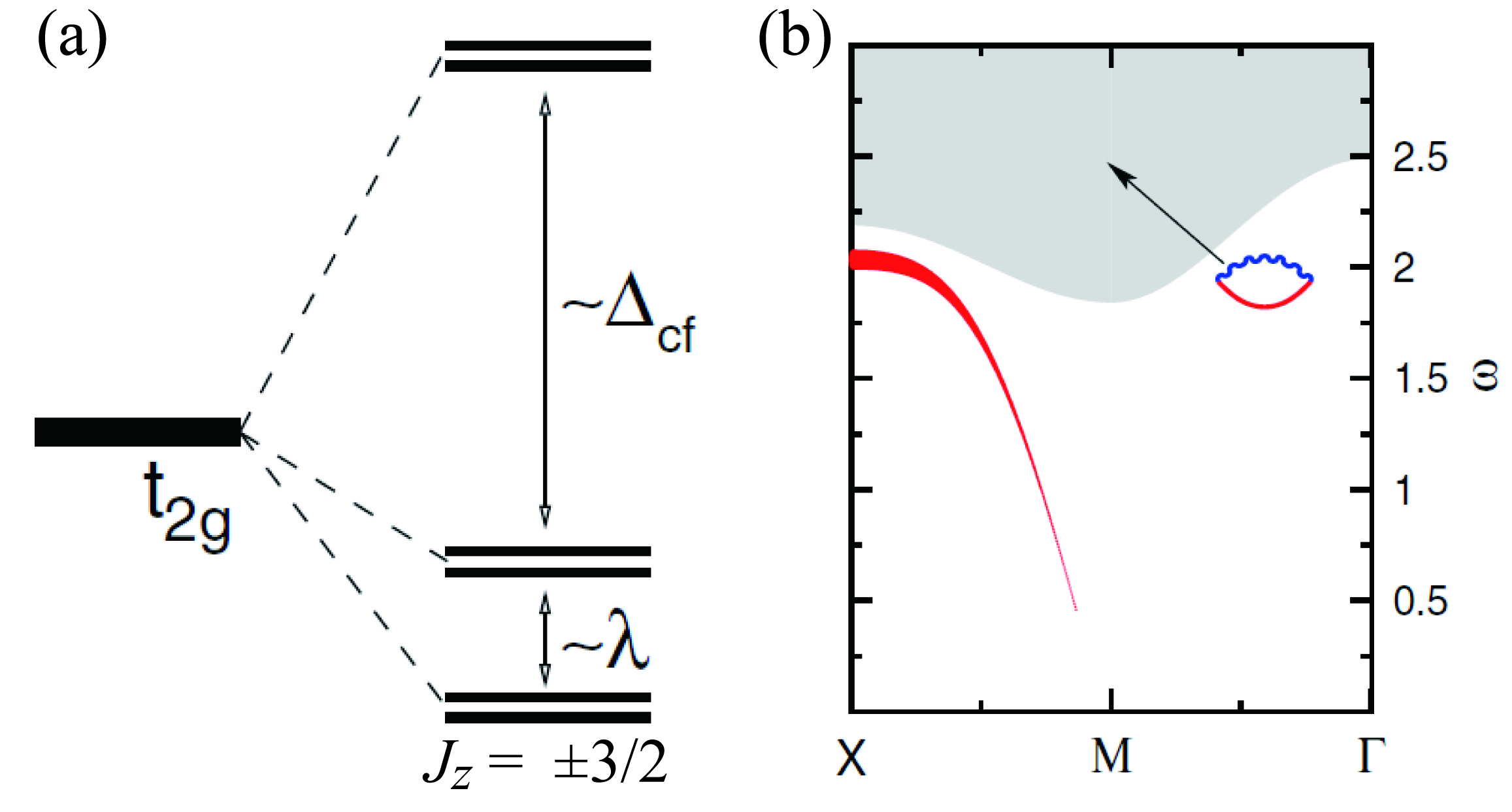}
\caption{
Local states and collective excitations in Sr$_2$VO$_4$.
(a) Splitting of the V$^{4+}$ $t_{2g}$ levels by a tetragonal crystal field $\Delta_{\rm cf}$ and spin-orbit coupling $\lambda$ results in a $J_z=\pm 3/2$ lowest-energy doublet hosting octupolar moment.
(b) Prediction for the magnetic response associated with octupolar order.
There is a sharp dispersive band of octupolar-wave excitations below the continuum, whose spectral weight in magnetic channel (shown by line width) disappears approaching the octupolar Bragg point at $M=(\pi,\pi)$, reflecting the absence of dipolar order in the ground state. The energy $\omega$ is in units of $J=t^2/U$.
The figures are taken with permission from Ref.~[\onlinecite{jackeli09}] ($\copyright$2009 the American Physical Society).
}
\label{fig:Sr2VO4}
\end{figure}

A candidate to realise octupolar order in the absence of any concomitant dipolar order is the material Sr$_2$VO$_4$ with perovskite structure.
Although V$^{4+}$ ($3d^1$) is not usually thought of as a strongly spin-orbit coupled ion, the combination of SOC and a tetragonal elongation of the oxygen octahedra conspire to select a $J_z = \pm 3/2$ lowest-energy doublet from the $t_{2g}$ manifold, as can be seen in Fig.~\ref{fig:Sr2VO4}(a) \cite{jackeli09}.
Projection of the usual exchange Hamiltonian for $t_{2g}$ electrons into this ground state doublet reveals a checkerboard ground state of alternating $|\psi \rangle = (| 3/2 \rangle \pm | -3/2 \rangle)/\sqrt{2}$ states, which corresponds to a staggered octupolar order.
One possible signature of this octupolar order would be a Goldstone mode with vanishing magnetic response at low energies, potentially visible in inelastic neutron scattering, as shown in Fig.~\ref{fig:Sr2VO4}(b).
Experimental studies are consistent with the local level scheme proposed for the V ions, but the question of whether the ground state is octupolar ordered remains open \cite{zhou07,zhou10,teyssier11,teyssier16}.

%

%%%%%%%%%%%%%%%%
\subsubsection{Octupoles in $d^2$ systems}
%%%%%%%%%%%%%%%%

Octupolar order has been suggested to be realised in the $d^2$ double perovskite family Ba$_2${\it M}OsO$_6$ ({\it M} = Zn, Mg, Ca) \cite{paramekanti20,maharaj20}. 
While phase transitions are observed at approximately 30 K (Zn) and 50 K (Ca, Mg), there is no associated development of dipolar magnetic order \cite{thompson14, marjerrison16_2,maharaj20}.
Furthermore, the development of quadrupolar order is incompatible with the absence of detectable lattice distortion.
At the same time the recovery of only $R\ln 2$ of entropy at temperatures considerably above the transition is indicative of a low-lying doublet, and matches the expected $E_g$-$T_{2g}$ splitting shown in Fig.~\ref{fig:d2split}.

From a theoretical point of view, projection of the interactions between $t_{2g}$ electrons into the $J=2$ quintuplet shows the importance of bitriatic interactions, such as $(J^z_i)^3(J^z_j)^3$ \cite{chen11,paramekanti20}.
These can be rewritten as interactions between octupoles, and, if large enough compared to competing bilinear and biquadratic interactions, can drive the formation of octupolar order.
This may provide a mechanism for selecting octupolar order with a ferro-octupolar ground-state wavefunction that is a complex mix of the $E_g$ states as shown in Fig.~\ref{fig:d2split}(c), and in terms of $J^z$ states is given by $|\psi \rangle = \frac{1}{2}|2\rangle + \frac{i}{\sqrt{2}}|0\rangle +\frac{1}{2}|-2\rangle$ \cite{paramekanti20,maharaj20}. The breaking of time-reversal symmetry at the transition supports this scenario \cite{thompson14, marjerrison16_2}.

%%%%%%%%%%%%%%%%%%%%%%%%%%%%%%%%%%%%%%%%%%%%%%%%%%%%%%%%%%

\subsection{Multipoles and frustration}

%%%%%%%%%%%%%%%%%%%%%%%%%%%%%%%%%%%%%%%%%%%%%%%%%%%%%%%%%%

Often more interesting than those systems that show robust multipolar order, are those that combine multipolar  order with spin-liquid behaviour, or those that avoid multipole ordering and instead form spin liquids with multipolar correlations.
This type of behaviour is associated with frustration, which arises in myriad ways in strongly spin-orbit-entangled systems due to the interplay of lattice geometry with directional-dependent exchange and higher-order biquadratic and bitriatic interactions.

%%%%%%%%%%%%%%%%
\subsubsection{$d^1$ on the FCC lattice}
%%%%%%%%%%%%%%%%

\begin{figure}[h]
\centering
\includegraphics[width=0.4\textwidth]{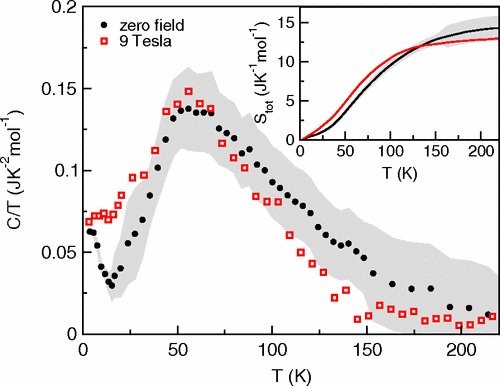}
\caption{
Valence-bond glass formation in Ba$_2$YMoO$_6$.
Heat capacity measurements show no evidence of a phase transition, but do show evidence for a gradual freezing, centred on a broad maximum at about 50 K.
This suggests the formation of an amorphous valence-bond state, with a distribution of triplet excitation energies.
The figure is reproduced with permission from Ref.~[\onlinecite{deVries10}] ($\copyright$2010 the American Physical Society).
}
\label{fig:Ba2YMoO6}
\end{figure}

While not a spin liquid, the double perovskite Ba$_2$YMoO$_6$ does form a valence-bond glass, in which a disordered pattern of spin-singlet dimers freezes at temperatures below about 50 K, as can be seen in Fig.~\ref{fig:Ba2YMoO6} \cite{deVries10,aharen10,carlo11,deVries13}.

One suggestion is that this could be associated with a hidden $SU(2)$ symmetry that can emerge from the complicated and apparently unsymmetric Hamiltonian between $J=3/2$ states \cite{chen10,romhanyi17}.
Solving this Hamiltonian for 2 sites, $i$ and $j$, gives a lowest-energy singlet state $|\psi\rangle = (|1/2\rangle_i |\!\!-\!\!1/2\rangle_j - |\!\!-\!\!1/2\rangle_i |1/2\rangle_j)/\sqrt{2}$, and, extending this to the FCC lattice, results in a degenerate set of singlet dimer coverings with lower energy than any magnetically ordered state \cite{romhanyi17}.
The idea is that in the material a small disorder is responsible for selecting one of the many degenerate dimer configurations, resulting in a dimer glass.
This idea is appealing, and, due to the nature of the excitations above the dimer states, gives an explanation for the experimentally determined soft gap, but there remains the question of whether Jahn-Teller interactions, active in $d^1$ systems, play an important role.

%

%%%%%%%%%%%%%%%%
\subsubsection{$d^2$ on the pyrochlore lattice}
%%%%%%%%%%%%%%%%

\begin{figure*}[tb]
\begin{center}
\includegraphics[width=0.8\textwidth]{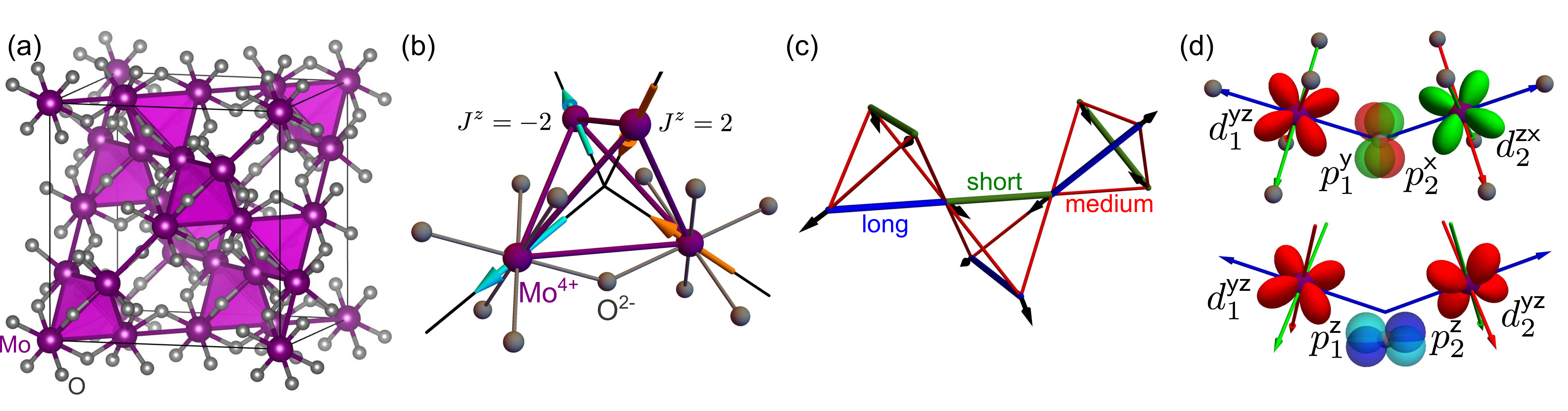}
\caption{
The interplay of spin and lattice degrees of freedom in Y$_2$Mo$_2$O$_7$.
(a) Average positions of Mo and O ions, showing the pyrochlore lattice of Mo ions.
(b) $J_z=\pm2$ states represented as Ising spins pointing along the in/out (local $z$) axes of the Mo tetrahedra.
(c) Mo ions displace into or away from tetrahedral centres, creating long, short and medium length Mo-Mo separations.
(d) Superexchange paths in the neighbouring MoO$_6$ octahedra: (upper part) ``$\pi$-type'' superexchange path that dominates when the Mo's form an undistorted pyrochlore lattice;
(lower part) additional ``$\sigma$-type'' superexchange path that becomes increasingly important the more the Mo-O-Mo bond angle is changed from its average value.
The figures are reproduced with permission from Ref.~[\onlinecite{smerald19}] ($\copyright$2019 the American Physical Society).
}
\label{fig:dpdhopping}
\end{center}
\end{figure*}

The material Y$_2$Mo$_2$O$_7$ provides an example of how spin-glass and potentially spin-liquid physics can emerge out of a quadrupolar phase \cite{greedan86}.
The Mo$^{4+}$ ions form a pyrochlore sublattice and sit in oxygen octahedra that have a large trigonal distortion at all temperatures, with band structure calculations suggesting that the characteristic energy scale of the trigonal splitting is more than 100 meV \cite{shinaoka13} [see Fig.~\ref{fig:dpdhopping}(a)].
When combined with SOC this results in a low energy $J_z=\pm2$ doublet with quadrupolar symmetry, well separated from higher energy states, and with the $z$ axis orientated along the local in/out axes of the Mo tetrahedra, as shown in Fig.~\ref{fig:dpdhopping}(b) \cite{smerald19}.
Since there are no interactions that can transform $J_z=\pm2$ states into one another, the Hamiltonian must be  dominated by Ising interactions between the effective spins \cite{shinaoka13,Shinaoka2019,smerald19}.
This would suggest that either an all-in-all-out ordered state or a spin-ice-like disordered 2-in-2-out configuration should be realised.
However, neutron scattering experiments suggest that spin degrees of freedom alone are insufficient to describe the low-temperature behaviour of the system \cite{silverstein14}.

Evidence for what else needs to be taken into account comes from x-ray and neutron pair distribution analyses, which show that the Mo ions are not forming a perfect pyrochlore lattice, but instead their positions are shifted towards or away from the tetrahedral centres in a disordered 2-in-2-out pattern [see Fig.~\ref{fig:dpdhopping}(c)] \cite{thygesen17}.
The experiments further show that the oxygen octahedra are dragged along by the Mo ions, resulting in very little change in the local crystal-field environment, but large variations in the Mo-O-Mo bond angles, with individual bond angles dependent on the details of the 2-in-2-out lattice displacements.
Deviations from the average Mo-O-Mo bond angle are expected to result in large changes to both the strength and sign of the exchange interactions [see Fig.~\ref{fig:dpdhopping}(d)], resulting in a large coupling between the lattice and spin degrees of freedom and a resulting distribution in the exchange interactions \cite{shinaoka13,smerald19}.
As such, these materials are nice examples of the interplay of SOC with strong magneto-elastic coupling.

At low temperatures Y$_2$Mo$_2$O$_7$ shows spin-glass behaviour \cite{greedan86,gingras97}, and a number of explanations have been put forward to explain this \cite{shinaoka13,silverstein14,smerald19,Shinaoka2019,mitsumoto20}.
One possibility is that the low-temperature spin-glass state freezes out of an intermediate-temperature spin-lattice-liquid state, in which the strong magneto-elastic coupling ties together the spin and lattice degrees of freedom, but the system remains dynamic and explores an extensive set of low-energy configurations \cite{smerald19}.

\section{Spin-orbit-coupled exotic metals and non-trivial topological phases}

In the previous sections, we discussed the spin-orbit-entangled electronic phases in Mott insulators. However, the Mott insulating state of 4$d$ and 5$d$ transition-metal oxides is not so robust and often close to a metal-insulator transition. In fact, metallic ground states are also frequently observed. In the itinerant limit, the spin-orbit-entangled states form bands which may be understood in the framework of $jj$-coupling. The strong SOC of $4d$ and $5d$ electrons can drastically modify the band structure and may give rise to exotic metallic states, potentially with nontrivial topological character. The emergence of exotic phases such as nodal-line semimetals, Weyl semimetals, and topological Mott insulators has been theoretically discussed. Compared to typical topological semimetals composed of $s$ and $p$ electrons, the presence of electron correlations in these oxides with $d$-electrons is expected to provide a distinct physics of correlated topological materials. We review in this section the exotic metallic states in perovskite and pyrochlore iridates, as well as in the doped hyperkagome. In addition to iridates, the recently verified hidden multipolar phase and possible unconventional superconductivity in the pyrochlore rhenate will be discussed.

\subsection{Orthorhombic perovskite AIrO$_3$ (A = Ca, Sr) with Dirac line node}

As discussed in Sec.III.A, the layered perovskites Sr$_2$IrO$_4$ and Sr$_3$Ir$_2$O$_7$ are Mott insulators with localized $J$ = 1/2 pseudospins. In this series of Ruddlesden-Popper perovskite Sr$_{\rm n+1}$Ir$_{\rm n}$O$_{\rm 3n+1}$ (n = 1, 2, $\ldots$), the Ir 5$d$ bandwidth is expected to increase as a function of the number of IrO$_2$ planes, n. SrIrO$_3$, which corresponds to the limit of n = $\infty$, crystallizes in an orthorhombic perovskite with rotation and tilting distortion of IrO$_6$ octahedra (space group $Pbnm$), illustrated in Fig.~\ref{fig:perovskite}(b) \cite{Longo1971}. This orthorhombic perovskite is a metastable phase stabilized under high-pressure or in a thin-film form; at ambient pressure, SrIrO$_3$ crystallizes in a distorted 6H-type perovskite structure \cite{Longo1971}.

The orthorhombic perovskite SrIrO$_3$ was shown to be metallic from the transport and optical properties \cite{JGZhao2008, Moo08}. It is in fact a semimetal with a small carrier density, which is produced by an interplay of crystalline symmetry and strong SOC. If there were no rotations and tilts of IrO$_6$ octahedra, cubic SrIrO$_3$ would have a half-filled $J$ = 1/2 band with a moderate bandwidth. When the rotations and the tiltings of IrO$_6$ are incorporated, the bands are back-folded, and many crossing points in the $J$ = 1/2 bands show up. The incorporation of SOC opens a gap at many of the crossing points, which makes the system close to a band insulator with 20 $d$-electrons per unit cell with four Ir atoms. In reality, the presence of symmetry-protected band crossing and the overlap of split bands give rise to a semimetallic state \cite{Nie2015}.

\begin{figure}[t]
\begin{center}
\includegraphics[scale=0.65]{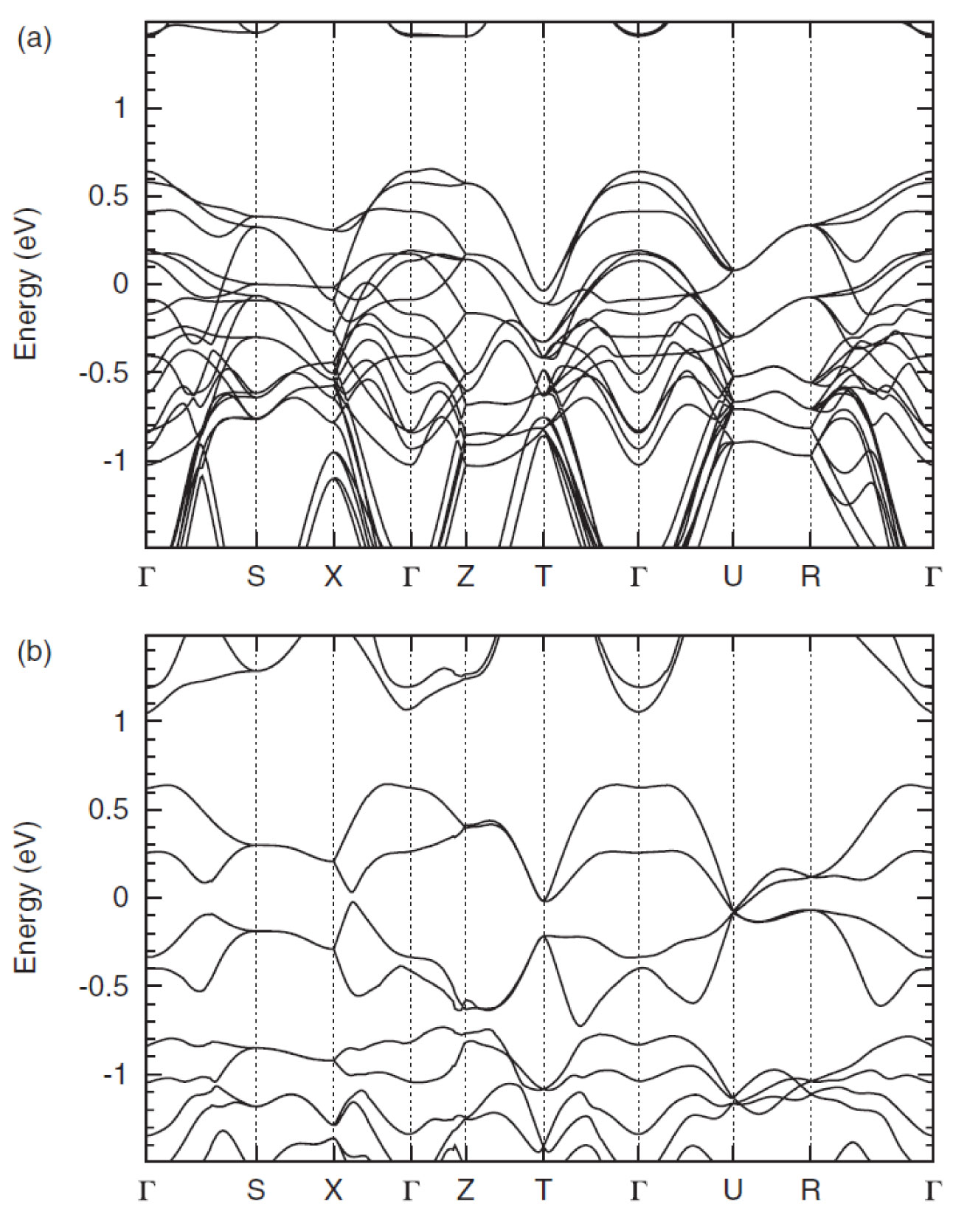}
\caption{Band structures of SrIrO$_3$ obtained from LDA + $U$ calculation with Hubbard $U$ = 2 eV: (a) without SOC, and (b) with SOC $\zeta = 2\zeta_{at}$ ($\zeta_{at}$ is atomic spin-orbit coupling). The figure is reproduced with permission from Ref.~[\onlinecite{Carter2012}] ($\copyright$2012 the American Physical Society).
}
\label{fig:SrIrO3_band}
\end{center}
\end{figure}

The semimetallic band structure of SrIrO$_3$ hosts the Dirac bands near the Fermi energy $E_{\rm F}$, which prevents a gap opening. A density functional theory calculation and a tight-binding analysis showed the two interpenetrating Dirac dispersions around the U-point of the Brillouin zone, which yield a nodal-line [Fig.~\ref{fig:SrIrO3_band}(b)] \cite{Carter2012, Zeb2012}. The Dirac nodes are protected by the nonsymmorphic symmetry of the space group $Pbnm$, which contains two glide symmetries, in addition to space- and time-reversal symmetries \cite{YChen2016, YChen2015}. The Dirac points are located slightly below $E_{\rm F}$, and there are other heavy hole bands crossing $E_{\rm F}$ to retain the charge neutrality. The presence of linearly dispersive electron bands was confirmed by an ARPES measurement of thin-films \cite{Nie2015}. We note that the ambient pressure phase of SrIrO$_3$, crystallizing in a monoclinic $C2/c$ structure, is also a Dirac semimetal protected by the nonsymmorphic symmetry ($c$-glide) \cite{Takayama2018}.

\begin{figure}[t]
\begin{center}
\includegraphics[scale=0.45]{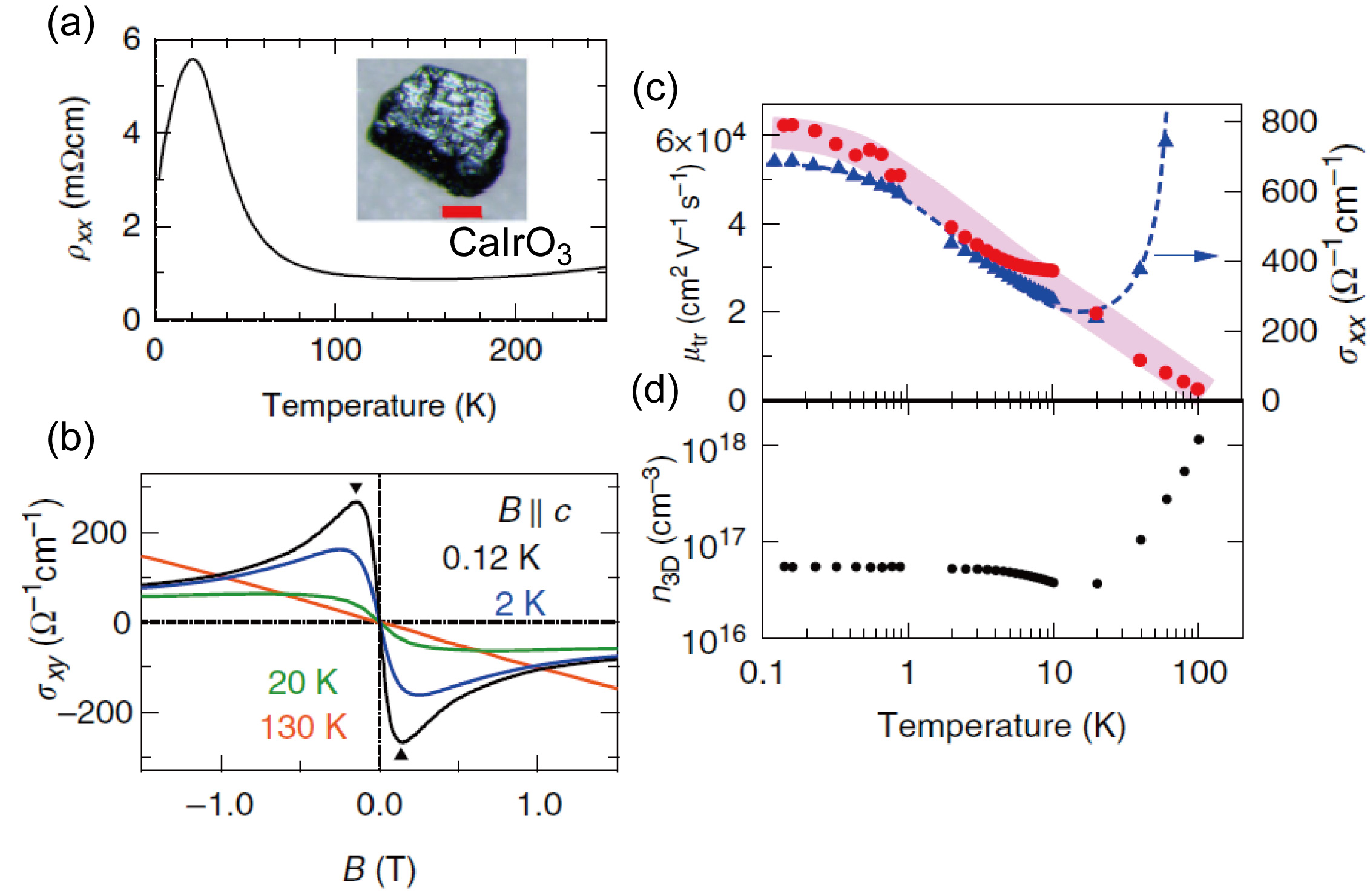}
\caption{Transport properties of orthorhombic perovskite CaIrO$_3$. (a) Temperature dependence of longitudinal resistivity $\rho_{xx}$ (b) Hall conductivity $\sigma_{xy}$ as a function of magnetic field at several temperatures. (c), (d) Temperature dependence of carrier mobility $\mu_{\rm tr}$ and carrier density $n_{/\rm 3D}$, respectively. The huge mobility as large as 60,000 cm$^2$/V$\cdot$s is seen below 1 K. The figure is reproduced from Ref.~[\onlinecite{Fujioka2019}], CC-BY-4.0 (\url{http://creativecommons.org/licenses/by/4.0/}).
}
\label{fig:CaIrO3}
\end{center}
\end{figure}

One of the characteristic features of Dirac semimetals is the presence of highly mobile carriers, which have been indeed identified in the perovskite CaIrO$_3$, isostructural to SrIrO$_3$. A carrier mobility as large as 60,000 cm$^2$/V$\cdot$s is observed at low temperatures, as shown in Fig.~\ref{fig:CaIrO3} \cite{Fujioka2019}. The remarkably high mobility is discussed to be attributed to the proximity of Dirac nodes to $E_{\rm F}$ \cite{Fujioka2019}. Because of the smaller ionic radius of Ca$^{2+}$ as compared to that of Sr$^{2+}$, CaIrO$_3$ inherits larger rotation and tilting of IrO$_6$ octahedra, which reduces the bandwidth and enhances electron correlations. The strong correlation renormalizes the band structure and places the Dirac nodes near $E_{\rm F}$. 

\begin{figure}[t]
\begin{center}
\includegraphics[scale=0.7]{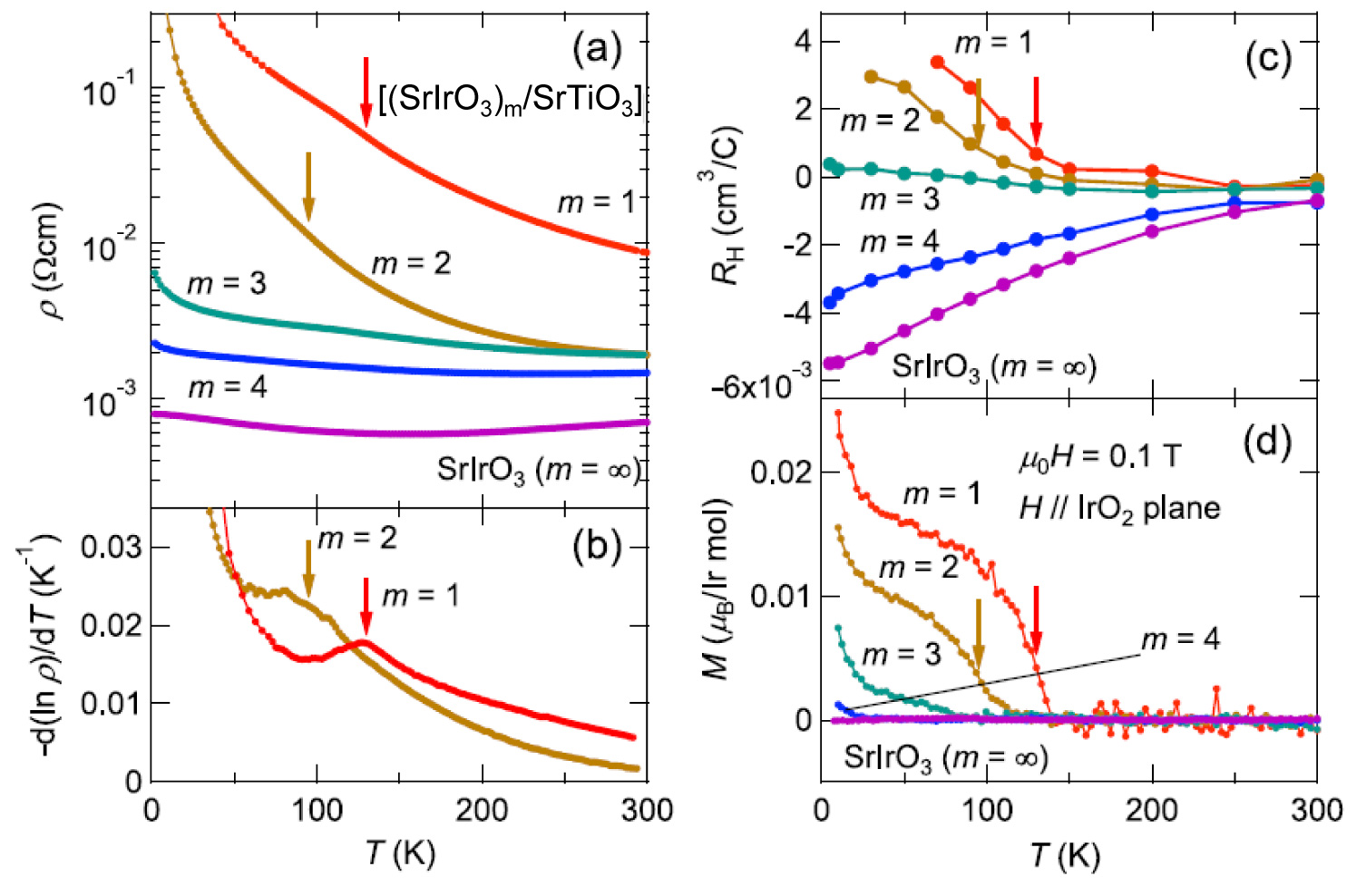}
\caption{Temperature dependent (a) resistivity $\rho(T)$,  (b) $-d({\rm ln}\rho)/ dT$, (c) Hall constant $R_{\rm H}$, and (d) in-plane magnetization $M(T)$ of  (001) superlattice [(SrIrO$_3$)$_m$/SrTiO$_3$] with $m$ = 1, 2, 3, 4 and $\infty$. The figure is reproduced with permission from Ref.~[\onlinecite{Matsuno2015}] ($\copyright$2015 the American Physical Society).
}
\label{fig:SrIrO3_SL}
\end{center}
\end{figure}

It is important to unravel the key factor determining the evolution from a 3D Dirac semimetal to a magnetic insulator in the series of Sr$_{\rm n+1}$Ir$_{\rm n}$O$_{\rm 3n+1}$. In bulk form, Sr$_{\rm n+1}$Ir$_{\rm n}$O$_{\rm 3n+1}$ with n $\geq$ 3 is not stable at ambient pressure and difficult to grow. As an alternative approach to track the metal-insulator transition, a (001) superlattice comprising SrIrO$_3$ and nonmagnetic SrTiO$_3$ layers, i.e.  [(SrIrO$_3$)$_{\rm m}$/SrTiO$_3$], has been designed \cite{Matsuno2015}. By increasing the number of SrIrO$_3$ layers m, the dimensionality, and thus the bandwidth, of SrIrO$_3$ layers can be controlled. The metal-insulator transition takes place at around m = 3 as shown in Fig.~\ref{fig:SrIrO3_SL}(a). The insulating samples with m = 1 and 2 show a magnetic transition with weak-ferromagnetic moments, which are induced by the rotations of IrO$_6$ octahedra about the [001] axis and the resultant DM interaction [Fig.~\ref{fig:SrIrO3_SL}(d)]. The intimate correlation between the metal-insulator transition and the appearance of magnetic order suggests that magnetism is essential for the occurrence of a metal-insulator transition with reducing m.

The nodal-line Dirac semimetallic state of SrIrO$_3$ can be potentially exploited as a platform for other correlated topological phases by the application of symmetry-breaking perturbations such as magnetic field and strain \cite{YChen2015}. In particular, a variety of superlattice structures has been proposed to realize novel topological phases. By introducing a staggered potential that breaks the mirror-symmetry, for example the (001) superlattice of [(SrIrO$_3$)/(SrRhO$_3$)], the appearance of a topological insulator phase is anticipated \cite{Carter2012}. The superlattice of [(SrIrO$_3$)$_2$/(CaIrO$_3$)$_2$] has been predicted to be a topological semimetal hosting a double-helicoid surface state \cite{CFang2016}.

In addition to the (001) superlattices, a topological insulator phase was also predicted from fabricating a bilayer of SrIrO$_3$ along the [111] direction \cite{Xiao2011, Okamoto2018, Lado2013}. In a bilayer of SrIrO$_3$, the IrO$_6$ octahedra form a buckled honeycomb lattice. As in the celebrated graphene, electron hopping on a honeycomb lattice gives rise to Dirac bands. When a trigonal crystal field is incorporated, it opens a gap at the Dirac points, giving rise to a $Z_2$ topological insulator \cite{Xiao2011}. 

In fact, the fabrication of a [111] oriented thin-film is technically challenging in perovskite oxides A$^{2+}$B$^{4+}$O$_3$, since the (111) surfaces, AO$_3$ or B planes, are polar, in contrast to the (001) surfaces of AO or BO$_2$ \cite{TJAnderson2016}. Additional difficulties arise from a size mismatch of SrIrO$_3$ with a standard substrate like SrTiO$_3$ and the stability of monoclinic SrIrO$_3$ with a hexagonal motif on a [111] substrate \cite{Sumi2005}. The stabilization of the orthorhombic phase by optimizing the A-site ion through Ca substition for Sr is quite useful to overcome this difficulty. (111) superlattices of [(Ca$_{0.5}$Sr$_{0.5}$IrO$_3$)$_{\rm 2m}$/(SrTiO$_3$)] with m = 1, 2 and 3 have been successfully fabricated \cite{Hirai2015}. In contrast to the prediction of a topological insulator, the (111) superlattices with m $\leq$ 3 were found to be magnetic insulators and likely trivial. This again points to the importance of magnetism in the superlattices of iridates \cite{Okamoto2018}.

\subsection{Potential topological semimetallic state in pyrochlore iridates}

Since soon after the discovery of spin-orbit-entangled phases in iridates, the pyrochlore iridates A$_2$Ir$_2$O$_7$ (A: trivalent cation) have been attracting tremendous interest, as they provide a unique interplay between SOC, electron correlation and frustration. There have been a plethora of theoretical proposals for non-trivial topological phases, including $Z_2$ topological insulators \cite{Guo2009}, topological Mott insulators \cite{Pesin2010}, Weyl semimetals \cite{XWan2011}, and axion insulators \cite{XWan2011, AGo2012}.

\begin{figure}[t]
\begin{center}
\includegraphics[scale=0.7]{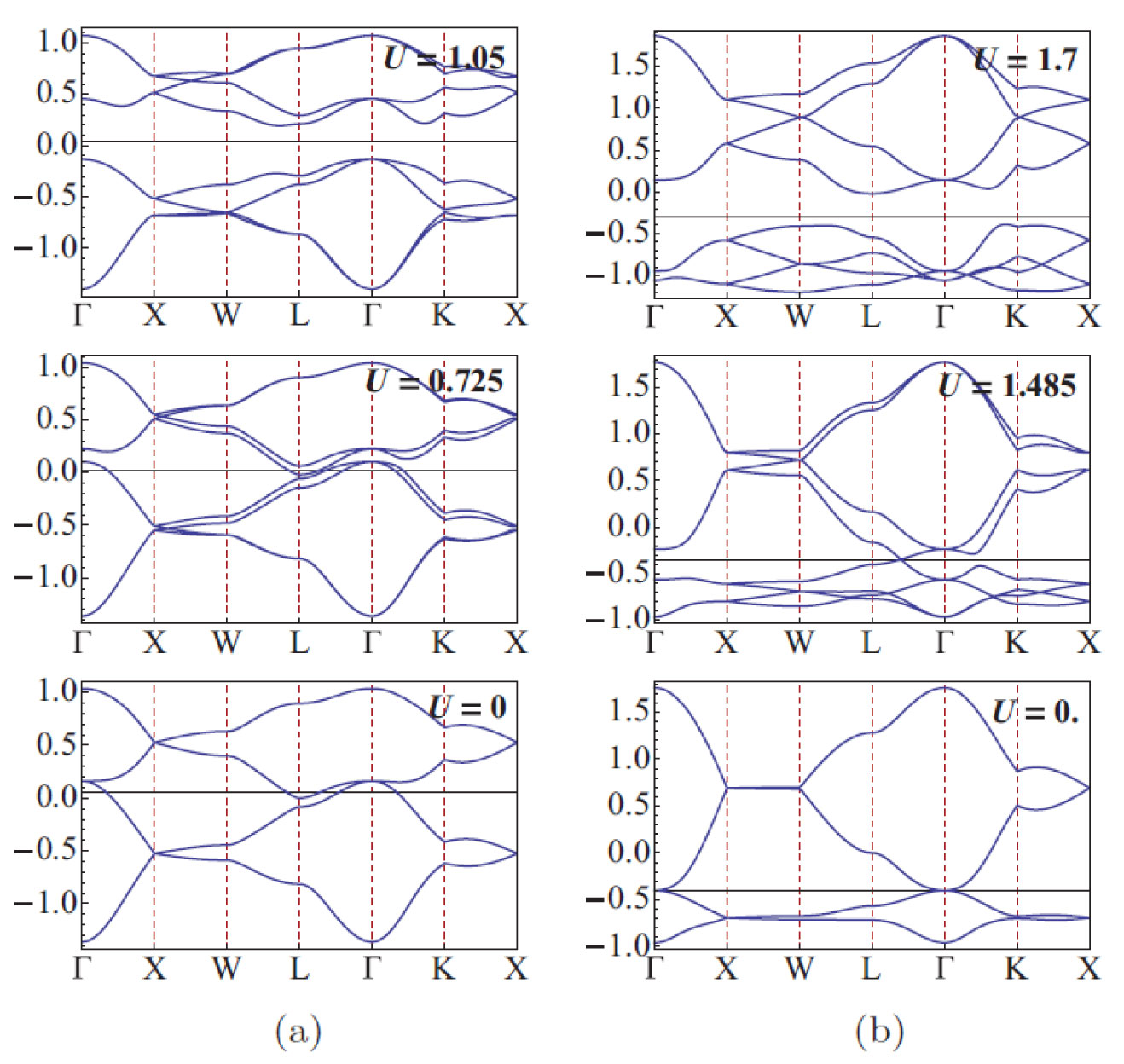}
\caption{Calculated band structures of pyrochlore iridate near Fermi energy with different on-site Hubbard repulsion $U$. The left and right columns show the results for the different magnitude of hopping parameters. The figure is taken with permission from Ref.~[\onlinecite{WKrempa2013}] ($\copyright$2013 the American Physical Society).
}
\label{fig:PI_band}
\end{center}
\end{figure}

The general trend for the electronic structure of pyrochlore iridates has been understood as follows \cite{WKrempa2013}. When the on-site electron correlation $U$ is weak, they show a semimetallic electronic structure. The semimetallic state may contain small pocket Fermi surfaces [Fig.~\ref{fig:PI_band}(a)] or a quadratic band touching point at the $\Gamma$ point near the Fermi energy [Fig.~\ref{fig:PI_band}(b)], depending on the hopping parameters. By increasing the electron correlation, AIAO order of Ir magnetic moments takes place. When the original nonmagnetic state is a semimetal with quadratic band touching, the AIAO order splits the degenerate bands and gives rise to crossings of linearly-dispersing non-degenerate bands. The resultant semimetallic phase is a Weyl semimetal with nodes of opposite chiralities. There are 4 pairs of Weyl nodes along the [111] or equivalent directions in the Brillouin zone. When $U$ is increased further, the Weyl nodes move to the high-symmetry point of the Brillouin zone and the distance between the pair of nodes increases. Eventually, the pair of Weyl nodes with different chiralities meets at the zone boundary and annihilates, which renders a gap over the whole Brillouin zone and makes the system a trivial AIAO antiferromagnetic insulator. 

\begin{figure}[t]
\begin{center}
\includegraphics[scale=0.5]{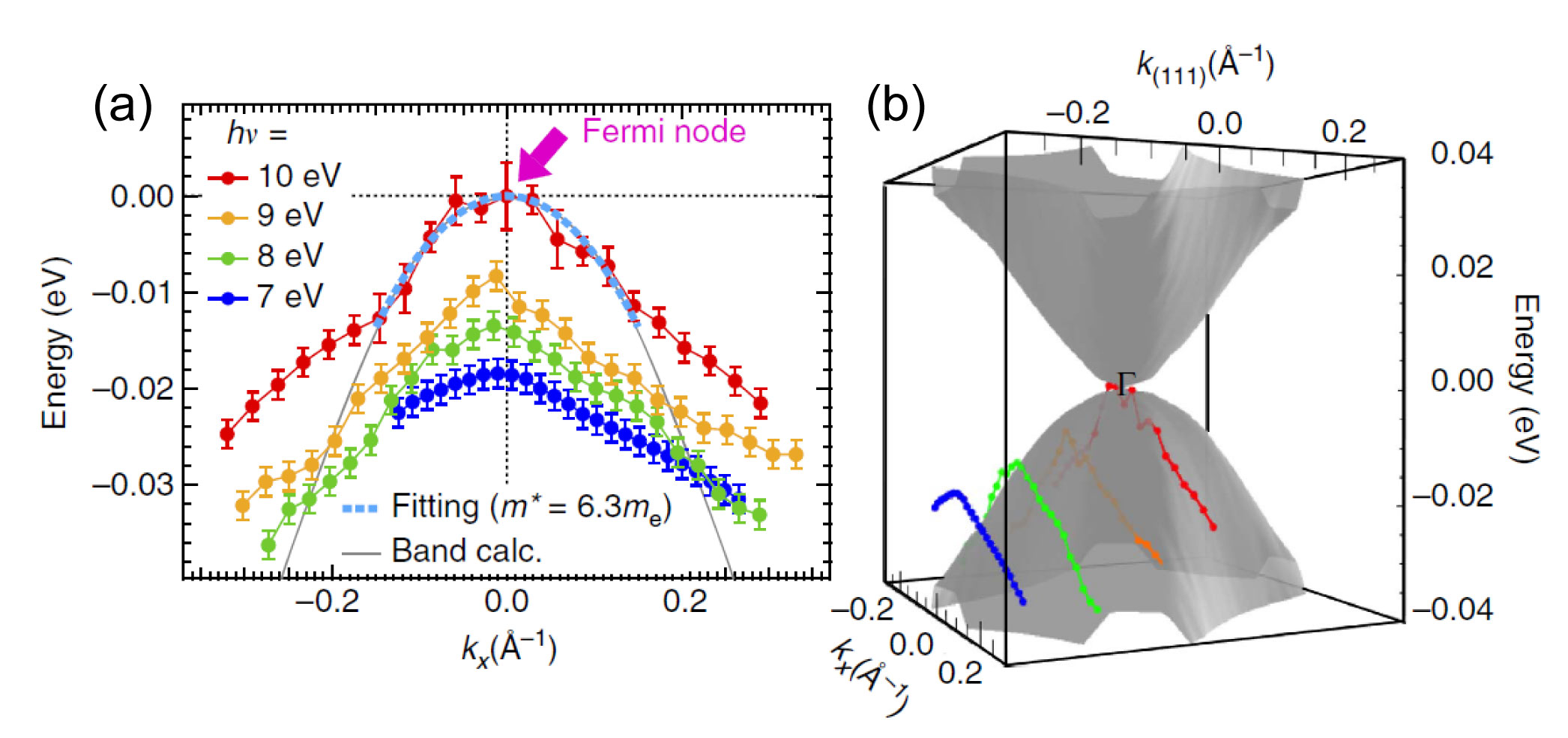}
\caption{Quadratic Fermi node of Pr$_2$Ir$_2$O$_7$ revealed by the ARPES measurement. (a) Energy dispersion along $k_x$ direction measured with different incidence photon energies. (b) The ARPES data in the $k_x - k_{(111)}$ sheet superposed on the calculated band dispersion. The figure is reproduced from Ref.~[\onlinecite{Kondo2015}], CC-BY-4.0 (\url{http://creativecommons.org/licenses/by/4.0/}).
}
\label{fig:Pr2Ir2O7_ARPES}
\end{center}
\end{figure}

In real materials, the relative strength of $U$ can be tuned effectively by changing the bandwidth of Ir 5$d$ states. As described in Sec.III.C.2, the bandwidth is reduced by decreasing the ionic radius of the A-cation, $r_{\rm A}$, i.e. changing the degree of trigonal distortion. Among the family of pyrochlore iridates A$_2$Ir$_2$O$_7$, Pr$_2$Ir$_2$O$_7$, which has the largest $r_{\rm A}$, remains metallic down to the lowest temperature measured. Pr$_2$Ir$_2$O$_7$ shows a poor metallic behavior with a small carrier density of $\sim10^{21}$ cm$^{-3}$ \cite{Machida2007}. The ARPES measurement revealed that Pr$_2$Ir$_2$O$_7$ has a quadratic band-touching at the $\Gamma$ point as shown in Fig.~\ref{fig:Pr2Ir2O7_ARPES} \cite{Kondo2015}. In this nodal semimetallic state, the density of states near $E_{\rm F}$ increases steeply since DOS($E$) $\propto \sqrt{E}$, which results in pronounced electron correlations and potentially leads to a non-Fermi liquid behavior \cite{EGMoon2013}. Another interesting behavior of Pr$_2$Ir$_2$O$_7$ is that Pr$^{3+}$ 4$f$ moments do not show a long-range magnetic order, but instead a spin-liquid-like behavior\cite{Nakatsuji2006}. A finite Hall conductivity was observed at zero magnetic field despite the absence of hysteresis in the magnetization curve, which has been discussed to originate from the chirality of the spin-liquid state \cite{Machida2010}.

With decreasing $r_{\rm A}$, a temperature-driven metal-insulator transition is observed for A = Nd, Sm, and Eu. The metal-insulator transition accompanies the AIAO magnetic order of Ir 5$d$ moments as discussed in Sec.III.C, and thus the low-temperature phase was expected as a possible realization of a Weyl semimetal. However, the presence of a charge gap has been seen at temperatures well below the magnetic ordering temperature $T_{\rm N}$ for Ir 5$d$ moments even in Nd$_2$Ir$_2$O$_7$ \cite{UedaPRL2012, Nakayama2016}, which is right next to Pr$_2$Ir$_2$O$_7$. This is incompatible with the Weyl semimetallic state.  A Weyl semimetal phase might be realized only in the critical vicinity of a metal-insulator transition and therefore hidden. Fine tuning of the metal-insulator transition using pressure or doping may help approaching a Weyl semimetal. Indeed, suppression of the metal-insulator transition was observed by the application of pressure \cite{Sakata2011, Tafti2012} or by doping a small amount of Rh atoms onto the Ir site \cite{UedaPRL2012}, which may stabilize the Weyl semimetallic state.

\begin{figure}[t]
\begin{center}
\includegraphics[scale=0.4]{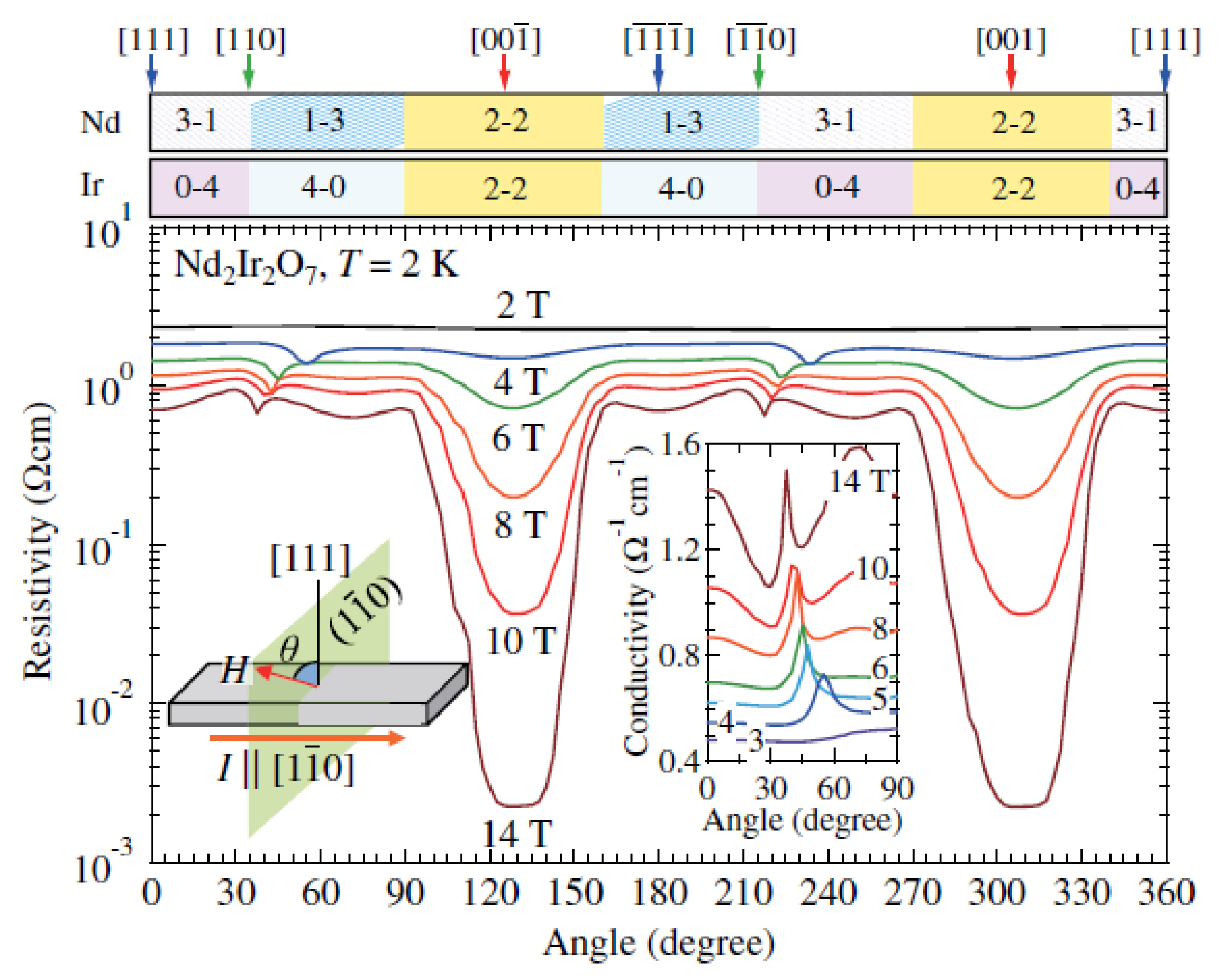}
\caption{Angle-dependent magnetoresistance in Nd$_2$Ir$_2$O$_7$. The tables on the top indicate the magnetic configuration of Nd and Ir sublattices such as AIAO (AOAI) order (0-4 and 4-0), 2-in-2-out configuration  (2-2) and 3-in-1-out state (3-1 or 1-3), respectively. The figure is reproduced with permission from Ref.~[\onlinecite{UedaPRL2015}] ($\copyright$2015 the American Physical Society).
}
\label{fig:Nd2Ir2O7}
\end{center}
\end{figure}

Although the ground state of Nd$_2$Ir$_2$O$_7$ is unlikely to be a Weyl semimetal at ambient conditions, a drastic magnetic-field-induced change of transport properties was discovered, reflecting the modification of Nd$^{3+}$ magnetic order \cite{UedaPRL2015, Tian2016}. When a magnetic field is applied along the [001] direction, the AIAO order of Nd$^{3+}$ 4$f$ moments is switched into the 2-in-2-out configuration above $\sim$10 T. Concomitantly, a drastic drop of resistivity was observed as shown in Fig.~\ref{fig:Nd2Ir2O7}, indicating an insulator to semimetal transition by suppressing the AIAO order of Ir 5$d$ electrons via the $f$-$d$ magnetic exchange. The high-field semimetallic state has been proposed to be a nodal-line semimetal \cite{UedaPRL2015,Ueda2017}. On the other hand, an application of magnetic field along the [111] direction induces the 3-in-1-out order of Nd$^{3+}$ moments, which is discussed to realize another Weyl semimetallic phase \cite{Ueda2017}. 

The putative Weyl semimetallic state in pyrochlore iridates is expected to show characteristic features such as surface Fermi arcs and anomalous Hall effect (AHE). The AHE is associated with the fact that the Weyl nodes can be regarded as a source/sink of Berry curvature. In a bulk pyrochlore iridate, the anomalous Hall conductivity is canceled because of the cubic symmetry \cite{KYYang2011}. However, in a strained thin-film, the cubic symmetry is broken and the emergence of an AHE has been predicted \cite{BJYang2014}. Experimentally, such an AHE was indeed observed in thin-film pyrochlore iridates. For the Pr$_2$Ir$_2$O$_7$ thin-film, this was argued to arise from the strain-induced Weyl semimetallic state with a magnetic order at the surface/interface as well as the breaking of cubic-symmetry \cite{Ohtsuki2019}. The AHE was observed also in the insulating pyrochlores such as Eu$_2$Ir$_2$O$_7$ and Nd$_2$Ir$_2$O$_7$, but was attributed to spin-chirality \cite{Fujita2015} or domain walls \cite{WJKim2018} of AIAO magnetic order, rather than the anomalous conductivity from Weyl nodes.

\subsection{Spin-orbit-coupled semimetal out of the competition with molecular orbital formation}

A metallic state is realized also by carrier doping into spin-orbit-entangled Mott insulators.  In particular, carrier-doping into Sr$_2$IrO$_4$ has been attempted intensively in the search for superconductivity, motivated by the cuprate physics, as discussed in Sec.III.A. A spin-orbit-coupled metallic state induced by carrier-doping was found also in the doped hyperkagome iridate Na$_4$Ir$_3$O$_8$. A sister compound Na$_3$Ir$_3$O$_8$, which shares the same hyperkagome sublattice of Ir atoms was synthesized \cite{Takayama2014}. The chemical formula indicates that Ir has a valence state of Ir$^{4.33+}$, i.e. 1/3-hole doped state of the Na$_4$Ir$_3$O$_8$ Mott insulator.  

\begin{figure}[t]
\begin{center}
\includegraphics[scale=0.45]{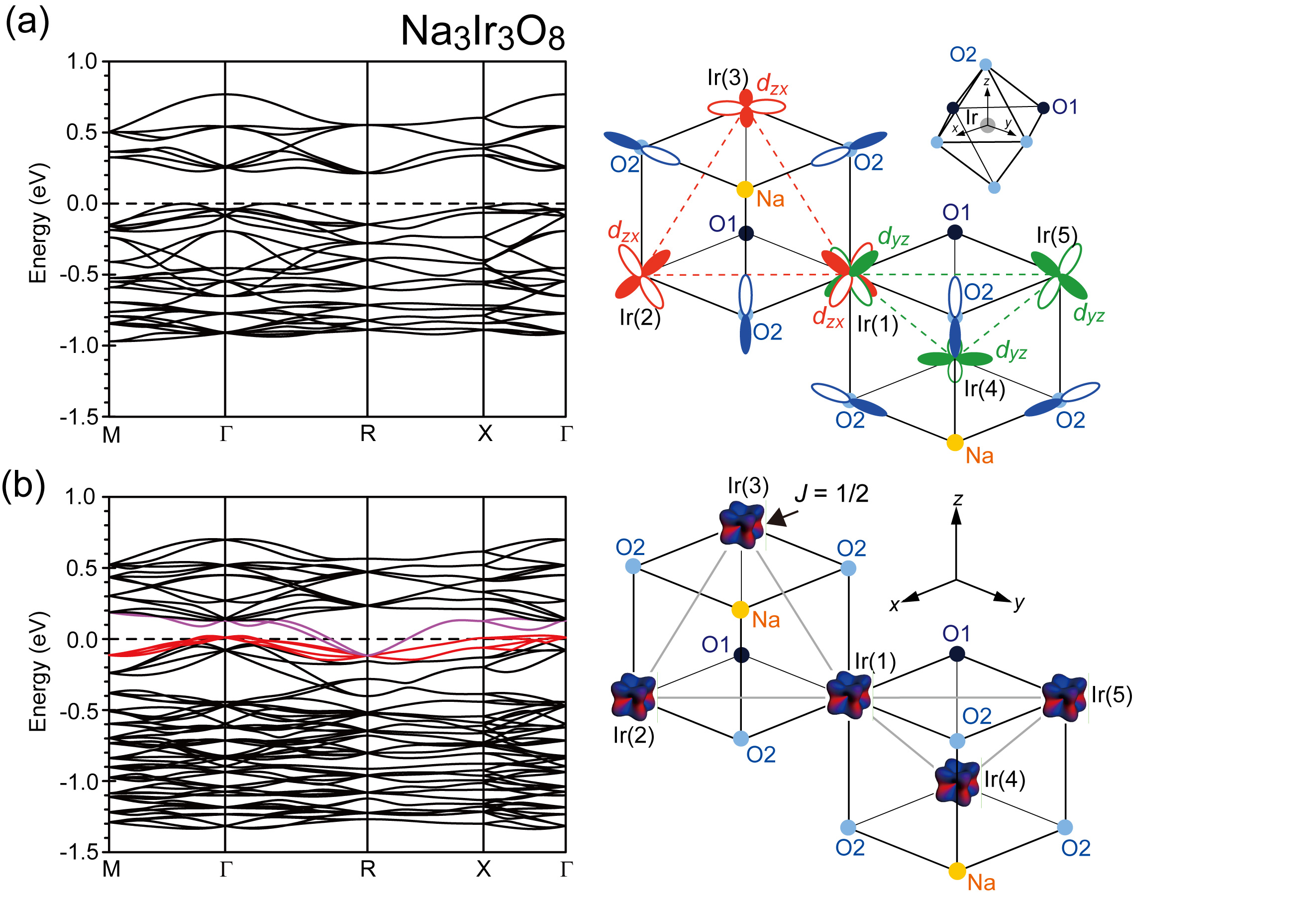}
\caption{Calculated band structure of Na$_3$Ir$_3$O$_8$. (a) Scalar relativistic band structure showing a band insulating state. The right panel illustrates the molecular orbital formation on the Ir hyperkagome lattice. (b) Relativistic band structure including SOC. The bands which form hole and electron
pockets are colored in red and magenta, respectively. The right panel schematically represents the suppression of molecular orbital formation by SOC. The figures are reproduced from Ref.~[\onlinecite{Takayama2014}], CC-BY-4.0 (\url{http://creativecommons.org/licenses/by/4.0/}).
}
\label{fig:Na3Ir3O8}
\end{center}
\end{figure}

Naively, we would expect the 1/3-hole-doped Mott insulator to be a correlated metal with a large Fermi surface. Na$_3$Ir$_3$O$_8$, as well as the sister compound Li$_3$Ir$_3$O$_8$ \cite{Takayama2020}, shows a metallic behavior, but turned out to be a semimetal with a small number of electrons and holes, rather than a large Fermi-surface metal \cite{Takayama2014}. The first-principle calculations indicate that the semimetallic electronic structure is produced by an interplay of molecular orbital formation and SOC. The calculation without SOC yields a band insulator as the ground state of Na$_3$Ir$_3$O$_8$, despite the non-integer number of $d$-electrons per Ir atom. The band insulating state can be understood as the formation of Ir$_3$ trimer molecules with 14 $d$-electrons on the triangular unit of the hyperkagome lattice [Fig.~\ref{fig:Na3Ir3O8}(a)]. The incorporation of SOC suppresses the formation of molecular orbitals by orbital mixing. The conduction and valence bands made out of the molecular orbitals get broader and overlap, giving rise to a semimetallic state with small pockets of Fermi surface. Such a competition between molecular orbital formation and SOC is likely a common feature of 4$d$ and 5$d$ transition-metal oxides with spatially extended $d$-orbitals. Indeed, $J$ = 1/2 magnets often switch into a dimerized state of transition-metal ions, which can be viewed as a molecular orbital formation, for example, in honeycomb-based iridates and in the ruthenium chloride under high pressure \cite{Veiga2017, Takayama2019, Hermann2018,  Biesner2018}.

\subsection{Spin-orbit-coupled metallic state in Cd$_2$Re$_2$O$_7$}

The pyrochlore material Cd$_2$Re$_2$O$_7$ has received much attention in recent years due to the spontaneous breaking of inversion symmetry, and its impact on superconductivity below $T_{\rm c}$ $\sim$ 1 K.
The multitudinous experimental studies of this compound have been well reviewed in Ref.~[\onlinecite{hiroi18}], and we briefly discuss here the basic physical ideas. %principles under investigation, while referring the reader to Ref.~\cite{hiroi18} for a more detailed description of the experimental situation.

\begin{figure}[t]
\centering
\includegraphics[width=0.5\textwidth]{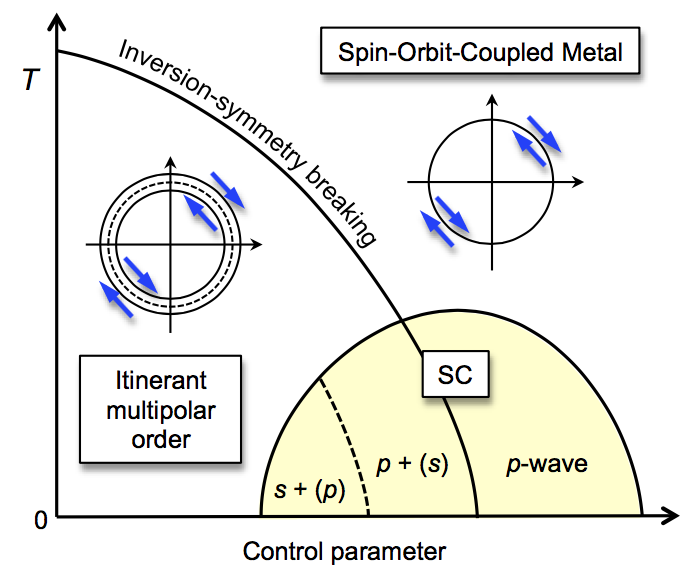}
\caption{
Schematic theoretical phase diagram for spin-orbit-coupled metals.
The breaking of inversion symmetry drives an itinerant multipolar-ordered phase with spin split bands, where the spin orientation is tied to the momentum \cite{fu15}.
As a function of a control parameter, such as pressure, a quantum critical point for inversion-symmetry breaking emerges. 
At around the quantum critical point, a dome-like superconducting phase may be anticipated.
The superconducting dome consists of pure $p$-wave region and mixed $s$- and $p$-wave region.
Exotic topological properties are expected when the $p$-wave pairing dominates \cite{kozii15,wang16}.
The figure is reproduced with permission from Ref.~[\onlinecite{hiroi18}] ($\copyright$2018 The Physical Society of Japan). 
}
\label{fig:SOC_metal_phasediag}
\end{figure}

The most popular theoretical framework for describing Cd$_2$Re$_2$O$_7$ is that of strongly spin-orbit-coupled metals with relatively weak electron-electron interactions \cite{fu15}. 
Starting from the high-temperature metal with intact time-reversal and inversion symmetries, one can consider the possible Fermi-surface instabilities.
These include phases in which inversion symmetry is spontaneously broken, while time-reversal symmetry remains intact, and the result is a deformation and splitting of the Fermi-surface into spin polarised bands, with momentum-dependent spin orientation (see Fig.~\ref{fig:SOC_metal_phasediag}).
Many of these electronic order parameters couple to the lattice, and should therefore drive a structural phase transition.
The instability that may be relevant to Cd$_2$Re$_2$O$_7$ results in a quadrupolar order parameter, and can be thought of as the electron analog of chiral nematic liquid crystals \cite{lubensky02}.

The inversion symmetry breaking instability may open up the possibility of unconventional, odd-parity, topological superconductivity in the vicinity of the associated quantum critical point \cite{kozii15,wang16}.
The superconductivity mediated by the fluctuations of the inversion-symmetry-breaking order parameter can be either pure $p$-wave or mixed $s$- and $p$-wave, where the $s$- and $p$-mixed state comes from distinct superconducting channels developing from the weakly-coupled, SOC-split bands (see Fig.~\ref{fig:SOC_metal_phasediag}).
In the case that the $p$-wave channel is dominant, a topologically non-trivial state is expected, and the topological transition between the $s$- and $p$-wave dominated regions is particularly interesting due to the presence of unusual vortex defects associated with the enlarged symmetry \cite{wang16}.

Experimentally, an inversion-symmetry-breaking structural transition has been observed in Cd$_2$Re$_2$O$_7$ at $T_{\sf s1} \approx 200$ K, while superconductivity sets in at $T_{\rm c} \approx 1$ K \cite{hiroi18}.
Analysis of second-harmonic generation experiments has been used to tease apart the lattice and electronic changes at $T_{\sf s1}$, and suggests that an inversion-symmetry-breaking electronic nematic phases is formed at the transition \cite{harter17}.
For lower temperatures, while at ambient pressure the superconductivity appears to be essentially $s$-wave, pressure can be used to tune the system, increasing both $T_{\rm c}$ and the upper critical field, $B_{\rm c2}$, with the significant increase of the latter taken to indicate the enhancement of the $p$-wave channel \cite{hiroi18}.
While the agreement between theory and experiment is very encouraging, the experimental phase diagram as a function of both temperature and pressure is considerably more complicated than the theoretical predictions, and much work remains on both the theoretical and experimental fronts.
\\

\section{Conclusion}
Correlated electrons in the presence of strong SOC form a rich variety of localized and itinerant spin-orbit-entangled phases in 4$d$ and 5$d$ transition metal compounds. Localized 4$d^5$ and 5$d^5$ systems with $J$ = 1/2 pseudospins have been explored extensively in the last decade, which has established the 4$d$ and 5$d$ transition metal oxides and related compounds as an emergent paradigm in the search for unprecedented quantum phases. The Kitaev model has been shown to be relevant in a family of $d^5$ $J$ = 1/2 honeycomb magnets. Partly motivated by the $J$ = 1/2 physics in the insulating $d^5$ systems, the research effort on $d^1$ - $d^4$ and itinerant systems has become quite active recently. Many attractive spin-orbit-entangled states are anticipated to emerge, including multipolar orderings, excitonic magnetism, a correlated topological insulator, and a topological superconductor. As seen in this review, their potential as a mine of novel electronic phases has not yet been explored fully, particularly for $d^1$ - $d^4$ and itinerant systems.  Concepts have been put forward, but their realization requires the development of novel materials and approaches. Unusual behaviors have been observed in experiments, but understanding the physics behind them requires more elaborate and realistic theories.  Besides, many yet unknown exotic phases likely remain hidden, and are waiting to be unveiled both theoretically and experimentally. We are convinced that the whole family of 4$d$ and 5$d$ correlated oxides and related compounds with strong SOC constitutes a rich mine of novel quantum phases and is worthy of further exploration.

\section*{Acknowledgment}

T.T., A.S. and H.T. were supported by Alexander von Humboldt Foundation. J.Ch. acknowledges support by Czech Science Foundation (GA\v{C}R) under Project No.~GA19-16937S. G.Kh. acknowledges support by the European Research Council under Advanced Grant 669550 (Com4Com).

\bibliographystyle{apsrev4-1}
\bibliography{paper_2021_01_28}

\end{document}